\documentclass[a4paper,12pt]{article}

\usepackage{graphicx}
\usepackage{amsfonts}
\usepackage{amsmath}
\usepackage[usenames]{color}
\usepackage{amssymb}
\usepackage[mathscr]{eucal} 
\usepackage{url}

\usepackage{hyperref}

\usepackage[all]{xy}

\def\N{\mathbb{N}}
\def\Z{\mathbb{Z}}

\def\R{\mathbb{R}}
\def\C{\mathbb{C}}
\def\P{\mathbb{P}}

\begin{document}

\baselineskip 0.6cm
\newcommand{\vev}[1]{ \left\langle {#1} \right\rangle }
\newcommand{\bra}[1]{ \langle {#1} | }
\newcommand{\ket}[1]{ | {#1} \rangle }
\newcommand{\Dsl}{\mbox{\ooalign{\hfil/\hfil\crcr$D$}}}
\newcommand{\nequiv}{\mbox{\ooalign{\hfil/\hfil\crcr$\equiv$}}}
\newcommand{\nsupset}{\mbox{\ooalign{\hfil/\hfil\crcr$\supset$}}}
\newcommand{\nni}{\mbox{\ooalign{\hfil/\hfil\crcr$\ni$}}}
\newcommand{\nin}{\mbox{\ooalign{\hfil/\hfil\crcr$\in$}}}
\newcommand{\Slash}[1]{{\ooalign{\hfil/\hfil\crcr$#1$}}}
\newcommand{\EV}{ {\rm eV} }
\newcommand{\KEV}{ {\rm keV} }
\newcommand{\MEV}{ {\rm MeV} }
\newcommand{\GEV}{ {\rm GeV} }
\newcommand{\TEV}{ {\rm TeV} }

\def\diag{\mathop{\rm diag}\nolimits}
\def\tr{\mathop{\rm tr}}

\def\Spin{\mathop{\rm Spin}}
\def\SO{\mathop{\rm SO}}
\def\SU{\mathop{\rm SU}}
\def\U{\mathop{\rm U}}
\def\Sp{\mathop{\rm Sp}}
\def\SL{\mathop{\rm SL}}

\def\change#1#2{{\color{blue}#1}{\color{red} [#2]}\color{black}\hbox{}}

\begin{titlepage}
  
\begin{flushright}
 IPMU16-0055
\end{flushright}
 
 \vskip 1cm
\begin{center}
 
 {\large \bf Heterotic--Type IIA Duality and Degenerations of K3 Surfaces}

\vskip 1.2cm
 
Andreas P. Braun$^1$ and Taizan Watari$^2$
  
\vskip 0.4cm
  
 {\it $^1$ Mathematical Institute, University of Oxford, \\ Andrew Wiles Building, Woodstock Rd, Oxford OX2 6GG, UK 
 \\[2mm]
     $^2$Kavli Institute for the Physics and Mathematics of the Universe, 
    University of Tokyo, Kashiwa-no-ha 5-1-5, 277-8583, Japan
 }

\vskip 1.5cm
   
\abstract{
We study the duality between four-dimensional ${\cal N}=2$ compactifications of heterotic and type IIA string theories. 
Via adiabatic fibration of the duality in six dimensions, type IIA string theory compactified on a K3-fibred Calabi-Yau 
threefold has a potential heterotic dual compactification. This adiabatic picture fails whenever the K3 fibre degenerates into multiple components over points in the base of the fibration. Guided by monodromy, we identify such degenerate K3 fibres as solitons generalizing the NS5-brane in heterotic string theory. The theory of degenerations of K3 surfaces can then be used to find which solitons can be present on the heterotic side. Similar to small instanton transitions, these solitons escort singular transitions between different Calabi-Yau threefolds. Starting from well-known examples of heterotic--type IIA duality, such transitions can take us to type IIA compactifications with unknown heterotic duals.
}

\end{center}
\end{titlepage}
 
 \tableofcontents

\section{Introduction}

The duality between heterotic string theory and Type II string theories hints at 
non-trivial relations between seemingly totally unrelated mathematical objects. For example, 
heterotic string compactifications involve gauge field moduli, whereas only the compactification geometry must be specified for
Type II compactifications. This article addresses some aspects of the duality dictionary of heterotic--type IIA duality in four dimensions. 

Heterotic--type IIA duality in 4D \cite{KV} is not only a historical precursor to heterotic--F-theory duality 
at 6D \cite{MV-1,MV-2}. The former can also be regarded as a more general version of the latter, and moreover, we expect to 
formulate the duality in terms of world-sheet string theory. 

The study of duality begins with finding a correspondence between discrete data of the compactifications on both sides, followed by an
identification of the moduli. At the level of discrete data, however, we must say that the correspondence 
remains to be understood very poorly even today, apart from a few cases that have been studied in the context of 
heterotic--F-theory duality in 6D. This article intends to provide a survey on the status of understanding on this 
problem, and also to make a little progress.

Starting with the heterotic--type IIA duality in six dimensions, the key principle in understanding the correspondence 
of discrete data is the idea of adiabatically fibering the dual six dimensional theories over a base $\P^1$ \cite{Ferrara:1995yx, KLM, VW-95}. 
Armed with this principle, the problem of discrete data correspondence roughly splits into two fronts. One is to find out the 
variety in fibering the duality 
at higher dimensions without violating the adiabaticity; this is the subject of Section \ref{sec:discr-dictnry} in this 
article. There are often multiple adiabatic fibrations of the duality for a given pair of lattices
$\widetilde{\Lambda_S} \oplus \Lambda_T \subset {\rm II}_{4,20}$. We introduce an approach to use the hypermultiplet moduli 
information to distinguish multiple adiabatic fibrations from one another. This approach is used, for example, to indicate
that the heterotic $ST$-model is dual to the type IIA compactification on $(12) \subset W\P^4_{[1:1:2:2:6]}$, 
while three other candidate compactifications\footnote{All three are qualified candidates, so far as the values of 
$h^{1,1}$, $h^{2,1}$ and the Gromov--Witten invariants are concerned.} of type IIA are excluded because of the hypermultiplet 
moduli information. 

The other front is to study how the adiabaticity condition can be violated, and how to maintain the duality correspondence 
in the presence of such a violation. We address this question in Sections \ref{sec:soliton-4D} 
and \ref{sec:6D}. Some background material from mathematics (Kulikov's theory of degenerations of K3 surfaces) is reviewed in 
Appendix \ref{sec:K3-degen-review}. A degeneration of a K3 fibre in a type IIA compactification should be regarded as a soliton 
in the heterotic dual. The variety in degenerations of a K3 fibre translates into the variety of generalizations of the NS5-brane
in heterotic string theory. The classification theory of degenerations of lattice-polarized K3 surfaces indicates which pairs 
of solitons can (co)exist in a BPS configuration. Furthermore, we can learn about the phase structure of the moduli space of
solitons from the phase structure of the K\"ahler cone of the Calabi-Yau threefolds on the type IIA side.

Apart from the issues discussed in Appendix \ref{sec:K3-degen-review}, we do not try to make this article strictly self-contained.  
The review article \cite{Aspinwall} contains a lot of useful material about K3 surfaces. A review of those parts of lattice 
theory which are heavily used in the study of K3 surfaces can be found e.g. in \cite{BKW-K3}. We use the same notation 
as in \cite{BKW-K3}, and mostly only offer brief explanations here. Similarly, we use the same notation 
as in \cite{BW-H22} for toric geometry. We refer to \cite{BW-H22} for definitions and explanations concerning the methods of toric 
geometry used in this article.

\section{A Quick Review}
\label{sec:quick}

Type IIA string theory compactified on certain Calabi--Yau threefolds $M$ are known to be dual 
to certain compactifications of heterotic string theory preserving ${\cal N}=2$ supersymmetry in four dimensions \cite{KV}. 
This D=4 heterotic--IIA duality is best understood as an adiabatic fibration \cite{Ferrara:1995yx,KLM, VW-95} of 
the D=6 heterotic--IIA duality, where a dual pair is formed of a Narain compactification of the heterotic string on $T^4$ and 
a K3 compactification of the type IIA string \cite{Het-IIA-early, Het-mod-grp, AM-94, modern-Het-IIA-6D-1, HS}.
For type IIA compactification, we hence focus on non-singular Calabi--Yau threefolds $M$ that admit a K3-fibration morphism,
\begin{align}
  \pi_M: M \longrightarrow \P^1_A \, .
\end{align}
A generic fibre $S_{t.A} := \pi_M^{-1}(t)$ for $t \in \P^1_A \backslash \Delta$ ($\Delta$ is a set containing a finite number of 
points in $\P^1_A$) is a non-singular K3 surface, the subscript $_A$ for the base $\P^1$ and the fibre K3 $S_t$ 
is a mnemonic for their use in type IIA compactification. 

To a Calabi--Yau threefold $M$ with a K3-fibration morphism $\pi_M$ we can naturally associate two lattices: 
the Neron--Severi (NS) lattice ${\rm NS}(S_{t.A})$ (rank $\rho$) of a generic K3 fibre, and the
lattice polarization $\Lambda_S$ (rank $r$) of the K3 fibration. A K3-fibration morphism 
$\pi_M$ is said to be $\Lambda_S$-polarized, when a set of divisors $\{D_1,\cdots, D_r\}$ of $M$ are restricted to 
a generic fibre $S_{t.A}$ to generate a sublattice $\Lambda_S$ of ${\rm NS}(S_{t.A})$.

The orthogonal complement of $\Lambda_S$ in $H^2(S_{t.A}; \Z) \cong {\rm II}_{3,19}$---denoted by $\Lambda_T$---is well-defined (regardless of $t \in \P^1_A$) up to lattice isometry. Since the generators of $H^0(S_{t.A})$ and $H^4(S_{t.A})$ of the fibre remain 
well-defined over the base $\P^1_A$, we can replace the lattice $\Lambda_S$ by $\widetilde{\Lambda}_S:=U[-1] \oplus 
\Lambda_S$. The pair of lattices  
\begin{equation}
  \widetilde{\Lambda}_S \oplus \Lambda_T \subset {\rm II}_{4,20}
\end{equation}
can be used to classify Calabi--Yau threefolds fibred by a lattice-polarized K3 surface used for type IIA compactification. 
Heterotic duals of such type IIA vacua are obtained by adiabatically fibering the Narain compactification over 
the base $\P^1_{\rm Het}$. The central charge 
\begin{align}
  (k^8+ik^9): {\rm II}_{4,20} \supset \widetilde{\Lambda}_S \longrightarrow \C
\end{align}
remains non-trivial and invariant over the base $\P^1_{\rm Het}$, while 
$(k^6+ik^7)$ takes values in $\Lambda_T \otimes \C$ and is allowed to 
undergo monodromy transformations over the base $\P^1_{\rm Het}$. 

An intuitive heterotic description is available for such dual pairs of vacua whenever 
\begin{align}
  \Lambda_S \cong U \oplus W^{(0,\rho-2)}, \qquad  \Lambda_T \cong R^{(0,18-\rho)} \oplus U \oplus U,
  \label{eq:Het-intu}
\end{align}
where $W$ and $R$ are some even negative-definite lattices. In particular, $R$ is regarded 
as an overlattice of an ADE root lattice.\footnote{We adopt the 'geometrical' convention in this article in which 
ADE lattices are negative definite. } When the first condition above is satisfied, the threefold for 
heterotic string compactification can be regarded as $T^2 \times {\rm K3}_{\rm Het}$. A supergravity 
description is available (for most of the moduli space), because we can take the 
volume of $T^2$ to be parametrically larger than $\alpha'$; even a lift to Het--F duality at 6D is possible in this case. 
When the second condition above is satisfied, we can think of the value of $(k^6+ik^7)$ varying over the base $\P^1_{\rm Het}$ as an
instanton in the gauge group corresponding to the algebra $R$.  

Even when those two conditions are not satisfied, however, it is common belief that such heterotic vacua do indeed exist. 
We can provide a zero-th order approximation of what those compactifications are by using the language of adiabatic 
fibrations of Narain moduli. When there are points in the base $\P^1$ where the adiabatic argument fails, extra care needs to be taken. 
This is the subject of Section \ref{sec:soliton-4D} in this article. 

\subsection{Examples of Algebraic K3 Surfaces}

In this section we collect some results about algebraic K3 surfaces, impatient readers 
are recommended to directly proceed to section \ref{sec:discr-dictnry}.

To an algebraic K3 surface, we can assign its Neron--Severi lattice ${\rm NS}_{\rm K3}$ (signature $(1,\rho-1)$) and its 
transcendental lattice $T_{\rm K3}$. Conversely, to see which pair of lattices $(NS_{\rm K3}, T_{\rm K3})$ 
can be realized for some algebraic K3 surface, Morrison's theorem \cite{Mor-k3-pic} is useful.
\begin{itemize}
\item Any even lattice ${\rm NS}_{\rm K3}$ with signature $(1,\rho-1)$ with $\rho \leq 10$ can be realized as the 
Neron--Severi lattice of an algebraic K3 surface, and furthermore, the corresponding lattice $T_{\rm K3}$ is 
determined uniquely (modulo isometry) for a given such ${\rm NS}_{\rm K3}$. 
\item Any even lattice $T_{\rm K3}$ with signature $(2,20-\rho)$ with $12 \leq \rho$ can be realized as the 
transcendental lattice of an algebraic K3 surface, and furthermore, 
the lattice ${\rm NS}_{\rm K3}$ is determined uniquely (modulo isometry) for a given such $T_{\rm K3}$.
\end{itemize}
Thus, lattice polarizations with low $\rho$ are best worked out by classifying even signature $(1,\rho-1)$ lattices 
$\Lambda_S$ modulo lattice isometry, and those with high $\rho$ by classifying even signature $(2,20-\rho)$ lattices 
$\Lambda_T$ modulo isometry. Clearly, only some algebraic K3 surfaces listed up in this way satisfy the 
conditions (\ref{eq:Het-intu}). 

Here, we list up a few choices of ${\rm NS}_{\rm K3} \oplus T_{\rm K3}$, some of which are used in the discussion later. 

{\bf Picard number 1 cases} are classified simply by the degree, 
\begin{align}
  {\rm NS}_{\rm K3} \cong \vev{2k}, \qquad k = 1,2,\cdots.
\end{align}
The signature of the Neron--Severi lattice is $(1,0)$. The degree $2k$ is an arbitrary even 
positive integer, without an upper limit in the value. The theorem above guarantees that 
\begin{align}
  T_{\rm K3} \cong \vev{-2k} \oplus U^{\oplus 2} \oplus E_8 \oplus E_8.  
  \label{eq:Tk3-rho=1}
\end{align}
Only the case $k=1$ satisfies the second condition of (\ref{eq:Het-intu}), because $\vev{-2} = A_1$.

The degree $2k=2$ case has a realization in the form of a double cover over $\P^2$, ramified over a sextic curve. The degree $2k=4$ case is realized by the quartic K3 in $\P^3$. The degree $2k=6$ and degree $2k=8$ cases are realized in the form of complete intersections, $(2) \cap (3) \subset \P^4$ and $(2) \cap (2) \cap (2) \subset \P^5$, respectively. For higher degrees, the construction becomes more involved, see \cite{Huy-Mukai} for further information. 

$E_8${\bf -elliptic} K3 is an example with $\rho=2$.
\begin{align}
 {\rm NS}_{\rm K3} = U, \qquad 
 T_{\rm K3} = U^{\oplus 2} \oplus E_8 \oplus E_8 \cong {\rm II}_{2,18}.
\end{align}
This family of K3 surfaces is elliptically fibred with the elliptic fibre given in the form of a Weierstrass model,
embedded in the ambient space $WP^2_{[1:2:3]}$. There is a unique choice of fibration of $WP^2_{[1:2:3]}$ over $\P^1$ realizing 
an elliptic Weierstrass $K3$ surface as a hypersurface.

There are infinitely many cases with {\bf Picard number 2}. The intersection form of the Neron--Severi lattice of any 
$\rho = 2$ case can be written as
\begin{align}
 {\rm NS}_{\rm K3} = \left[ \begin{array}{cc} 2a & b \\ b & 2c \end{array} \right], \quad a,b,c \in \Z, \quad 4ac - b^2 < 0,
\end{align}
once a basis is chosen. The determinant $(4ac-b^2)$ is independent of the choice of basis.\footnote{
The discriminant form $(G,q)$ is a better invariant for the classification of lattices modulo isometry. 
Even the discriminant group $G$ alone, taken in the standard form $G \cong \Z_{n_1} \times \Z_{n_2}$ with $n_1|n_2$ 
for the $\rho =2$ cases, provides more detailed information than just the value of their product $n_1n_2=b^2-4ac$. The $\rho =2$ cases 
with an $E_7$-elliptic K3 surface ($n=0,1$ in (\ref{eq:E7-ell-int-form-rho2})), however, provide an example where the discriminant form $q$ is necessary for distinction between them.} There is no upper limit for the value of $(b^2-4ac)>0$, 
although $(b^2-4ac) \equiv 0, 1$ mod 4. Among these infinitely many algebraic K3 surfaces with $\rho =2$, just nine are 
realized as a generic hypersurface of a 3D toric variety. Only the $(b^2-4ac)=1$ case corresponds to the $E_8$-elliptic K3 surface; more generally, an algebraic K3 surface with $\rho = 2$ admits an elliptic fibration morphism to $\P^1$ if and only if $(b^2-4ac)=D^2$ for some integer $D \in \N$; this elliptic fibration has a $D$-section (cf \cite{PSS}).

$E_7${\bf -elliptic} K3 surfaces come with a choice. For this class of elliptic fibrations, the elliptic fibre 
curve is embedded in $WP_{[1:1:2]}^2$ and there is a choice we can make in fibering this ambient space over $\P^1$. Let the toric 
vectors of the ambient space be 
\begin{align}
  (\nu^1, \nu^2, \nu^3, \nu^4, \nu^5) = 
  \left[ \begin{array}{ccc|cc}
     1 &   & -2 & -1 & (-1 -n) \\
       & 1 & -1 &  0 &    -n  \\
      \hline
       &   &    & -1 &   1 
  \end{array} \right] \in \Z^{\oplus 3}, \qquad n=0,1,2.
\end{align}
The intersection form of the NS lattice is 
\begin{align}
 \left( \begin{array}{cc} 
   -2n & 2 \\ 2 & \end{array} \right)
  \label{eq:E7-ell-int-form-rho2}
\end{align}
in the basis of $\{ D_{\nu_3}, D_{\nu_4} \}$ for $n = 0,1$. A basis change $D_{\nu_3} \rightarrow D_{\nu_3}+D_{\nu_4}$ changes 
the upper-left entry $-2n$ by $4$, so that the two cases $n=0,1$ cannot be the same. For these two cases, the K3 surface has 
the $E_7$-elliptic curve as the fibre and there is a 2-section realized by the divisor $D_{\nu_3}$. 

In the $n=2$ case, however, the $\nu_3$ vector is not a vertex of the polytope in $\Z^{\oplus 3} \otimes \R$, 
but an interior point of an edge $\vev{\nu^4, \nu^5}$ of the polytope. The dual face (an edge) has one interior point so that $\rho = 3$; 
the 2-section obtained as the divisor $D_{\nu_3}$ now consists of two irreducible pieces, each of which provides an ordinary section. 
The intersection form of the NS lattice is 
\begin{align}
 \left( \begin{array}{ccc} 
     & 1 & 1 \\
   1 & -2&   \\
   1 &   & -2 
  \end{array} \right)    \cong U \oplus A_1[2] \cong {\rm NS}_{\rm K3} , \qquad 
  T_{\rm K3} \cong  \vev{+4} \oplus U \oplus E_8\oplus E_8.
\end{align}

$E_6${\bf -elliptic} K3 also comes with a choice. These are characterized by using $WP_{[1:1:1]}^2 = \P^2$ as the ambient space of the 
elliptic fibre. When we choose the toric ambient space to be given by 
\begin{align}
 (\nu^1, \nu^2, \nu^3, \nu^4, \nu^5) =
   \left[ \begin{array}{ccccc}
      1 &   & -1 &    & a \\
        & 1 & -1 &    & b \\
   \hline
        &   &    & -1 & 1
   \end{array} \right] \in \Z^{\oplus 3}, 
\end{align}
the K3 surface will have $\rho =2$, and the divisor $D_{\nu_3}$ in the K3 surface provides a 3-section over 
the base $\P^1$ for most of the ten possible choices of $(a,b)$. When we choose $(a,b) = 2(1,0)$, $2(0,1)$ 
or $2(-1,-1)$, however, $\nu^1$ (or $\nu^2$ or $\nu^3$) is an interior point of an edge of the 3D polytope 
in $\Z^{\oplus 3} \otimes \R$ and its dual face (an edge) contains two interior points. We hence have a $\rho=4$ family of K3 surfaces. 
The intersection form of the NS lattice is given by 
\begin{align}
  \left( \begin{array}{cccc} 
       & 1 & 1 & 1 \\
     1 & -2&   &   \\
     1 &   &-2 &   \\
     1 &   &   & -2
  \end{array} \right)  \cong U \oplus A_2[2] \cong {\rm NS}_{\rm K3}, \qquad 
T_{\rm K3} \cong A_2[-2] \oplus E_8 \oplus E_8 .
\end{align}

A series of choices of $({\rm NS}_{\rm K3}, T_{\rm K3})$, which are discussed in \cite{Aldazabal, MV-1,MV-2} in the context of 
F-theory/heterotic duality, is 
\begin{align}
  {\rm NS}_{\rm K3} \cong U \oplus R_{\rm vis},  \qquad T_{\rm K3} \cong U^{\oplus 2} \oplus E_8 \oplus R_{\rm str}, 
\end{align}
where $(R_{\rm vis}, R_{\rm str})$ are (none, $E_8$), $(A_2, E_6)$, $(D_4,D_4)$, $(E_6,A_2)$, $(E_7,A_1)$ and 
($E_8$, none) and the conditions (\ref{eq:Het-intu}) are satisfied.  
When this series of $({\rm NS}_{\rm K3}, T_{\rm K3})$ is used for heterotic--type IIA duality in four-dimensions 
(as stated in section \ref{sec:quick}), the heterotic string description of the dual vacua is a $T^2 \times {\rm K3}_{\rm Het}$
compactification with $R_{\rm vis}$-valued Wilson lines on $T^2$ and $(R_{\rm str} + E_8)$-valued instantons 
on ${\rm K3}_{\rm Het}$. Note also that the $(R_{\rm vis}, R_{\rm str}) = (A_1, E_7)$ and $(A_2,E_6)$ choices in this series
are not the same as the $E_7$-elliptic and $E_6$-elliptic K3 cases above.

\section{Choices of Lattice-Polarized K3 Fibration and Duality}
\label{sec:discr-dictnry}

In this section, we focus our attention to K3-fibred Calabi--Yau threefolds where 
\begin{itemize}
 \item[a)] The lattice polarization of the fibration, $\Lambda_S$, is equal to ${\rm NS}_{\rm K3}$ (not a proper subset).
 \item[b)] The fibre K3 surface remains irreducible everywhere over the base $\P^1_A$.
\end{itemize}
This is where the adiabatic argument has full strength. To get started, we use toric hypersurface constructions to illustrate how often these two conditions are satisfied. Once a pair of lattices ${\rm NS}_{\rm K3}=\Lambda_S$ and $T_{\rm K3}=\Lambda_T$ is chosen, there are still 
discrete choices to be made in how to take the corresponding algebraic K3 surface into the fibre to form a threefold $M$ for 
type IIA compactification. We find, towards the end of section \ref{ssec:fibr-option-general}, that there are multiple 
choices for many pairs of $(\Lambda_S, \Lambda_T)$. This general phenomenon motivates a case study in section \ref{ssec:case-deg2}.

\subsection{Discrete Choices in K3 Fibrations} 
\label{ssec:fibr-option-general}

As a preparation for later in this article, let us first consider one of the best known cases:  
an $E_8$-elliptic K3 surface (${\rm NS}_{\rm K3}=U$) as the generic fibre. 
An $E_8$-elliptic K3 surface can be constructed as a hypersurface of a toric ambient space whose toric vectors 
are given by 
\begin{align}
 (\nu^1_F,  \nu^2_F,  \nu^3_F ,  \nu^4_F) = 
 \left( \begin{array}{cccc} 1 & & -2 & -2 \\ & 1 & -3 & -3 \\ & & -1 & 1 \end{array} \right), 
\qquad   \nu^{i=1,2,3,4}_F \in \Z^3 = N_F.
\end{align}
Let $\widetilde{\Delta}_F$ be the polytope in $N_F \otimes \R$ spanned by the four vertices above. 

In order to obtain a $K3$-fibred Calabi--Yau threefold $M$ with such a K3 surface in the fibre, 
we construct an appropriate toric ambient space as follows. Consider a toric variety given by the toric vectors 
\begin{align}
 \nu^{1,2,3,4} = \left( \begin{array}{c} \nu^{1,2,3,4}_F \\ 0 \end{array} \right), \qquad 
 \nu^5 = \left( \begin{array}{c} 0 \\ -1 \end{array} \right), \qquad 
 \nu^6 = \left( \begin{array}{c} \nu_F^6 \\ +1 \end{array} \right) \qquad 
  \nu^{i=1,\cdots,6} \in \Z^{\oplus 4} = N, 
\label{eq:fibre-toric}
\end{align}
where $\nu^6_F$ is chosen so that 
\begin{align}
  \nu^6_F \in (2 \widetilde{\Delta}_F) \cap N_F. 
\label{eq:cond-convex}
\end{align}
This choice secures that the polytope $\widetilde{\Delta}$ which forms the convex hull of $\nu^1 \cdots \nu^6$
is reflexive. A Calabi--Yau hypersurface $M=M_U$ of such a toric ambient space has the $E_8$-elliptic K3 surface 
over generic points in the base $\P^1_A$. In this article, we often use $\Lambda_S$ or ${\rm NS}_{\rm K3}$ 
in the subscript, as in $M_U$, to have the lattice polarization of the fibre manifest.\footnote{$\Lambda_S$ and 
${\rm NS}_{\rm K3}$ are not the same, in general, but this article does not deal with any explicit example 
where they are different. }

Not all the Calabi--Yau threefolds $M_U$ constructed in this way realize completely adiabatic fibrations over 
$\P^1_A$, however. In the polytope $\widetilde{\Delta}_F$ for the $E_8$-elliptic K3 surface in the fibre, there are 
two facets---$\vev{\nu^{1,3,4}_F} \cap \partial \widetilde{\Delta}_F$ and $\vev{\nu^{2,3,4}_F} \cap \partial \widetilde{\Delta}_F$---that contain interior points. The condition for the absence of extra vertical divisors (equivalent to 
the K3 fibre being irreducible everywhere) is equivalent to choosing $\nu^6_F$ such that those interior 
points in the facets of $\widetilde{\Delta}_F$ remain interior points of facets of $\widetilde{\Delta}$, i.e.
\begin{align}
\nu_F^6 \in \vev{2\nu^{1,3,4}_F} \cap  \vev{2\nu^{2,3,4}_F} \cap \partial (2\widetilde{\Delta}_F) \cap N_F \, ,
\label{eq:choice-Fn}
\end{align}
There are five points of this kind. The corresponding Calabi--Yau threefolds are denoted by $M_{U}^n$, with 
$-2 \leq n \leq 2$. They are known to be the same as $E_8$-elliptic fibrations over $F_n$, 
with $-2 \leq n \leq 2$. Type IIA compactification on $M_U^n$ is dual to heterotic string compactification on 
$T^2 \times $ K3, with instantons distributed by $12+n$ and $12-n$ among the two $E_8$'s.

It is straightforward to generalized this observation. Suppose that an algebraic K3 surface with a Neron--Severi lattice 
$({\rm NS}_{\rm K3}, T_{\rm K3})$ can be constructed as a generic toric hypersurface. There are 4319 toric hypersurface families 
of $K3$ surfaces realized via pairs of reflexive three-dimensional polytopes \cite{Kreuzer:1998vb}. Let $\widetilde{\Delta}_F$ be 
the polytope in $N_F \otimes \R = \R^3$. A toric ambient space for a threefold $M$ is obtained by fibering the toric ambient 
space for K3 over $\P^1_A$. When we choose a toric vector $\nu^5=(\vec{0},-1) \in \Z^{\oplus 4}$, we can take 
$\nu^6 = (\nu^6_F,+1)^T \in \Z^{\oplus 4}$, with any one of 
\begin{align}
 \nu^6_F \in (2 \widetilde{\Delta}_F) \cap N_F\, 
\label{eq:cond-convex-2}
\end{align}
to construct a reflexive four-dimensional polytope. 

Different choices of $\nu^6$ will, in general, result in different geometries, in particular, the Hodge numbers for the resulting threefolds can 
be different. As reviewed shortly, for some choices of the fibre K3 polytope, $\widetilde{\Delta}_F$, it depends 
on the choice of $\nu^6$ whether conditions a) and b) are satisfied.

The condition b) is at stake whenever we consider a fibre K3 polytope $\widetilde{\Delta}_F$ with a facet $\widetilde{\Theta}^{[2]}$ 
with an interior lattice point. To keep the condition b) in a threefold $M$, we need to choose $\nu^6$ so that the polytope 
$\widetilde{\Delta}$ has a facet $\widetilde{\Theta}^{[3]}$ that contains $\widetilde{\Theta}^{[2]}$ and its interior lattice 
points altogether (see section \ref{ssec:a17fibres} and Appendix \ref{sec:K3-fib-tor-review} for more explanations). The 
restriction (\ref{eq:choice-Fn}) in the example of $E_8$-elliptic K3 surface came about precisely for this purpose. To take a few other 
examples, consider degree-2 and degree-4 K3 surfaces (${\rm NS}_{\rm K3}=\vev{+2}$ and $\vev{+4}$, respectively); they are 
both realized as toric hypersurfaces. None of the facets of the polytope $\widetilde{\Delta}_F$ for the degree-4 (quartic) 
K3 surface contains an interior point, while just one facet of the polytope $\widetilde{\Delta}_F$ for the degree-2 K3 
surface has an interior point. In constructing a Calabi--Yau threefold $M_{\vev{+4}}$ that has a quartic K3 surface 
in the fibre over $\P^1_A$, one can therefore use any one of the lattice points in $2\widetilde{\Delta}_F$ for $\nu^6_F$. 
In the case of degree-2 K3 fibred Calabi--Yau threefolds, however, only lattice points in one facet of 
$2 \widetilde{\Delta}_F$ are permitted if we want to satisfy condition b). Such different choices of taking a given polarized K3 surface as 
a fibre have been a subject of study, for example, in \cite{Klemm:2004km}.

The condition a) is at stake when a fibre K3 polytope $\widetilde{\Delta}_F$ and its dual polytope $\Delta_F$ have a dual 
pair of 1-dimensional faces, $\widetilde{\Theta}^{[1]}$ and $\Theta^{[1]}$ such that $\ell^*(\widetilde{\Theta}^{[1]}) >0$ and 
$\ell^*(\Theta^{[1]}) > 0$. A divisor $D_\nu$ of the generic fibre $S_{t,A}$ corresponding to an interior point $\nu$ of 
$\widetilde{\Theta}^{[1]}$ is reducible and each one of the irreducible components of $D_\nu$ is an independent generator of 
${\rm Pic}(S_{t,A})$ and contributes to $\rho$. When we choose $\nu^6_F \in 2 \widetilde{\Delta}_F \cap M$ to construct a 
Calabi--Yau threefold $M$, however, the divisor $D$ remains to have $\ell^*(\Theta^{[1]})+1$ irreducible components only 
when the choice of $\nu^6_F$ is such that the point $\nu$ remains to be an interior point of some two-dimensional face 
of $\widetilde{\Delta}$. In this case, the contributions of this divisors to ${\rm NS}_{\rm K3}$ and $\Lambda_S$ agree.
If $\nu^6_F$ is chosen such that $\widetilde{\Theta}^{[1]}$ becomes a one-dimensional face\footnote{In this case, 
the toric hypersurface construction for a threefold $M$ fails to implement $\ell^*(\widetilde{\Theta}^{[1]}) \ell^*(\Theta^{[2]}) > 0$ complex structure deformation.} of $\widetilde{\Delta}$, on the other hand, the $\ell^*(\Theta^{[1]})+1$ irreducible components of the divisor 
$D_\nu$ in a generic K3 fibre undergo monodromy transformations over the base $\P^1_A$ and do not define 
separate independent divisors of the threefold $M$. Correspondingly, this lattice point $\nu$ leads to only one irreducible 
divisor in $h^{1,1}(M)$ and contributes only by a single class to the lattice polarization of fibration, $\Lambda_S$. 

Among the 4319 three-dimensional reflexive polytopes $\widetilde{\Delta}_F$ to be use for the fibre K3 
polytope \cite{Kreuzer:1998vb}, 131 do not allow a single choice of $\nu^6$ where condition b) is satisfied; 
for the remaining 4188 polytopes, there is at least one choice of $\nu^6$ such that condition b) is satisfied.  
Among these 4188 fibre K3 polytopes $\widetilde{\Delta}_F$, 1071 do not allow a choice of $\nu^6$ where 
condition a) is satisfied as well. For the remaining 3117 fibre K3 polytopes, however, there is at least one choice of 
$\nu^6$ such that both the conditions a) and b) are satisfied. 

For each one of these 3117 fibre K3 polytopes $\widetilde{\Delta}_F$, we must be able to use the adiabatic argument to discuss 
heterotic--type IIA duality. By a computer scan we have found that for 1134 of the polytopes among the 3117 options the choice 
of $\nu^6$ for which the conditions a) and b) are satisfied is unique. The remaining 1983 fibre K3 polytopes $\widetilde{\Delta}_F$ 
allow multiple choices\footnote{Choices of $\nu^6$ different as toric data are treated separately here, although some of 
them may be equivalent under symmetry of the geometry. A case study in \ref{ssec:case-deg2} takes care of this symmetry action, 
but we have not implemented anything like this in our simple scan. } of $\nu^6$ satisfying both conditions.
Multiple choices available in (\ref{eq:choice-Fn}), even after fixing $({\rm NS}_{\rm K3}, T_{\rm K3}) = (\Lambda_S, \Lambda_T)$, can be regarded 
as an example of this more general phenomenon.  

Focussing on the 1983 polytopes $\widetilde{\Delta}_F$ for which there are multiple choices of $\nu^6$ satisfying both conditions a) and b),
it turns out that the value of $h^{2,1}(M)$ depends on the choice of $\nu^6$ for some polytopes $\widetilde{\Delta}_F$, but remains
invariant for others. The example discussed above and the example presented in the nect section are among those $\widetilde{\Delta}_F$ for which there are different fibration options satisfying conditions a) and b) which all have the same $h^{2,1}(M)$.

\subsection{Duality Dictionary in a Case Study: Degree-2 K3 in the Fibre}
\label{ssec:case-deg2}

The moduli space of type IIA compactification on a K3-fibred Calabi--Yau threefold is therefore 
classified by the choice of $(\Lambda_S, \Lambda_T)$, and further by discrete choices of the fibration. 
The moduli space of heterotic string compactifications should also have the same structure, and there should 
be a duality map that translates discrete as well as 
continuous data of the two moduli spaces. The dictionary on the $(\Lambda_S, \Lambda_T)$ part simply 
descends from the dictionary of the heterotic--type IIA duality at higher dimensions (reviewed in section \ref{sec:quick}).
We can then ask the question how the discrete choices of fibration are mapped to heterotic string language.

In the case that the Calabi--Yau threefold $M$ for type IIA compactification has an $E_8$-elliptic K3 surface 
in the fibre, the dual background for heterotic string theory is well-known \cite{KV, MV-1, MV-2}.
Different choices of fibering an $E_8$-elliptic K3 surface over $\P^1_A$---$M_{U}^{n}$ with $-2 \leq n \leq +2$---correspond 
to the $(12+n, 12-n)$ distribution of 24 instantons into the two $E_8$'s in heterotic string theory. 
This is a rare example, however, where the heterotic string interpretation of the discrete data is known. 
Different choices of instanton number distribution in the heterotic string often result in different unbroken 
symmetries, which correspond to different choices of $(\Lambda_S, \Lambda_T)$, not the different choices of fibration 
with the same $(\Lambda_S, \Lambda_T)$. 

Here, we address this question for the $\Lambda_S=\vev{+2}$ case, where a degree-2 K3 surface is the fibre for a 
K3 fibred Calabi-Yau threefold used for type IIA compactification. This is not as easy as the $\Lambda_S = U$ 
($E_8$-elliptic) case, but still remains relatively tractable. We start off by listing up discrete choices of degree-2 K3 
fibred Calabi--Yau threefold $M_{\vev{+2}}$ (see also \cite{Klemm:2004km}).  

The toric vectors in $N_F \cong \Z^{\oplus 3}$ for the degree-2 K3 surface (${\rm NS}_{\rm K3}=\vev{+2}$, $\rho = 1$) 
can be chosen as follows\footnote{
The literature also contains a complete intersection construction of degree-2 
K3 surfaces. The authors consider, however, that such ``degree-2 K3 surfaces'' should be regarded as $E_8$-elliptic 
K3 surfaces with the zero-section blown-down to an $A_1$ singularity point; it is essentially a $\rho =2$ case.}
\begin{equation}
 (\nu^1_F,  \nu^2_F,  \nu^3_F,  \nu^4_F) = 
\left( \begin{array}{cccc} 1 & & & -3 \\ & 1 & & -1 \\ & & 1 & -1 \end{array} \right). 
\end{equation}
The polytope $\widetilde{\Delta}_F \subset N_F \otimes \R$ has four facets, only one of which, 
$\vev{\nu^2_F, \nu^3_F, \nu^4_F}$, contains an interior point $\nu^7_F := (-1,0,0)^T = -\nu^1_F$.
The hypersurface equation is in the form of 
\begin{align}
 (X_1)^2 + F^{(6)}(X_2,X_3,X_4) = 0,
\end{align}
where $X_{1,2,3,4}$ are the homogeneous coordinates associated with the $\nu^i_F$. This provides the picture 
of a double cover over $\P^2$ (homogeneous coordinates $[X_2:X_3:X_4]$) ramified over a sextic curve $\{F^{(6)} =0\} \subset \P^2$.

\subsubsection{Four branches with \texorpdfstring{$h^{1,1}(M)=\rho+1$}{Lg}}
\label{sssec:rho+1}

\begin{figure}[tbp]
\begin{center}
\begin{tabular}{c}
  \includegraphics[width=.35\linewidth]{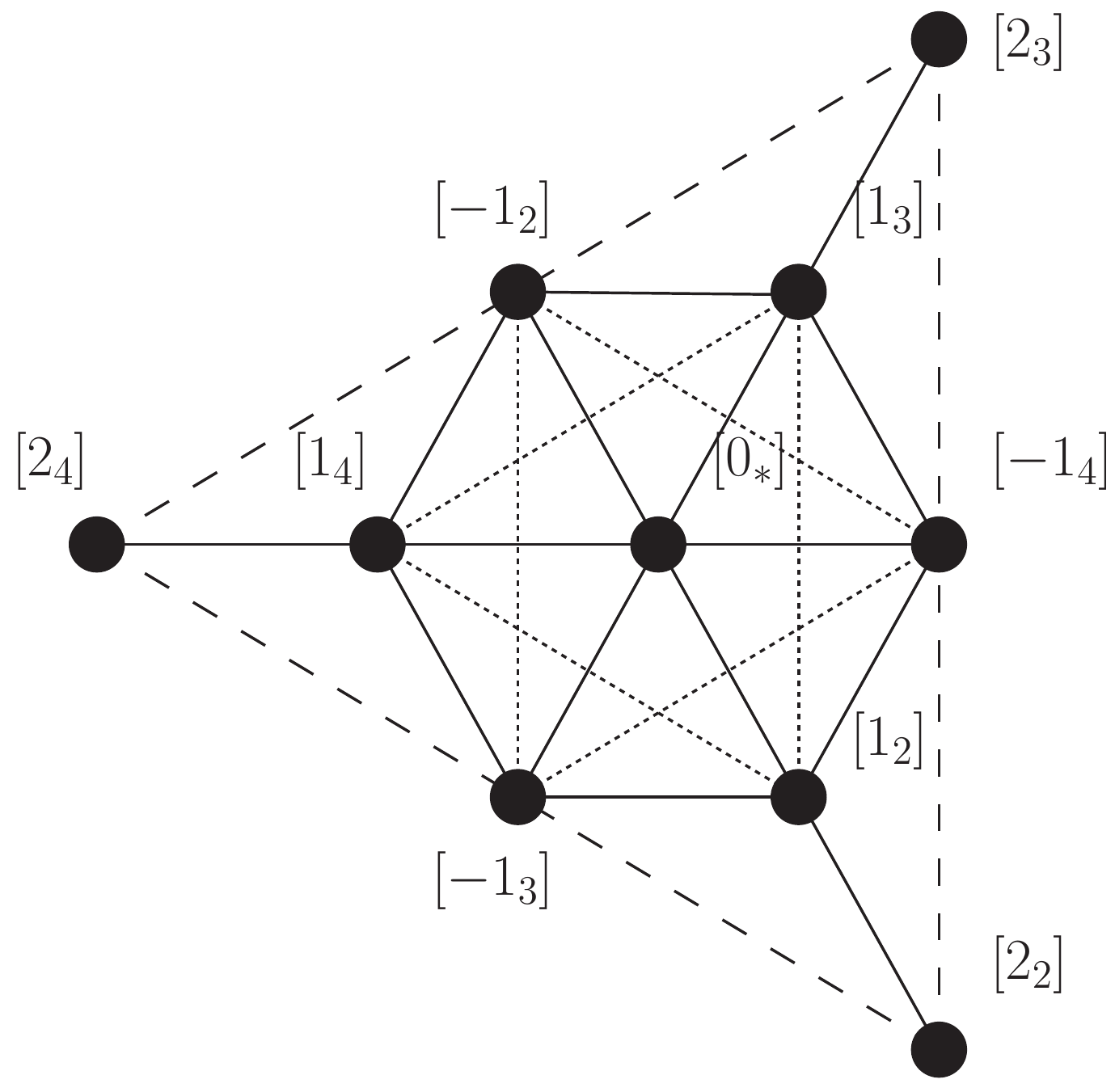} 
\end{tabular}
\caption{\label{fig:choicesofabc}The ten lattice points $\vev{2 \nu^{2,3,4}_F} \cap \partial(2\widetilde{\Delta}_F) 
\cap N_F$, which can be used for a toric vector $\nu^6_F$ in constructing a Calabi--Yau threefold with degree-2 
K3 fibration satisfying conditions a) and b). Lattice points parametrized by $(k_2,k_3,k_4)$ in (\ref{eq:new-choices}) are labelled $[n_a]$ 
when we can choose $k_a=n$ and two other $k$'s zero. Due to the $S_3$ symmetry acting on this graph, at most 
four of them (maybe only three as discussed in the main text) define mutually non-isomorphic complex geometries, however.} 
\end{center}
\end{figure}

A toric hypersurface Calabi--Yau threefold $M$ with a $\Lambda_S=\vev{+2}$-polarized K3 surface 
in the fibre can be constructed by using a toric ambient space constructed as in (\ref{eq:fibre-toric}). 
Any one of the choices of $\nu^6_F$ satisfying (\ref{eq:cond-convex-2}) can be used to construct a non-singular 
threefold $M_{\vev{+2}}$ for type IIA compactification. In this section, we focus on the choices where the condition b) 
is satisfied (condition a) is automatic), such that $h^{1,1}(M_{\vev{+2}})=\rho+1=2$ and
the resulting D=4 ${\cal N}=2$ effective theory has $h^{1,1}(M)=2$ vector multiplets. This narrows down the choice of $\nu^6_F$ to 
$\vev{2\nu^{2,3,4}_F} \cap \partial(2\widetilde{\Delta}_F) \cap N_F$. There are ten choices for the integers 
$(k_2,k_3,k_4)$ in 
\begin{align}
  \nu^6_F = 2 \nu^7_F + k_4 (\nu^4_F-\nu^7_F)+k_2(\nu^2_F-\nu^7_F)+k_3(\nu^3_F-\nu^7_F),
   \label{eq:new-choices}
\end{align}
as shown in Figure~\ref{fig:choicesofabc}.
The ambient space for $M_{\vev{+2}}$ is  
\begin{align}
 \P_{[3:1:1:1]} \left[ {\cal O}_{\P^1}(2-k_2-k_3-k_4) \oplus {\cal O}_{\P^1}(-k_2) \oplus 
  {\cal O}_{\P^1}(-k_3) \oplus {\cal O}_{\P^1}(-k_4) \right].
\end{align}

There is an $S^3$ symmetry transformation acting on the lattice $N$ which keeps $\widetilde{\Delta}_F$ invariant. 
It acts as a permutation on the toric vectors $\nu^{2,3,4}_F$ as well as the corresponding homogeneous coordinates 
$[X_2:X_3:X_4]$. The ten choices of $\nu^6_F$ are grouped into four orbits under this $S^3$ symmetry and we can choose 
$\{ \nu^{6,n}_F := \nu^6_F|_{(k_2,k_3,k_4) = (0,0,n)} \}_{n=-1,0,1,2}$ as representatives of those orbits.  
A (family of) Calabi--Yau threefold(s) obtained as a hypersurface of such a toric ambient space 
is denoted by $M^n_{\vev{+2}}$. The ambient space for the choice $[\nu^{6,n=2}_F]$ can be regarded as (a resolution of) the weighted 
projective space $W\P^4_{[1:1:2:2:6]}$, while the three others ($n=1,0,-1$) cannot be regarded as such. 

Wall's theorem states that the diffeomorphism class of a threefold $M$ is characterized up to a finite number of possibilities by 
$H^{3}(M,\mathbb{Z})$, $H^{2}(M,\mathbb{Z})$, the intersection ring and the second Chern class. Furthermore, if
$H^{3}(M,\mathbb{Z})$ is torsion free, the diffeomorphism class is characterized uniquely. 
The intersection rings of $M_{\vev{+2}}^{n}$ with $n=2,1,0$ do not agree for any identification of integral cohomology groups $H^2(M_{\vev{+2}}^n;\Z)$,
so that these manifolds cannot be diffeomorphic. This means that type IIA compactifications on $M_{\vev{+2}}^n$ with $n=2,1,0$ each have their own 
separate moduli spaces \cite{Klemm:2004km}. There is such an identification between $H^2(M_{\vev{+2}}^{n=2};\Z)$ 
and $H^2(M_{\vev{+2}}^{n=-1};\Z)$, on the other hand. This indicates that $M_{\vev{+2}}^{n=2}$ and $M_{\vev{+2}}^{n=-1}$ 
are the same as real manifolds. It is not known, however, whether the complex structure 
moduli space of this real manifold has just one connected component.\footnote{An example of this phenomenon is discussed e.g. in \cite{Thomas:1998uj}.} We therefore have not ruled out the possibility that type IIA compactifications on $M_{\vev{+2}}^n$ with $n=2$ and $n=-1$ describe physically different vacua, and we treat them separately in the rest of this article.\footnote{The K\"{a}hler cones of $M_{\vev{+2}}^{n=2}$ and 
$M_{\vev{+2}}^{n=-1}$ are mapped to each other under the identification $\phi$. The genus zero Gromov--Witten 
invariants of $M_{\vev{+2}}^n$ with $n=2$ and $n=-1$ also seem to agree under the identification $\phi$. The 
result in section \ref{sssec:10-10-4-interpret}, however, makes us hesitate from saying that $M_{\vev{+2}}^2$ and $M_{\vev{+2}}^{-1}$ are the same.} 

At the very beginning of the study of heterotic--type IIA duality \cite{KV}, type IIA compactification on 
$M^{n=2}_{\vev{+2}} = [(12) \subset W\P^4_{[1:1:2:2:6]}]$ was pointed out as the dual of a heterotic compactification, 
the $ST$-model.\footnote{The $ST$-model is a heterotic string compactification on ``$T^2 \times {\rm K3}$'' with one 
$S^1 \subset T^2$ at the self-dual radius. The $S^1$ at the self-dual radius gives rise to an extra gauge group $A_1$ and
the 24 instantons are distributed as 4, 10 and 10 between $A_1$, $E_8$ and $E_8$, respectively.} Primary evidence 
for this duality claim is comprised of i) the agreement of the pair of lattices $(\widetilde{\Lambda}_S, \Lambda_T)$ and 
ii) the agreement of the number of vector and hypermultiplets in the D=4 ${\cal N}=2$ effective theory, on both 
sides of heterotic and type IIA descriptions. It turns out, however, that all of the Calabi--Yau threefolds 
$M_{\vev{+2}}^n$ with $n=2,1,0,-1$---sharing the lattices $(\widetilde{\Lambda}_S,\Lambda_T)$---have \cite{Klemm:2004km}
\begin{align}
  h^{1,1}(M_{\vev{+2}}^n) = \rho + 1 = 2, \qquad h^{2,1}(M_{\vev{+2}}^n)=128, \qquad \chi(M_{\vev{+2}}^n) = -252.
  \label{eq:h11-h21-deg2-gen}
\end{align}
Hence from observations i) and ii) not just type IIA compactification on $M_{\vev{+2}}^{n=2}$, but on any one 
of the $M_{\vev{+2}}^n$'s must be regarded as an eligible candidate for the dual of the heterotic $ST$-model.

Let $D_i$ be the divisors of the toric ambient space corresponding to the toric 
vector $\nu^i$ ($i=1,\cdots,6$), and $\bar{D}_i := D_i|_{M^n_{\vev{+2}}}$. 

Let us focus on $M^n_{\vev{+2}}$ with $n=2,1,0$. Then we can choose two curves $C_2$ and $C_5$ in $M^n_{\vev{+2}}$ to 
generate the cone of effective curves (Mori cone) of $M^n_{\vev{+2}}$; here, 
$C_2 \cdot \bar{D}_2 = C_5 \cdot \bar{D}_5 = 1$ and 
$C_2 \cdot \bar{D}_5 = C_5 \cdot \bar{D}_2 = 0$. Complexified K\"{a}hler parameters $(t_2, t_5)$ are introduced 
(${\rm Im}(t_2) > 0$ and ${\rm Im}(t_5)>0$), and the complexified K\"{a}hler form is given by 
$(B+iJ)= t = t_2 \bar{D}_2 + t_5 \bar{D}_5$. The Gromov--Witten invariants of vertical curve 
classes, namely, $\beta = n_2 C_2 + n_5 C_5$ with $n_5 = 0$, remain independent of the discrete choices 
($n=2,1,0$) of the fibration \cite{Klemm:2004km}. This type IIA information corresponds to the 1-loop threshold 
correction in heterotic computations. Thus, the experimental evidence so far allows an interpretation 
that all of type IIA compactifications with $M^{n=2,1,0}_{\vev{+2}}$ are dual to the heterotic string $ST$-model 
defined at the perturbative level (in the $g_s$ expansion of the heterotic string) and the presence of multiple choices 
of $M^n_{\vev{+2}}$ in type IIA ($n=2,1,0$) is an indication that multiple non-perturbative completions are possible 
in heterotic string theory \cite{Klemm:2004km}. 

This is an attractive interpretation, but we must say that it still sounds odd. Certainly the third term
in the $D=4$ $N=2$ prepotential 
\begin{align}
  {\cal F} = \frac{2}{2} t_5 t_2^2 + \frac{2n}{6} (t_2)^3
    + \frac{1}{(2\pi i)^3} \sum_{\beta; n_5=0} d_\beta {\rm Li}_3(e^{2\pi i \vev{t,\beta}}) 
    + \frac{1}{(2\pi i)^3} \sum_{\beta; n_5 > 0} d_\beta {\rm Li}_3(e^{2\pi i \vev{t,\beta}})  
\label{eq:prepotential-GW}
\end{align}
come from 1-loop threshold correction in a heterotic string computation, but the second term---computed from 
the intersection ring of $M_{\vev{+2}}^n$ in type IIA compactifications---is also supposed to come 
from heterotic string 1-loop threshold correction. A given heterotic string compactification (say, the $ST$-model)
cannot take multiple values. If type IIA compactification on one of the Calabi-Yau threefolds $M_{\vev{+2}}^n$ is dual 
to the $ST$-model, the IIA compactifications on the other $M_{\vev{+2}}^n$ cannot be dual to the $ST$-model. 

Reference \cite{HM} indicates how to extract the coefficients of such tri-linear term in the prepotential, 
for some examples of heterotic string compactifications to four-dimensions.\footnote{The $ST$-model was 
not chosen as an example there. The study in \cite{KapLusThs} is not sensitive enough to the coefficient of the $(t_2)^3$ 
term, either.} Hence, it is possible to pursue this approach, which exploits information on the vector multiplet moduli 
space on both sides of the duality. In this article, we provide an alternative method to study the duality dictionary 
of the choices of fibration, which uses the hypermultiplet moduli space. It is better to have more tools than less!

\subsubsection{\texorpdfstring{$E_8 \oplus E_8$}{Lg} degeneration of \texorpdfstring{$E_8$}{Lg}-elliptic K3 surfaces}
\label{sssec:AetaBK}

To get started, let us go back to the case where we choose the generic fibre to be the $E_8$-elliptic K3 surface, because 
a lot more is known in the physics literature. In this case, ${\rm NS}_{\rm K3} = U$ and 
$T_{\rm K3} = E_8^{\oplus 2} \oplus U^{\oplus 2}$. Our discussion in the following is valid in the 
weak coupling regime of heterotic vacua, or equivalently, the large volume region of $\P^1_A$ of 
type IIA vacua. The Narain moduli of heterotic string theory and the period integral of the fibre K3 surface for 
type IIA can be compared fibrewise in this situation. In order to study the duality map of 
discrete data (such as the instanton number distributions and the choice of fibration of a given lattice 
polarized K3 surface), it is enough to use any corner of moduli space that is continuously connected. 

The hypersurface equation of an $E_8$-elliptic K3 surface is written down as in 
\begin{align}
  0 = y^2 + x^3 + x \left( \sum_{k=-4}^4 f_k z^{4+k} \right) + \left( \sum_{m=-6}^6 g_m z^{6+m} \right).
\label{eq:E8-elliptic}
\end{align}
For a specific fibre, $f_k$ and $g_m$ are complex numbers. This K3 surface is to be used for type IIA compactification. 
On the heterotic side, we have the Narain moduli $(\tilde{\rho}, \tau, a_{I=1,\cdots,16})$ corresponding to 
the volume $\tilde{\rho}$ and complex structure $\tau$ of $T^2$, as well as Wilson lines $a_{I=1,\cdots,16}$. 
To establish a dictionary between $(f_k,g_m)$ and $(\tilde{\rho}, \tau, a_I)$ is simple, at least conceptually.
One merely needs to compute period integrals for transcendental cycles, and express them in terms of $(f_k,g_m)$. 
 
In practice, it is not a simple task to determine period integrals depending on 18 complex variables, but even 
knowing their qualitative behaviour goes a long way.\footnote{This is similar to the use of the logarithmic 
singularity in the vector multiplet moduli space as a test of duality.} At the qualitative level, 
there are well-known constraints on the $(f_k,g_m)$ for the $E_8 \oplus E_8$ gauge symmetry of the heterotic string 
to remain unbroken, and furthermore, it is known how to scale $(f_k,g_m)$ such that the $T^2$ volume of 
the heterotic description is large (i.e., $\tilde{\rho} \rightarrow i \infty$) \cite{MV-2}. In this way, 
we learn which part of the complex coefficients of the hypersurface equation corresponds to the 
moduli controlling geometric aspects of the heterotic compactification. 
This dictionary has been extended to some extent to include the moduli controlling the breaking of the 
$E_8 \oplus E_8$ symmetry. Such a map has been discussed in the context of local mirror symmetry for 
any ABCDE group \cite{Katz:1997eq, Berglund:1998ej}. For a compact K3 surface, the duality map has been discussed 
for the case the symmetry breaking stays within $\SU(5) \times \SU(5) \subset E_8 \times E_8$ 
\cite{Curio_Donagi,Donagi:2008ca,Hayashi-A, Hayashi:2009bt}. See \cite{KSChoi} for symmetry breaking in $\SU(6) \subset E_8$, and \cite{Anderson:2015cqy}
for more general cases. For other aspects of hypermultiplet moduli map, 
see e.g. \cite{Aspinwall:2000fd,Halmagyi:2007wi,Louis:2011aa,Alexandrov:2012pr} and references therein.

Consider taking the coefficients $(f_k,g_m)$ in (\ref{eq:E8-elliptic}) to be 
\begin{align}
 g_m = g'_m \times \epsilon_\eta^{|m|} \epsilon_K^{6(|m|-1)}, \qquad 
 f_k = f'_k \times \epsilon_\eta^{|k|} \epsilon_K^{6|k|-4}, 
\label{eq:gen-scaling-E8E8}
\end{align}
with $|g'_m|, |f'_k| \sim {\cal O}(1)$ and $\epsilon_\eta \ll 1$ and $\epsilon_K \ll 1$. This is a 
generalization of the scaling in \cite{MV-2, Hayashi:2010zp}. Under such a choice of complex coefficients, 
10 out of the 24 discriminant points of this $E_8$-elliptic K3 surface are found in the region 
$z \sim (\epsilon_\eta \epsilon_K^{6})$, 2 are in the region $z \sim \epsilon_\eta$, 2 more are in the region $z \sim \epsilon_\eta^{-1}$ 
and the remaining 10 are found in the region $z \sim (\epsilon_\eta \epsilon_K^6)^{-1}$. 
The $\epsilon_K$-scaling power of individual coefficients above is determined such that the hypersurface 
equation (\ref{eq:E8-elliptic}) around $(x,y,z)=(0,0,0)$ is well approximated by a deformation of an
$E_8$ singularity. Indeed, we only need to rewrite (\ref{eq:E8-elliptic}) by using a set of local 
coordinates $(\xi, \eta,\zeta)$ in 
\begin{align}
(x,y,z) = (\epsilon_\eta^2 \epsilon_K^{10}\xi, \; \epsilon_\eta^3 \epsilon_K^{15} \eta, \; 
    \epsilon_\eta \epsilon_K^6\zeta),
\end{align}  
and drop all the terms with positive powers in $\epsilon_\eta$ or $\epsilon_K$.

It is also possible (though not necessary) to consider the (reducible) K3 surface associated with the stable degeneration limit corresponding to 
$\epsilon_\eta \rightarrow 0$ and focus on one of the two irreducible components. We then have a rational 
elliptic surface \cite{FMW}
\begin{align}
 0 = y^2 + x^3 + x \left( f_0 \zeta^4 + \sum_{k=1}^4 \epsilon_K^{6k-4} f_k \zeta^{4-k} \right)
   + \left( g_0 \zeta^6 + \sum_{m=1}^6 \epsilon_K^{6m-6} g_m \zeta^{6-m} \right).
\end{align}
The $f_k$ and $g_m$ here correspond to $\beta_k$ and $\alpha_m$ of $dP_8$ in \cite{FMW}. 
 
This description (parametrization) of $E_8$ Wilson lines in $T^2$ is redundant. This is due 
to the fact that we have not fixed the automorphisms acting on the base $\P^1$ of the $E_8$-elliptic 
K3 surface. For two constants $c_1$ and $c_2$, it is
\begin{align}
  z' = \frac{z + c_1}{c_2 z + 1}. 
\end{align}
By allowing this redundancy in the parametrization, however, the collection of $\epsilon_\eta$ 
scaling powers, $\{1,2,3,4,5,6,1,2,3,4\}$, contains the full list of Dynkin labels of the extended Dynkin 
diagram of $E_8$. The collection of $\epsilon_K$ scaling powers, $\{0,6,12,18,24,30,2,8,14,20\}$, contains 
degrees of all of the independent deformation parameters of the $E_8$ singularity in \cite{KM}.

With this preparation, let us now consider a Calabi--Yau threefold $M$ that has an $E_8$-elliptic 
K3 surface in the fibre. Here, the coefficients $f_{\pm k}$ and $g_{\pm m}$ are promoted to sections of line bundles over the 
base $\P^1_A$. When the fibration corresponds to a choice of $\nu^6_F$ in (\ref{eq:choice-Fn}) with 
$-2 \leq n \leq 2$, 
\begin{align}
 g_m \in \Gamma(\P^1_A; {\cal O}(12+n \cdot m)) =
         \Gamma(\P^1_A; {\cal O}(|m| \eta_{\pm} +(6|m|-6)K_{\P^1})). \label{eq:gm-E8ell} \\
 f_k \in \Gamma(\P^1_A; {\cal O}(8+n \cdot k)) =
         \Gamma(\P^1_A; {\cal O}(|k| \eta_{\pm} +(6|k|-4)K_{\P^1})),
\end{align}
where $\eta_+ = 12+n$ for $m>0$ and $k>0$, and $\eta_- = 12-n$ for $m<0$ and $k<0$. 
Therefore, we see that any one of $(f_k, g_m)$ which is required to have a scaling 
$\epsilon_\eta^A \epsilon_K^B$ for the Het-sugra and near-symmetry-restoration takes its 
value in a line bundle ${\cal O}_{\P^1_A}(A \eta + B K_{\P^1})$. It is also known, based on
the study of chains of Higgs cascades and singular transitions among various branches of 
the moduli spaces of F-theory (or type IIA) compactifications, that 
$M$ is dual to heterotic string theory compactified on K3 (K3 x $T^2$) with the instantons 
distributed as $\eta_+$ and $\eta_-$ to $E_8 \oplus E_8$. 

For a given branch of the moduli space of type IIA compactifications on a Calabi--Yau manifold $M$ with a 
lattice-polarized K3 fibration $\pi_M: M \longrightarrow B$, we are hence motivated to ask 
if there is an assignment of scalings $\epsilon_\eta^A \epsilon_K^B$ of the complex coefficients 
that leads to a decoupling of gravity and symmetry restoration. When there is such an assignment of the 
scalings, one can then further ask if there is a divisor $\eta$ on the base $B$ such that 
a section $f$ under the scaling $\epsilon_\eta^A \epsilon_K^B$ is a section of ${\cal O}_B(A \eta + B K_B)$.
If such a divisor $\eta$ is found, then we can take it to be the instanton number (more generally second Chern character ${\rm ch}_2$) defined purely in terms of (hypermultiplet moduli of) the type IIA compactification. In cases where the heterotic--IIA duality dictionary is not understood well enough, it is important that we can extract such information intrinsically.

This is a generalization of the same idea that has been known to hold\footnote{There is an alternative 
idea for an intrinsic definition of the instanton number in F-theory compactifications. 
Let $\pi_M: M \rightarrow B$ be a K3-fibration, and $\pi_M':M \rightarrow B'$ be an elliptic fibration. $B'$ is a $\P^1$-fibration 
over $B$. When there is an unbroken non-Abelian symmetry, the corresponding discriminant locus in $B'$
will appear as a section of the $\P^1$ fibration over $B$. Let $S \subset B'$ be the image of this section (often referred to as the GUT divisor). Then in case the heterotic dual involves an ${\rm SU}(N)$ bundle in $E_8$, its instanton number can be extracted from $\eta = c_1(N_{S|B'}) - 6K_S$ in the F-theory geometry \cite{Ragesh}. Unfortunately we cannot rely on this idea in this article, since neither a $\P^1$-fibration 
$B' \rightarrow B$ nor GUT divisor $S$ is available in heterotic--IIA duality in general.}  
in the case of symmetry breaking with the structure group $\SU(N)$. When sections $a_r \in \Gamma(B; {\cal O}_B(\eta + r K_B))$ 
($r=0,2,\cdots, N$) for some divisor $\eta$ on $B$ in a hypersurface equation has the scaling 
$\epsilon_\eta \epsilon_K^r$ for symmetry restoration of ${\rm SU}(N)$, the heterotic dual involves 
a vector bundle with the instanton number (second Chern character ${\rm ch}(2)$) specified by $\eta$.  

\subsubsection{\texorpdfstring{$E_8 \oplus E_8$}{Lg} degeneration of degree-2 K3 surfaces}
\label{sssec:10-10-4-interpret}

In order to argue what is the distribution of instanton numbers in the heterotic dual of type IIA compactifications on 
$M_{\vev{+2}}^{n}$ ($n=2,1,0,-1$), where the fibre is a degree-2 K3 surface, we first need to find a scaling behaviour 
of the complex coefficients of $M_{\vev{+2}}^n$ that leads to symmetry restoration. 

\begin{figure}[tbp]
\begin{center}
\begin{tabular}{ccc}
   \includegraphics[width=.2\linewidth]{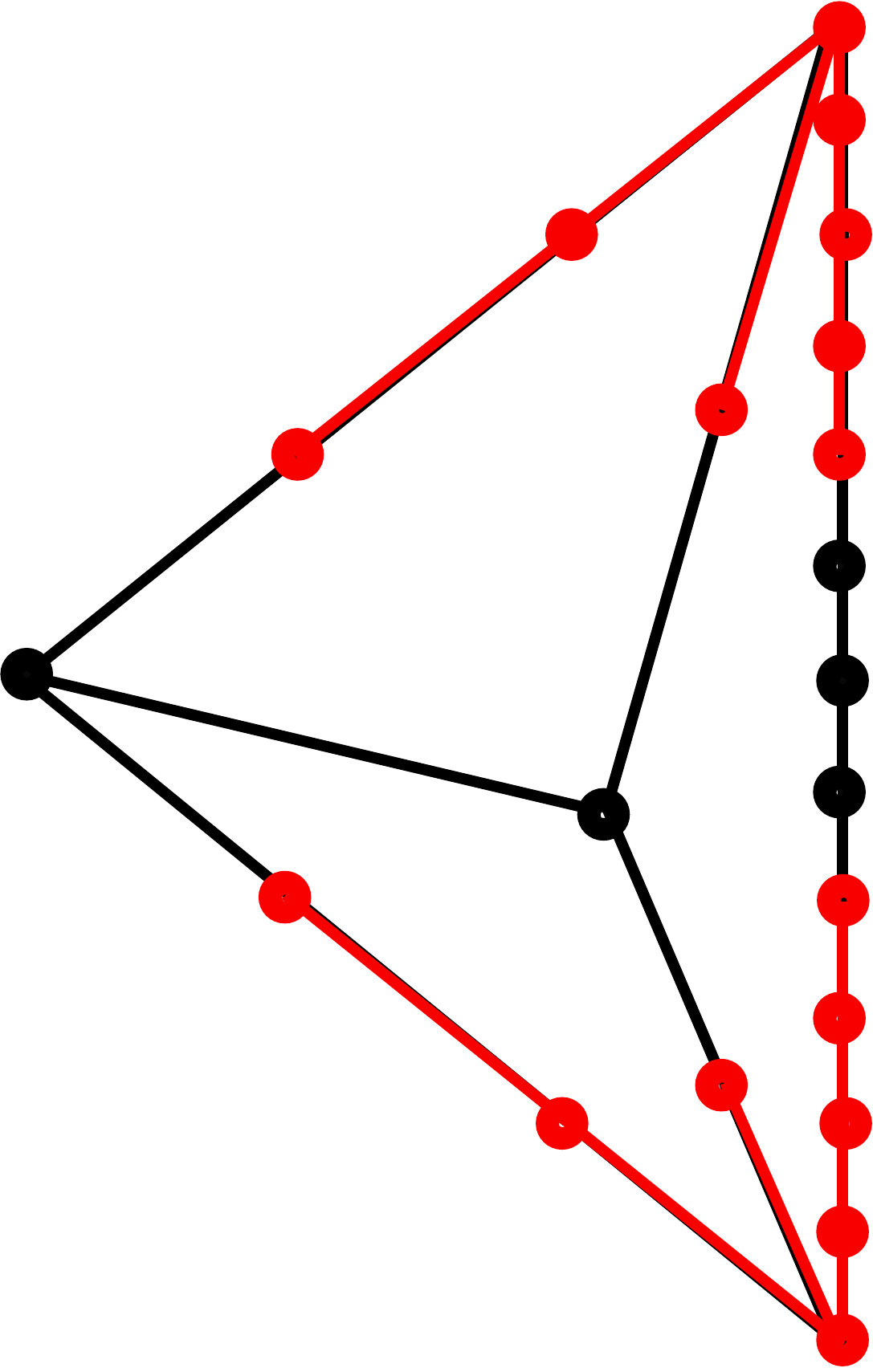} & $\quad$ & 
   \includegraphics[width=.4\linewidth]{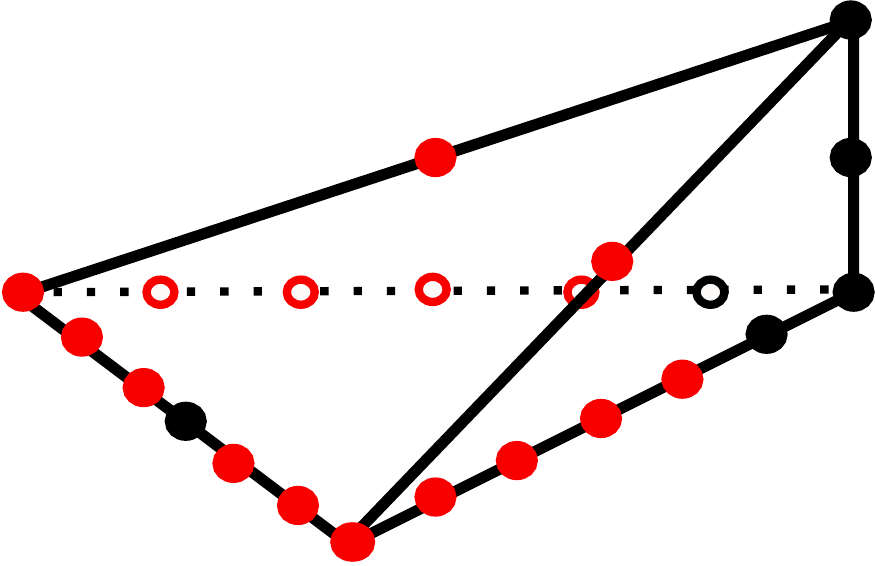} \\
  (a) & & (b) 
\end{tabular}
\caption{\label{fig:2DdualPolytope}(colour online) Dual polytopes $\Delta_F$ of $E_8$-elliptic K3 surface (a) 
and degree-2 K3 surface (b). Transcendental 2-cycles forming the $E_8 \oplus E_8$ lattice are indicated 
by red (light grey) nodes. }
\end{center}
\end{figure}

The $E_8 \oplus E_8 \oplus A_1$ part of two-cycles remains in the transcendental lattice $T_{\rm K3}$ 
for a generic fibre. Since transcendental cycles of a K3 surface 
correspond to divisors of its mirror K3 surface (up to a sublattice $U$), lattice points of the dual polytope $\Delta_F$ 
can be used to capture those two-cycles. The dual polytope $\Delta_F$ of degree-2 K3 surfaces 
is shown in Figure~\ref{fig:2DdualPolytope}~(b) along with that of $E_8$-elliptic K3 surface. 
To each lattice point marked in Figure~\ref{fig:2DdualPolytope}~(b), there is a corresponding 
transcendental two-cycle. They are all isomorphic to $S^2$, except for the cycle corresponding to the 
lattice point at the top of the polytope, which is isomorphic to a $T^2$.
There are three linear (topological) relations among them, and after a computation of the intersection form, 
a rank-18 lattice of transcendental two-cycles, 
\begin{align}
  \overline{T}_{\rm K3}(S_{t.A}) = E_8 \oplus E_8 \oplus \vev{-2} \oplus U, 
\end{align}
is obtained. Two more transcendental cycles are missing here, because they correspond to $H^0$ and $H^4$ 
of the mirror of $S_{t.A}$. The entire transcendental lattice is $\overline{T}_{\rm K3}(S_{t.A}) \oplus U$. 
There are three different ways in identifying the $E_8 \oplus E_8$ lattice among those two-cycles, which can be traced back to 
the $S^3$ symmetry on the moduli space of degree-2 K3 surface. One of the three identifications is already 
shown in Figure~\ref{fig:2DdualPolytope}~(b).

With this picture in mind, it is now easy to figure out how to assign the scaling behaviour for the 
approximate restoration of the $E_8 \oplus E_8$ symmetry in the moduli space of degree-2 K3 surfaces. 
Let us write down the hypersurface equation of a degree-2 K3 surface in a set of affine coordinates, 
$(y,x_4,x_3,x_6) = (X_1/X_2^3, \; X_4/X_2, \; X_3/X_2, \; X_6/X_5)$:
\begin{align}
 y^2 =
  & \quad \quad (a'_1 x_4^5 x_3+ a'_2 x_4^4 x_3^2 + a'_3 x_4^3 x_3^3+ a'_4 x_4^2x_3^4 + a'_5 x_4 x_3^5+ a'_6 x_3^6)  \nonumber \\
  & \quad  + (b'_1 x_4^3 x_3^2 + b'_2 x_4^2 x_3^3 + b'_3 x_4 x_3^4 + b'_4 x_3^5)
      + (c'_1 x_4 x_3^3 + c'_2 x_3^4) \nonumber \\
  & + x_4^6 + b_0 x_4^4 x_3 + c_0 x_4^2 x_3^2 + d_0 x_3^3 \nonumber \\
  & \quad + (b_1 x_4^3 + b_2 x_4^2 + b_3 x_4 + b_4) x_3 + (c_1 x_4 + c_2) x_3^2 \nonumber \\
  & \quad + (a_1 x_4^5 + a_2 x_4^4 + a_3 x_4^3 + a_4 x_4^2 + a_5 x_4 + a_6),  
\end{align}
where we used an affine patch $(x_3,x_4) = (X_3/X_2, X_4/X_2)$ of the $\P^2[X_2:X_3:X_4]$, 
The scaling is 
\begin{align}
 a_r = a_{r*} \times \epsilon_\eta^r \epsilon_K^{6r-6}, \qquad 
 b_r = b_{r*} \times \epsilon_\eta^r \epsilon_K^{6r-4}, \qquad 
 c_r = c_{r*} \times \epsilon_\eta^r \epsilon_K^{6r-2},
\end{align}
where $\epsilon_\eta$ and $\epsilon_K$ are taken to be small, while $a_{r*}$, $b_{r*}$ and $c_{r*}$ are ${\cal O}(1)$.
When the value of $\epsilon_\eta$ is small, one set of $E_8$ transcendental cycles (visible sector) 
are found close to the point $X_4=X_3= 0$ in $\P^2$, while the other set of $E_8$ transcendental cycles 
(hidden sector) are located near the point $X_4=X_2=0$. When $\epsilon_K$ is set to zero, while 
$\epsilon_\eta$ remains small but non-zero, we have an $E_8$ singularity, 
$y^2 + a_1 \epsilon_\eta x_4^5 + d_0 x_3^3=0$. The scaling behaviour assigned for $a_m$ and $b_k$ 
agrees with those for $g_m$ and $f_k$ in the case of an $E_8$-elliptic K3. 

When the degree-2 K3 surface is fibred over the base $\P^1_A$, the coefficients $a_r$, $b_r$ and 
$c_r$ for the visible $E_8$ and those for the hidden $E_8$ are promoted to sections of certain 
line bundles. We can work out the degree of those line bundles for any given choice of fibration 
in Figure \ref{fig:choicesofabc}.
\begin{table}[tbp]
\begin{center}
\begin{tabular}{|c|ccccccc|}
\hline
& $[2_4]$ & $[1_4]$ & $[0_*]$ & $[-1_4]$ & $[-1_2]$ & $[2_3]$ & $[1_3]$ \\
\hline
$a_r$ & $(12-2r)$ & $(8-r)$ & 4 & $r$ & 6 & $2r$ & $2+r$ \\
$b_r$ & $(8-2r)$ & $(6-r)$ & 4 & $2+r$ & 5 & $(2+2r)$ & $3+r$ \\
$c_r$ & $(4-2r)$ & $(4-r)$ & 4 & $(4+r)$ & 4 & $(4+2r)$ & $4+r$ \\
\hline
$a'_r$ & $(12-2r)$ & $(8-r)$ & 4 & $r$ & $(6-r)$ & 0 & 2 \\
$b'_r$ & $(8-2r)$ & $(6-r)$ & 4 & $(2+r)$ & $(5-r)$ & 2 & 3 \\
$c'_r$ & $(4-2r)$ & $(4-r)$ & 4 & $(4+r)$ & $(4-r)$ & 4 & 4 \\
\hline
\end{tabular}
\caption{\label{tab:E8-instanton-count}
Each column corresponds to a Calabi--Yau threefold hypersurface $M$ in the toric ambient space 
corresponding to one of the choices of $\nu^6_F$ shown in Figure~\ref{fig:choicesofabc}.
$a_r$ ($r=1,\cdots, 6$), $b_r$ ($r=1,\cdots, 4$), $c_{r=1,2}$, and $a'_r$, $b'_r$, $c'_r$ are 
sections of line bundles on $\P^1_A$ whose degrees are indicated in the 1st--6th rows in this table. }
\end{center}
\end{table}
The results are shown in Table \ref{tab:E8-instanton-count}. 
If the complex structure moduli of a threefold $M_{\vev{+2}}^n$ are to be interpreted as $E_8+E_8$ instanton 
moduli in heterotic string, we expect that 
\begin{eqnarray}
 a_{r=1,\cdots,6} \in \Gamma(\P^1_A; {\cal O}(12+r(I_v-12))), & \qquad & 
 a'_{r=1,\cdots,6} \in \Gamma(\P^1_A; {\cal O}(12+r(I_h-12))), \nonumber \\
 b_{r=1,\cdots,4} \in \Gamma(\P^1_A; {\cal O}(8+r(I_v-12))), & \qquad &
 b'_{r=1,\cdots,4} \in \Gamma(\P^1_A; {\cal O}(8+r(I_h-12))), \\
 c_{r=1,2} \in \Gamma(\P^1_A; {\cal O}(4+r(I_v-12))) , & \qquad & 
 c'_{r=1,2} \in \Gamma(\P^1_A; {\cal O}(4+r(I_h-12)))  \nonumber 
\end{eqnarray}
for some choice of instanton numbers $I_v$ and $I_h$. It turns out that only the $[2_4]$ choice of $\nu^6_F$ allows for
an interpretation of $E_8+E_8$ bundle moduli, where $I_v = I_h = 10$ as in the heterotic string $ST$ model. 
The choice $[2_4]$ of $\nu^6_F$ corresponds to taking the ambient space to be the weighted projective space, 
$W\P^4_{[1:1:2:2:6]}$. For any other choice, the degrees of the relevant line bundles cannot have the right pattern to 
even define\footnote{As remarked earlier, there are three different identifications of 
the $E_8 + E_8$ transcendental cycles in a degree-2 K3 surface. 
Choosing an appropriate identification of the $E_8 + E_8$ transcendental cycles and the corresponding 
assignment of the scaling behaviour of the coefficients of the hypersurface equation (different 
from the one we adopted in the text), the threefolds for the choices $[2_3]$ and $[2_2]$ can also be 
regarded as the $I_v = I_h=10$ case. This should be obvious due to the $S^3$ symmetry of degree-2 K3 surface.
For other choices, such as $[-1_a]$, $[1_a]$ and $[0_*]$, however, the three options of $E_8+E_8$ 
identifications do not help in consistently defining instanton number assignments.} 
the instanton numbers of $E_8 + E_8$ intrinsically in terms of the threefolds $M_{\vev{+2}}^n$ ($n=1,0, -1$). 

We therefore conclude that only the type IIA compactification on $M_{\vev{+2}}^{n=2}$ is dual to the heterotic $ST$-model (where 
the instantons numbers are distributed by 4+10+10 in $A_1+E_8+E_8$). Type IIA compactifications on $M_{\vev{+2}}^n$ with 
$n=1,0,-1$ are not, although the effective theories with $D=4$ ${\cal N}=2$ supersymmetry have the same number of 
vector and hypermultiplets, and the special geometry passes highly non-trivial tests of duality (the third term of 
(\ref{eq:prepotential-GW})). The hypermultiplets, however, do not seem to reproduce the instanton moduli expected in the heterotic 
$ST$-model and the heterotic dual of type IIA compactifications on $M_{\vev{+2}}^n$ with $n=1,0,-1$ must be something other than 
the $ST$-model. 

A case study for $\Lambda_S = \vev{+2}$ was presented above, but this is a very small subset of 
all the ${\cal O}(2000)$ choices of $\Lambda_S$, where there are multiple choices of fibering 
$\Lambda_S$-polarized K3 surface over $\P^1_A$. The method described above may be applied to cases 
where $\Lambda_T$ contains $U \oplus U \oplus (\oplus_a R_a)$ with an ADE root lattice $R_a$; 
after assigning the height $A$ and degree $B$ for a symmetry $R_a$ to monomials in the defining equation 
of $M_{\Lambda_S}$, one can ask whether an appropriate divisor $\eta_a$ for $R_a$ is found. 
In the case of $\Lambda_S = \vev{+4}$ (where we have a quartic K3 surface as the fibre in IIA language), for example, it is at least 
possible to talk about the instanton number assignment in the $E_8 \oplus E_8$ part of $\Lambda_T$, if not for 
the $\vev{-4}$ part. It will be difficult, however, to apply this method to Calabi--Yau manifolds without a complete 
intersection construction, or to choices of $\Lambda_T$ without a $U \oplus U$ component, or a single factor of $R_a$. 

\section{Degenerations of K3 Surfaces and Soliton Solutions}
\label{sec:soliton-4D}

In the last section, we restricted our attention to K3-fibred Calabi--Yau threefolds where the K3 fibre remains irreducible everywhere over the base $\P^1_A$ and furthermore the entire ${\rm NS}_{\rm K3}$ lattice 
of a generic fibre becomes the lattice polarization of the fibration, ${\rm NS}(S_{t.A})=\Lambda_S$. The method of 
construction was limited to using a toric polytope $\widetilde{\Delta}$ spanned by \emph{just one} toric vector 
$\nu^6$ in addition to $\nu^{1,2,3,4,5}$. The spirit was to focus on situations where the adiabatic argument can be used. 

In this section, we explore fibrations where the adiabatic argument does not hold at isolated points in the base $\P^1_A$, 
and discuss their heterotic dual descriptions. In particular, we will relax the condition that the K3 fibre remains irreducible everywhere
over the base $\P^1_A$ for a smooth threefold $M$.  Examples in section \ref{ssec:a17fibres} are such that only a single extra vertex $\nu^6$ is introduced besides $\nu^{1,2,3,4,5}$ in the toric polytope. In sections \ref{ssec:corridor-II} and \ref{ssec:corridor-III} we also relax this condition, and the construction in \cite{short_tops} (and its obvious generalization) is exploited. We will see a rich variety of branches of the type IIA compactification moduli space, even for a single choice of the lattice $\Lambda_S = {\rm NS}_{\rm K3}$. After examining the reducible fibre geometries and how those branches are connected, we will study their heterotic string interpretation in section \ref{ssec:Het-interpret}.

\subsection{A Simple Fibration with Reducible Fibre(s)}
\label{ssec:a17fibres}

Let us continue to work out discrete fibration choices of degree-2 K3 fibred Calabi--Yau threefolds $M_{\vev{+2}}$
realized as toric hypersurfaces corresponding to a polytope spanned by $\nu^{1,2,3,4,5,}$ and just one point $\nu^6$. Contrary to before, however,
we do not require that the fibre K3 surface remains irreducible everywhere over the base $\P^1_A$. 

This means that we can choose any one of (\ref{eq:cond-convex-2}), not just 
those in $\vev{\nu^{2,3,4}_F} \cap \partial(2\widetilde{\Delta}_F) \cap N_F$. 
For the choices of toric vectors $\nu^6 := (v_1-2,v_2,v_3,1)^T$ that are now allowed, there is one lattice point $\nu^7 = (\nu^7_F,0)^T \in N$ 
placed in the interior of a two-dimensional face $\widetilde{\Theta}^{[2]}$ of $\widetilde{\Delta}$, but not interior to any one 
of the facets of $\widetilde{\Delta}$. The contribution to $h^{1,1}(M_{\vev{+2}})$ is 
\begin{equation}
  h^{1,1}(M_3) - 2 = 1 + \ell^*(\Theta^{[1]}) = (v_1-v_2-v_3)\, ,
\end{equation}
where $\Theta^{[1]}$ is the dual face of $\widetilde{\Theta}^{[2]}$. The possible values of $(v_1-v_2-v_3)$ for $\nu^6_F$ 
in (\ref{eq:cond-convex-2}) range from 0 to $4$. The cases with $(v_1-v_2-v_3)=0$---those appearing in 
Figure~\ref{fig:choicesofabc}---have been studied in section \ref{sssec:rho+1}. 

Interior points to facets of $\widetilde{\Delta}_F$ which are not interior to facets of $\widetilde{\Delta}$ likewise give rise to reducible divisors for any K3-fibred Calabi-Yau threefold realized as a toric hypersurface. A brief explanation for this statement is given in Appendix \ref{sec:K3-fib-tor-review}, but this is well-known already in the case of having an $E_8$-elliptic K3 surface in the fibre \cite{Candelas:1996su}. When $\nu^6_F$ is chosen from (\ref{eq:cond-convex}), but not from (\ref{eq:choice-Fn}), then the two lattice points interior to the two-dimensional face $\vev{\nu^{2,3,4}_F} \cap (\partial \widetilde{\Delta}_F)$ result in a singular fibre (or even several singular fibres) with three components, whereas $\vev{\nu^{1,3,4}_F} \cap (\partial \widetilde{\Delta}_F)$ potentially gives rise to singular fibres with two components. 

Coming back to the case of a degree-2 K3 surface as the fibre, let us take $(v_1-v_2-v_3)= 1$ as an example.\footnote{
In the cases with $(v_1-v_2-v_3)=2,3,4$, we just have the same reducible fibre geometry, $V_0+V_1$, at $(v_1-v_2-v_3)$ isolated 
points in the base $\P^1_A$.} This is e.g. realized 
for $\nu^6 = (-\nu^1_F,1) =  (-1,0,0,1)$; we denote the resulting threefold by $M_{\vev{+2}}^{-\nu^1_F}$. 
Its defining equation is of the form
\begin{equation}\label{eq:hssexticfiba17}
 0 = X_1^2 F^{(0,1)} + X_7 \left(X_1 F^{(3,2)} + X_7 F^{(6,3)}\right) \, ,
\end{equation}
where the $F^{(i,j)}$ are homogeneous polynomials of degree $i$ in $[X_2:X_3:X_4]$ and degree $j$ in the $[X_5:X_6]$ coordinates of 
the base $\P^1_A$. At the point $t_0 \in \P^1_A$ defined by $F^{(0,1)}(X_5,X_6)=0$ the fibre geometry $S_{t_0}$ is singular and consists 
of the two irreducible components
\begin{align}
V_{0;t_0}:& \hspace{.5cm} X_7 = 0 \, ,\\
V_{1;t_0}:& \hspace{.5cm} X_1 F^{(3,2)} + X_7 F^{(6,3)} = 0 \, .
\label{eq:geom-A17-defeq}
\end{align}
We can think of either one of them as the $(v_1-v_2-v_3) = 1$ extra contribution to $h^{1,1}(M_{\vev{+2}}^{-\nu^1_F})$; their sum is
homologous to the class of the generic fibre. 

The K3-fibration degenerates at the point $t_0 \in \P^1_A$. This is an example of a Type II degeneration; background material on 
the theory of degeneration of K3 surface is summarized in Appendix \ref{sec:K3-degen-review} for the convenience of readers. 
In this particular example of Type II degeneration of a lattice-polarized K3 surface ($\Lambda_S = \vev{+2}$), 
$V_{0;t_0}=\P^1$ and $V_{1;t_0}$ is $\P^2_{[X_2:X_3:X_4]}$ blown-up at eighteen points ($F^{(3,2)}|_{t_0} = F^{(6,3)}|_{t_0}=0$).
The two surface components intersect along the elliptic curve $\{ F^{(3,2)}|_{t_0}=0 \} \subset \P^2$, see also \cite{friedman:new_proof}.

Let us parametrise the K\"ahler cone of this threefold $M_{\vev{+2}}^{-\nu^1_F}$ by 
\begin{align}
 J = \bar{D}_2 t_{2,I} + \bar{D}_7 t_{{\rm vrt},I} + \bar{D}_5 t_{5,I},
\end{align}
where $t_{2,I}$, $t_{{\rm vrt},I}$ and $t_{5,I}$ are real valued; the subscripts $I$ are a reminder that they are meant 
to be the imaginary part of the complexified K\"ahler parameter $B+iJ$. This parametrization respects the filtration 
structure in the space of divisors and curves associated with fibration. 
The polytope $\widetilde{\Delta}$ for the ambient space of $M_{\vev{+2}}^{-\nu^1_F}$ has a unique triangulation and 
the K\"ahler cone of the toric ambient space is bounded by three walls:
\begin{align}
 0 < t_{{\rm vrt},I}, \qquad 0 < t_{2,I}-3t_{{\rm vrt},I}, \qquad 0 < t_{5,I} - t_{2,I}.
\end{align}
At the wall $t_{{\rm vrt},I}=0$, the eighteen $(-1)$ curves in $V_1 = {\rm Bl}^{18}(\P^2)$ in the central fibre shrink to zero 
volume, while the volumes of $\P^1 \subset \P^2=V_0$ and $V_0=\P^2$ itself go to zero at the wall $t_{2,I}-3t_{{\rm vrt},I}=0$. 
The last inequality does not concern us, as we will stay within the large base $\P^1_A$ regime 
\begin{align}
 t_{5,I} \gg | t_{2,I}|, \; | t_{{\rm vrt},I} |
\end{align}
of type IIA compactification (which is dual to the weak 4D dilaton regime in heterotic compactification) in this article. 

\begin{figure}[tbp]
\begin{center}
\begin{tabular}{cc}
     \scalebox{.6}{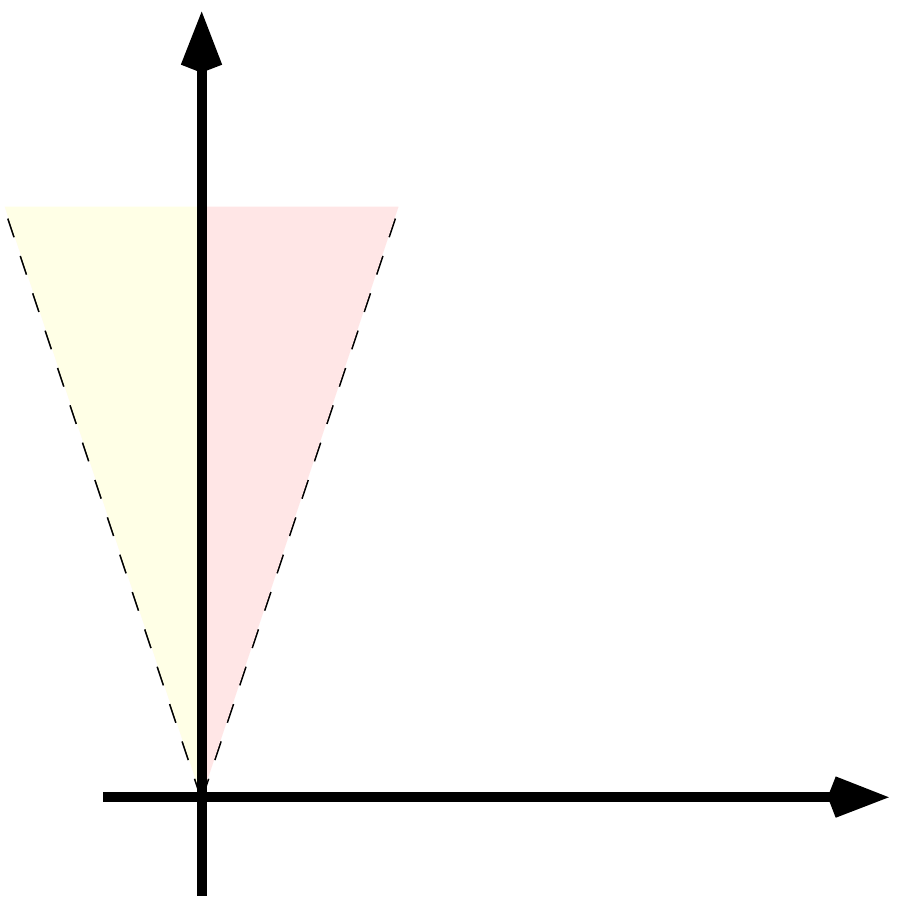} & 
     \scalebox{.6}{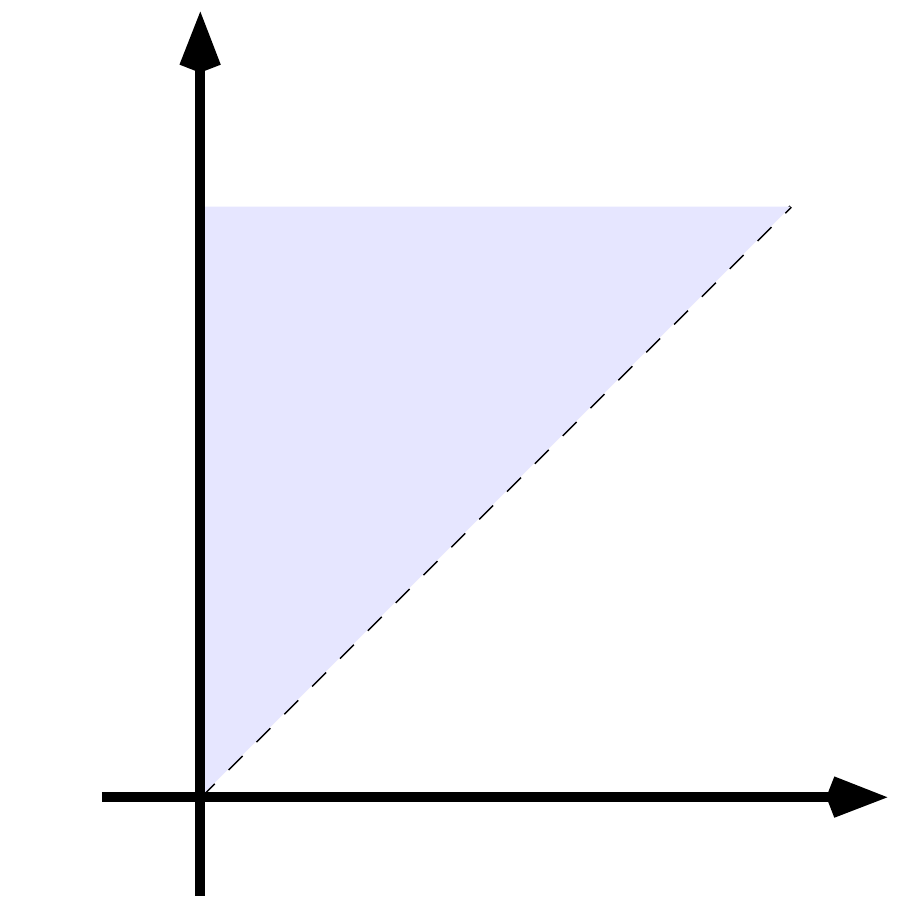} \\
 \\
  (a) &  (b) 
\end{tabular}
\caption{\label{fig:kahler-cone}
Geometric phases of the K\"ahler cone in the large base regime a) for $M_{\vev{+2}}^{-\nu^1_F}$ in section \ref{ssec:a17fibres} and b) for $M_{\vev{+2}}^{\{0,-1\}}$ in section \ref{sssec:kahler-E7D10}, presented in the space of K\"ahler parameters of the fibre K3. 
}
\end{center}
\end{figure}

At the wall $t_{{\rm vrt},I}=0$, a flop transition\footnote{\label{fn:extnd-kahler}
It is not a straightforward task to find a toric construction for the geometry on the other side of the flop transition 
of a Calabi--Yau hypersurface. Different choices of triangulation of the polytope $\widetilde{\Delta}$ sometimes do the 
job (just as they do for the transitions of the toric \emph{ambient} space), but that is not always the case. See e.g. 
\cite{Braun:2014kla,Braun:2015hkv} for more examples.} turns $V_1 = {\rm Bl}^{18}(\P^2)$ into $\P^2$, and $V_0=\P^2$ into 
${\rm Bl}^{18}(\P^2)$. The phases found at the two sides of the wall can be regarded as the two small resolutions of a geometry given by 
\begin{align}
 \left( \xi - \frac{F^{(3,2)}}{2} \right)\left( \xi + \frac{F^{(3,2)}}{2} \right) + F^{(6,3)}F^{(0,1)} = 0.
\end{align}
From the perspective of the gauged linear sigma model (type IIA string theory), therefore, the K\"ahler parameter phase diagram 
is like Figure~\ref{fig:kahler-cone}~(a) in the large base $\P^1_A$ regime. From the perspective of classical geometry, on the other hand,
there is a holomorphic biregular map from $M_{\vev{+2}}^{-\nu^1_F}$ in the $t_{{\rm vrt},I} < 0$ phase to that in the $t_{{\rm vrt},I}>0$ 
phase so that the singular fibre components $V_0={\rm Bl}^{18}(\P^2)$ and $V_1=\P^2$ in the $t_{{\rm vrt},I}<0$ phase are identified 
with $V_1={\rm Bl}^{18}(\P^2)$ and $V_0 = \P^2$, respectively. This isomorphism effectively cuts out the $t_{{\rm vrt},I}<0$ part 
of the K\"ahler moduli space. This is consistent with the fact that only one triangulation is found for the polytope 
$\widetilde{\Delta}$ under consideration. 

Applying the adiabatic argument of duality to the fibre over generic points around the degeneration point $t_0$, we find 
that the heterotic interpretation is to have a soliton (defect) localized at real codimension-two in the base $\P^1_{\rm Het}$. 
We cannot say much about what happens at the centre of the soliton (as the adiabatic argument breaks down there), but duality 
indicates that there is a U(1) vector multiplet associated with this soliton, at least for generic choice of moduli. 
An extended discussion on the heterotic string interpretation is provided in section \ref{ssec:Het-interpret}. Before we 
get there, we study a few more examples of degeneration in lattice-polarized K3-fibration in 
sections \ref{ssec:corridor-II} and \ref{ssec:corridor-III}.

\subsection{Corridor Branches and Reducible Fibres}
\label{ssec:corridor-II}

Once we allow the K3 fibre to degenerate and become reducible at isolated points in the base $\P^1_A$, 
we do not need to restrict to a construction where we choose to include \emph{just one} vector $\nu^6_F$ 
from $2\widetilde{\Delta}_F \cap N_F$ in the polytope $\widetilde{\Delta}$.
The moduli spaces of type IIA compactifications for this broader class of threefolds $M_{\Lambda_S}$ form 
bridges (or corridors) between the branches of the moduli space corresponding to the multiple fibration choices 
of a given algebraic K3 surface $\Lambda_S \sim {\rm NS}_{\rm K3}$ discussed in the previous section. 

This section will cover the geometry of threefolds where the reducible fibre is 
a Type II degeneration (as in section \ref{ssec:a17fibres}). Examples with a reducible fibre other than 
a Type II degeneration are postponed to section \ref{ssec:corridor-III}.

\subsubsection{Warm-up}

In the context of heterotic--type IIA duality, the best-known example of a K3-fibred Calabi--Yau threefold 
with a reducible fibre corresponding to a Type II degeneration is the case with $\Lambda_S = U$ (i.e., $E_8$-elliptic K3 
is in the fibre in type IIA compactification). When we include all the vectors $\nu^{6,n}_F$ 
in (\ref{eq:choice-Fn}) within a range $(-n_v) \leq n \leq n_h$ for some integers $-2 \leq (-n_v) < n_h \leq +2$, 
then the heterotic dual is a ${\rm K3} \times T^2$ compactification with $12-n_v$ and $12-n_h$ instantons in 
$E_8 \oplus E_8$ on K3, along with $n_h+n_v$ NS5-branes in the $S^1/\Z_2$ interval of the heterotic-M theory. 
The moduli space of this threefold $M_U^{ \{ n_h, \cdots, -n_v\} }$ forms a branch with more vector multiplets and 
fewer hypermultiplets 
\begin{align}
  h^{1,1}(M_U^{ \{n_h,\cdots, -n_v \} }) &= \rho+1+(n_h+n_v),    \label{eq:E8-ell-h11-h21-tradeoff-11} \\
  h^{2,1}(M_U^{\{ n_h,\cdots, -n_v \}}) &= 244 - 29(n_h+n_v),    \label{eq:E8-ell-h11-h21-tradeoff-21}
\end{align}
than any one of the moduli spaces for $M_{U}^{n}$ with $-n_v \leq n \leq n_h$. The moduli of $M_U^{\{ n_h, \cdots, -n_v\} }$ 
connects the moduli spaces of $M_U^n$ with $-n_v \leq n \leq n_h$ by a trade-off between 
the Coulomb and Higgs branch degrees of freedom \cite{MV-2}. 

\begin{figure}[tbp]
\begin{center}
\begin{tabular}{ccc}
    \includegraphics[width=.25\linewidth]{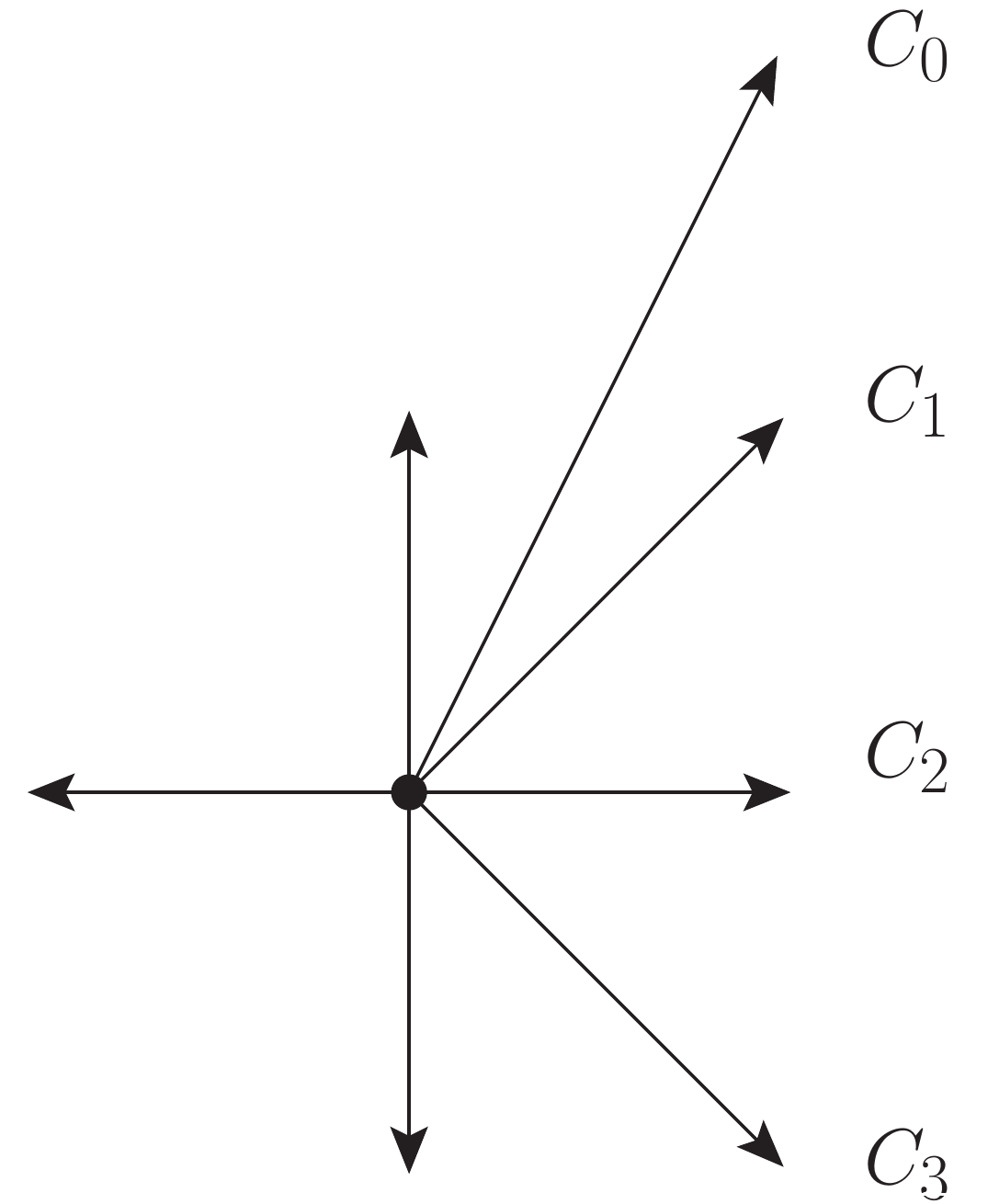} & $\quad$ &
    \includegraphics[width=.55\linewidth]{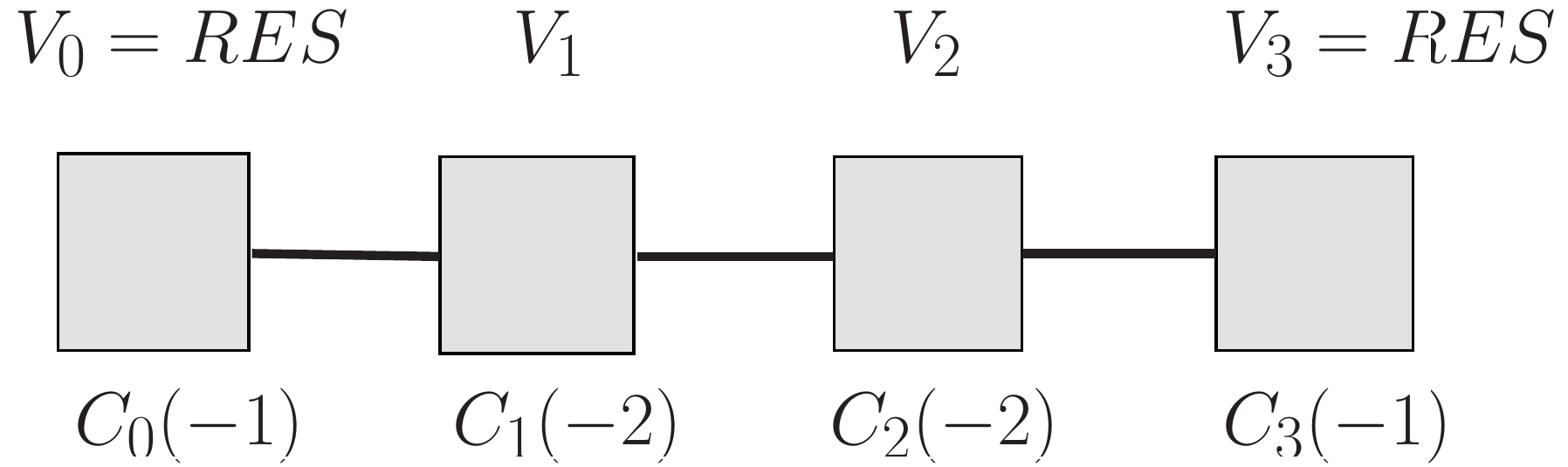} \\
  (a) &  & (b) 
\end{tabular}
\caption{\label{fig:NS5-Ftheory}
The toric vectors for the base $B'$ (a), and the graph of intersection of curves $C_0 \cup \cdots \cup C_{n_h+n_v}$ in $B'$ (b). 
The latter can also be regarded as the dual graph of a Type II degeneration of an $E_8$-elliptic K3 surface.}
\end{center}
\end{figure}

In the context of F-theory compactification, this threefold geometry $M_U^{\{n_h,\cdots, -n_v \}}$ is understood 
as an elliptic fibration $\pi'_M: M_U^{\{n_h, \cdots, -n_v \} } \rightarrow B'$ over a base surface 
$B' = {\rm Bl}^{n_h+n_v}(F_{n_h})$. A Hirzebruch surface $F_{n_h}$ is a $\P^1$ fibration over $\P^1_A$. Blowing up 
a Hirzebruch surface to $B'$, the fibration is modified in such a way that the fibre \emph{curve} $\P^1$ degenerates 
into $(n_h+n_v+1)$ curves, $C_0 \cup \cdots \cup C_{n_h+n_v}$; the graph of intersection of those curves in $B'$ 
is shown in Figure \ref{fig:NS5-Ftheory} (b).
\label{pg:E8E8-ex}
For the more general heterotic--type IIA duality, however, it is more suitable to describe the 
geometry of threefold $M_U^{\{ n_h, \cdots, -n_v\}}$ in terms of a degeneration processes of the 
K3 \emph{surface} $\pi_M: M_U^{\{n_h, \cdots, -n_v\}} \rightarrow \P^1_A$.
The complex structure of $M_U^{\{n_h,\cdots, -n_v\}}$ has been tuned so much (relatively to that of $M_U^{n}$ for 
any $-n_v \leq n \leq n_h$) that the fibre K3 surface---$E_8$-elliptic K3 surface generically---is forced to 
degenerate to a collection of $(n_h+n_v+1)$ irreducible non-singular surfaces 
$V_0 \cup V_1 \cup \cdots \cup V_{n_h+n_v}$ over one point in the base $\P^1_A$. 
The surfaces over the curves $C_0$ and $C_{n_h+n_v}$ are rational elliptic surfaces, $V_0=V_{n_h+n_v}={\rm RES}$ (also 
known as $dP_9$ in physics community), while the surfaces over $C_1, \cdots, C_{n_h+n_v-1}$ are all $T^2 \times \P^1 = 
V_{1,\cdots,n_h+n_v-1}$. This is an example of a Type II degeneration of a $(\Lambda_S = U)$-polarized K3 surface. 
This is the language suitable for heterotic--type IIA duality. 

\subsubsection{Corridor branches among models with a degree-2 K3 surface}
\label{sssec:corridor-E7D10}

Examples of this kind are also available in the case of Calabi--Yau threefolds with a degree-2 K3 surface 
($\Lambda_S = \vev{+2}$) as the fibre over $\P^1_A$. 
As a first group of examples, consider the threefolds $M_{\vev{+2}}^{ \{n,n-1,\dots,m \}}$ labelled by a choice of integers 
$n, m$ satisfying $2 \geq n \geq m \geq -1$. They are obtained as toric hypersurfaces for which the 
polytope $\widetilde{\Delta}$ contains all of the $\nu^{6}$'s with $n \geq k_4 \geq m$ (and 
$k_2=k_3=0$). An example of a ``short top'' \cite{short_tops} is found as a part of this polytope 
$\widetilde{\Delta}$. It turns out that Hodge numbers of those threefolds are as follows:
\begin{equation}\label{eq:singtransM3}
 \begin{array}{ccccccc|cc}
  M_{\vev{+2}}^2 & & M_{\vev{+2}}^1 & & M_{\vev{+2}}^0 & & M_{\vev{+2}}^{-1} &  h^{1,1} = 2 & h^{2,1} = 128\\
  & M_{\vev{+2}}^{\{2,1\}} & & M_{\vev{+2}}^{\{1,0\}} & & M_{\vev{+2}}^{\{0,-1\}} & & h^{1,1}=3 & h^{2,1} = 111\\
  & & M_{\vev{+2}}^{\{2,1,0\}} & & M_{\vev{+2}}^{\{1,0,-1\}} & & & h^{1,1} = 4 & h^{2,1} = 94\\
  & & & M_{\vev{+2}}^{\{2,1,0,-1\}} & & & & h^{1,1}=5 & h^{2,1} = 77
 \end{array}
\end{equation}
We observe, in these examples, that the value of $h^{1,1}$ and $h^{2,1}$ of $M_{\vev{+2}}^{\{n,n-1, \cdots, m\} }$ 
depend \emph{only} on $(n-m)$, just like they do on $(n_h+n_v)$ in (\ref{eq:E8-ell-h11-h21-tradeoff-11}, 
\ref{eq:E8-ell-h11-h21-tradeoff-21}). 

Let us first focus on the geometry of $M_{\vev{+2}}^{\{n,n-1\} }$. A generic fibre $S_{t.A}$ ($t \in \P^1_A$) in those 
threefolds is a degree-2 K3 surface, but this K3 surface degenerates at one point ($X_6=0$) in the base $\P^1_A$.
The singular fibre at the degeneration, referred to as the central fibre and denoted by $S_0$, consists 
of two irreducible pieces, $V_0 \cup V_1$. Let $V_0$ be the divisor $\bar{D}_{6,n}$ in $M_{\vev{+2}}^{\{n,n-1 \} }$ 
for definiteness; $V_1=\bar{D}_{6,n-1}$ then. Both of these surfaces combined, $\bar{D}_{6,n}+\bar{D}_{6,n-1} = S_0$ 
are linearly equivalent to the generic fibre class $\sim \bar{D}_5$. We found, by using computation techniques 
available for toric hypersurfaces \cite{DK} (plus additional formula in Appendix \ref{ssect:hodge-num-div}), 
that those two irreducible surfaces satisfy 
\begin{align}\label{eq:hodgenumtrans0-1}
h^{1,1}(\bar{D}_{6,n-1}) = 8, \qquad 
h^{1,1}(\bar{D}_{6,n}) = 12, \qquad  h^{i,0}(\bar{D}_{6,n-1})=h^{i,0}(\bar{D}_{6,n})=0 \quad (i=1,2).
\end{align}
Those two components meet along a curve of genus 1. This information is summarized in the first line 
of Table \ref{tab:central-fibre}.
\begin{table}
\begin{center}
\begin{tabular}{|cc|c|c|c|c|c|}
\hline
 $V_0$ & $V_1$ & $h^{1,1}(M)$ & $h^{2,1}(M)$ & $-\Delta h^{2,1}(M)$ & $h^{1,1}(V_0)$ & $h^{1,1}(V_1)$ \\
\hline 
 $[n_a]$&$[n-1_a]$ & 3 & 111 & 17 & 12 & 8 \\
 $[-1_2]$&$[1_4]$ & 3& 111 & 17 & 12 & 8 \\
 $[-1_4]$&$[-1_2]$ & 3& 99 & 29 & 9/11 & 11/9 \\
 $[1_4]$& $[1_2]$ & 3& 99 & 29 & 9/11 & 11/9 \\
 $[2_4]$& $[-1_2]$ & 3& 99 & 29 & 9/11 & 11/9 \\
  \hline
\end{tabular}
\caption{\label{tab:central-fibre} Data of irreducible components of the central fibres in threefolds $M_{\vev{+2}}$ 
obtained by using two ``neighbouring'' points in Figure~\ref{fig:choicesofabc} for $\nu^6_F$. A pair of 
points corresponding to the first two rows are connected by a solid line in Figure~\ref{fig:choicesofabc}, 
while a pair corresponding to the next two rows by a dotted line in Figure~\ref{fig:choicesofabc}. A pair corresponding to the last 
row are connected by a dashed line in Figure~\ref{fig:choicesofabc}. In the last three rows, the value of 
$h^{1,1}(V_0)$ and $h^{1,1}(V_1)$ can be 9 and 11, or 11 and 9, respectively, depending on the choice of triangulation 
of the polytope $\widetilde{\Delta}$ (see the text for more information).}
\end{center}
\end{table}

The moduli space of $M_{\vev{+2}}^{\{n,n-1\}}$ and that of $M_{\vev{+2}}^n$ are connected. The transition 
locus between these two branches is reached from $M_{\vev{+2}}^{\{n,n-1\}}$ by tuning a K\"{a}hler parameter such that 
the surface $V_1 = \bar{D}_{6,n-1}$ collapses to a point, and it is reached from $M_{\vev{+2}}^n$
by tuning 17 complex structure parameters. At the transition, the geometry has a point-like singularity of type $\widetilde{E}_7$, which is captured by
\begin{align}
 X_1^2 + F^{(4)}(X_2,X_3,X_6) \simeq 0
\label{eq:E7-tilde}
\end{align}
($[n_a]$-$[n-1_a]$ with $a=4$ is used for this expression). As discussed in \cite{reid_canonical,MV-2}, this type of singularity is reached by collapsing a $dP_7$. This can also be seen explicitly by observing that $V_1 = \bar{D}_{6,n-1}$ is described as a hypersurface of degree $4$ in $\mathbb{P}^3_{2111}$, which is a well-known realization of $dP_7$. 

The moduli space of $M_{\vev{+2}}^{\{n,n-1\}}$ is also connected to that of $M_{\vev{+2}}^{n-1}$.
Here, we can reach the transition point from $M_{\vev{+2}}^{\{n,n-1\}}$ by collapsing the surface $\bar{D}_{6,n}$ to zero 
volume to form a singular threefold, whereas we need to tune $17$ complex structure moduli of $M_{\vev{+2}}^{n-1}$ to reach it. At the transition, $M_{\vev{+2}}^{n-1}$ develops an $A_1$ singularity along a curve $X_1=X_4=X_6=0$, and the surface 
$V_0 = \bar{D}_{6,n}$ comes out as the exceptional divisor when this singularity is resolved.\footnote{
At the transition point, the gauge group is enhanced to $\SU(2) \times \U(1)^2$, and $N_F=10$ $\SU(2)$-doublet 
hypermultiplets emerge in the massless spectrum, in the 4D N=2 effective theory.} 
It turns out that $V_0$ can be regarded as ${\rm Bl}^{10}(F_2)$, see appendix \ref{ssec:E7D10-CS} for more information.

Therefore, type IIA compactification on $M_{\vev{+2}}^{\{n,n-1\}}$ has transitions both to a compactification on $M_{\vev{+2}}^n$ and a compactification on $M_{\vev{+2}}^{n-1}$. The moduli spaces of all of the $M_{\vev{+2}}^n$ with $n=2,1,0,-1$ are connected in this way.

Similarly, the threefold $M_{\vev{+2}}^{\{ 1_4, -1_2\}}$ provides a branch of moduli space connecting $M_{\vev{+2}}^{1}$ and $M_{\vev{+2}}^{-1}$. In fact, it turns out that the geometry of the degenerate singular fibre in $M_{\vev{+2}}^{\{ 1_4, -1_2\}}$ is the same as in the $M_{\vev{+2}}^{\{n,n-1\}}$ branch connecting those of $M_{\vev{+2}}^{n-1}$ and $M_{\vev{+2}}^n$. Although the $S_3$ symmetry of the graph of Figure~\ref{fig:choicesofabc} does not explain why the central fibre geometry is the same for those two transition channels, certainly the $[-1_2]$-$[1_4]$ pair is a nearest neighbour link in the graph in Figure~\ref{fig:choicesofabc} (just like the pairs $[n_a]$-$[n-1_a]$ are), and this agreement of the central fibre geometry may be just a trivial consequence of $M_{\vev{+2}}^{2}=M_{\vev{+2}}^{-1}$ as complex geometry.

By tuning more complex structure moduli and subsequent resolution, we can also reach the manifolds $M_{\vev{+2}}^{\{n,n-1,\cdots, m\} }$. Here, the degree-2 K3 surface in the fibre degenerates over one point ($X_6=0$) of the base $\P^1_A$ to a central fibre $S_0 = V_0 \cup V_1 \cup \cdots \cup V_{n-m}$. This is another example of Type II degeneration of degree-2 K3 surface. The dual graph of the irreducible components, $V_0, \cdots, V_{n-m}$ is a chain starting from a node for $V_0$ and ending with a node for $V_{n-m}$. This graph comes directly from an edge (at the height= +1) of the 
polytope $\widetilde{\Delta}$. This is an example of a theorem in \cite{short_tops}.

The rational surfaces at the end of the chain remain unchanged, $V_0={\rm Bl}^{10}(F_2)$, $V_{n-m}=dP_7$, 
while the surface components in the middle, $V_1, \cdots, V_{n-m-1}$ are all identical surfaces that are ruled over 
the elliptic curve $C \cong (V_{i} \cap V_{i+1})$. They are isomorphic\footnote{
Here, we follow the conventions of Chap.II.7 of \cite{Hartshorne} for $\P[{\cal E}]$ for some vector bundle ${\cal E}$, as opposed to
the convention often used in physics literature.} to $\P[{\cal O}_C \oplus {\cal L}]$ 
for some degree $(-2)$ line bundle ${\cal L}$ on $C$ (cf Chap.V.2 of \cite{Hartshorne}). The value $(-2)$ is tied to 
the self-intersection of the double curves $C = V_i \cap V_{i+1}$ (and the degree of $dP_7$). This ruled surface 
does not admit an elliptic fibration morphism, see e.g. \cite{MOflow}.
More information is provided in the appendices \ref{sec:K3-degen-review} and \ref{ssec:E7D10-CS}.

\subsubsection{The K\"ahler moduli space of \texorpdfstring{$M_{\vev{+2}}^{\{0,-1\} }$}{Lg}}
\label{sssec:kahler-E7D10}

Let us focus on $M_{\vev{+2}}^{\{0,-1\} }$ and have a closer look at the K\"{a}hler moduli space. Recall that the K3 
fibration has a single reducible fibre with irreducible surface components corresponding to $\overline{D}_{6,0}$ and 
$\overline{D}_{6,-1}$ with $\chi(\overline{D}_{6,0}) = 14$ and $\chi(\overline{D}_{6,-1}) = 10$. There are 5 different fine, 
regular, star triangulations of $\widetilde{\Delta}$. 
\begin{figure}[h]
\begin{center}
    \scalebox{.45}{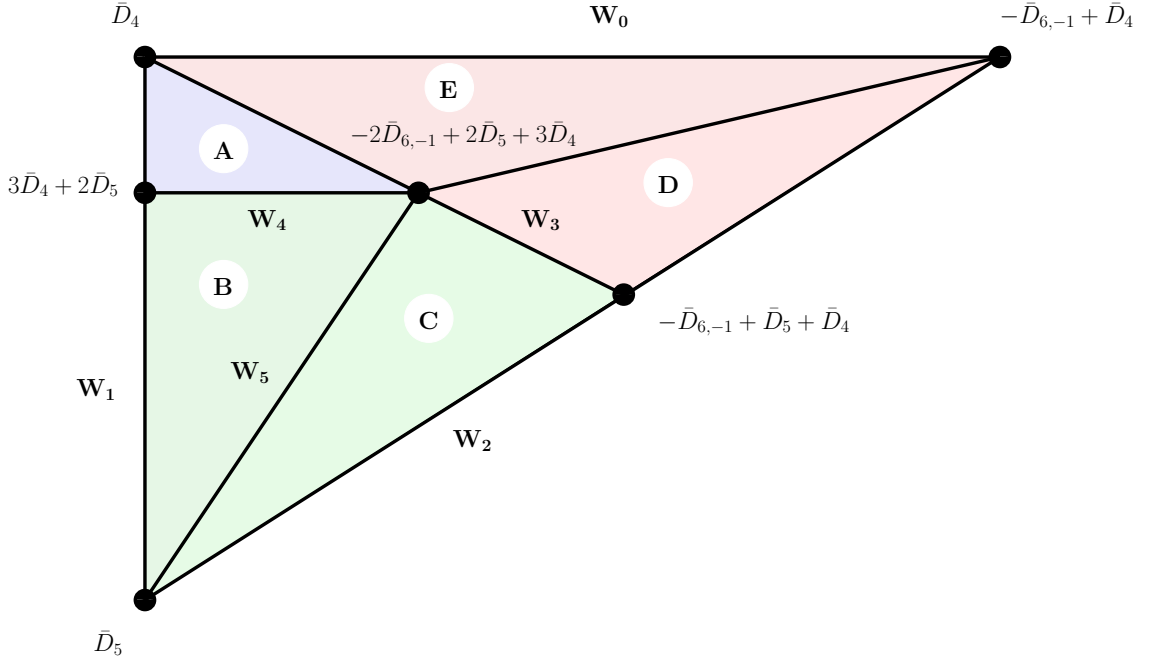}
 \caption{\label{fig:five-chambers}
The chamber structure of the K\"ahler parameters of the toric ambient space of $M_{\vev{+2}}^{\{0,-1\}}$. The figure shows a projection preserving the relative location of the rays of the five three-dimensional cones corresponding to different triangulations. We have labelled the five phases by ${\bf A}$ to ${\bf E}$ and indicated where the various walls mentioned in the text are located.
}
\end{center}
\end{figure}
Accordingly, there are 5 chambers in the K\"ahler moduli space of the toric ambient space, see Figure~\ref{fig:five-chambers}. 
Only two of those---phase B and C in Figure~\ref{fig:five-chambers}---result in 
a toric ambient space with an (obvious) toric fibration morphism to $\P^1_A$ from which the K3 fibration follows. At the wall 
between these two phases, there are surfaces and curves of the ambient space which collapse, but it seems the Calabi-Yau hypersurface 
we are interested in stays perfectly smooth.\footnote{While it is not hard to work out that none of the curves obtained by intersecting a surface in the ambient space with $M_{\vev{+2}}^{\{0,-1\}}$ collapses at this wall, it is much harder to exclude that there is no collapsing curve of the ambient space sitting entirely inside $M_{\vev{+2}}^{\{0,-1\}}$. An example of this phenomenon is given by the 18 $(-1)$ curves contained in one of the fibre components for the model discussed in Section \ref{ssec:a17fibres}. While we do not have a candidate for a similar behaviour in this case, this is of course not enough to rigorously exclude such a thing.} In such a case, we can glue these two cones together and treat them as a single phase \cite{Cox_Katz_book}. 

Let us parametrize the K\"ahler moduli space by
\begin{equation}
J = t_{4,I} \overline{D}_4 + t_{{\rm vrt},I} \overline{D}_{6,-1} + t_{5,I} \overline{D}_5 \, .
\end{equation}
The five geometric phases as a whole are delineated by the walls 
\begin{align}
 W_0:& \hspace{.5cm} t_{5,I}> 0, \\
 W_1:& \hspace{.5cm} (-t_{{\rm vrt},I}) > 0 \\
 W_2:& \hspace{.5cm} t_{4,I}+t_{{\rm vrt},I} > 0. 
\end{align}
The wall 
\begin{align}
W_3: \hspace{.5cm} (t_{{\rm vrt},I} + t_{5,I}) > 0
\end{align}
separates the phases A and B+C from the phases D and E. The phase B is distinguished \footnote{The phase B is 
distinguished from the phase C by the wall $W_5: 2t_{4,I}+3t_{{\rm vrt},I} > 0$.} from the phase A 
by the wall 
\begin{align}
 W_4:& \hspace{.5cm} 3t_{5,I}-2t_{4,I} > 0 .
\end{align}
Similarly to $W_5$, there seems to be no curve inside $M_{\vev{+2}}^{\{0,-1\} }$ which collapses at $W_4$, so that we expect this wall to be 
fictitious and phase A should be combined with the phase B+C. At the level of the ambient space, however, 
the triangulation in the phase A is not compatible with the projection to $P^1_A$, and hence we cannot obtain 
K3-fibration morphism $M_{\vev{+2}}^{\{2,1\}} \rightarrow \P^1_A$ as restriction of toric fibration morphism of the 
ambient space. While the Calabi-Yau $M_{\vev{+2}}^{\{0,-1\} }$ is most likely still K3 fibred after we cross $W_4$, 
we cannot confidently speak about the projection to $\P^1_A$ in phase A. 

The walls $W_1$ and $W_2$ are dual to curve classes $C_1$ and $C_2$, where $C_1$ is represented by one of seven $(-1)$ 
curves in $\bar{D}_{6,-1} = dP_7$, and the class $C_2$ by one of ten $(-1)$ curves in $\bar{D}_{6,0} = {\rm Bl}^{10}(F_2)$. 
The volume of $\bar{D}_{6,-1} = dP_7$ also vanishes at the wall $W_1$, and that of $\bar{D}_{6,0} = {\rm Bl}^{10}(F_2)$ 
at the wall $W_2$. There is no flop available in the compact manifold $M_{\vev{+2}}^{\{0,-1\}}$ acting on those $(-1)$ 
curves in the central fibre $S_0 = \bar{D}_{6,0} + \bar{D}_{6,-1}$. There is not even a limit of K\"ahler parameters 
where the volume of those $(-1)$ curves vanish while keeping the volume of $\bar{D}_{6,-1}$ and $\bar{D}_{6,0}$ 
non-zero. 

In the large base regime 
\begin{equation}
t_{5,I} \gg |t_{4,I}|, |t_{{\rm vrt},I}|,  
\label{eq:large-base-E7D10}
\end{equation}
only the phases B and C can be realized.\footnote{Here, we exclude A as we cannot rigorously establish the existence of a K3 fibration in this phase.} The phase diagram in this context is given by Figure~\ref{fig:kahler-cone}~(b). 

It is worth noting that the geometric phases D and E are available only outside of the large base regime (\ref{eq:large-base-E7D10}). Given the fact that one can take a detour around the wall of K\"ahler cone by turning on $B$-fields, the heterotic--type IIA duality map should extended (at least via analytic continuation) to the geometric phases which are not compatible with a K3-fibration, at least not in an obvious way. Because the large base regime (\ref{eq:large-base-E7D10}) corresponds to the weak coupling regime in heterotic string compactifications, the geometric phases D and E should be mapped to strongly coupled phase of heterotic string compactifications. It would be interesting to explore this territory, but this is beyond the scope of this article. 

\subsubsection{More transitions and degenerate fibres}
\label{sssec:root3}

Besides the singular transitions we have discussed, there are others which connect different threefolds along the ``links of length $\sqrt{3}$'' in Figure \ref{fig:choicesofabc}. Let us first discuss an example where we include the lattice points $[2_4]$ and $[-1_2]$ as vertices of $\widetilde{\Delta}$. 
The resulting threefold $M_{\vev{+2}}^{\{2_4,-1_2\}}$ has Hodge numbers
\begin{equation}
 h^{1,1}(M_{\vev{+2}}^{\{2_4,-1_2\}}) = 3 \, , \hspace{.5cm} h^{2,1}(M_{\vev{+2}}^{\{2_4,-1_2\}}) = 99 \, ,
\end{equation}
which signals a single reducible fibre with two irreducible components. By construction, this Calabi-Yau threefold sits in between the threefolds $M_{\vev{+2}}^{n=2}$ and $M_{\vev{+2}}^{n=-1}$. As before, these can be reached by blowing down one of the fibre components, followed by a subsequent deformation.

\begin{figure}[tbp]
 \begin{center}
  \includegraphics[width=4cm]{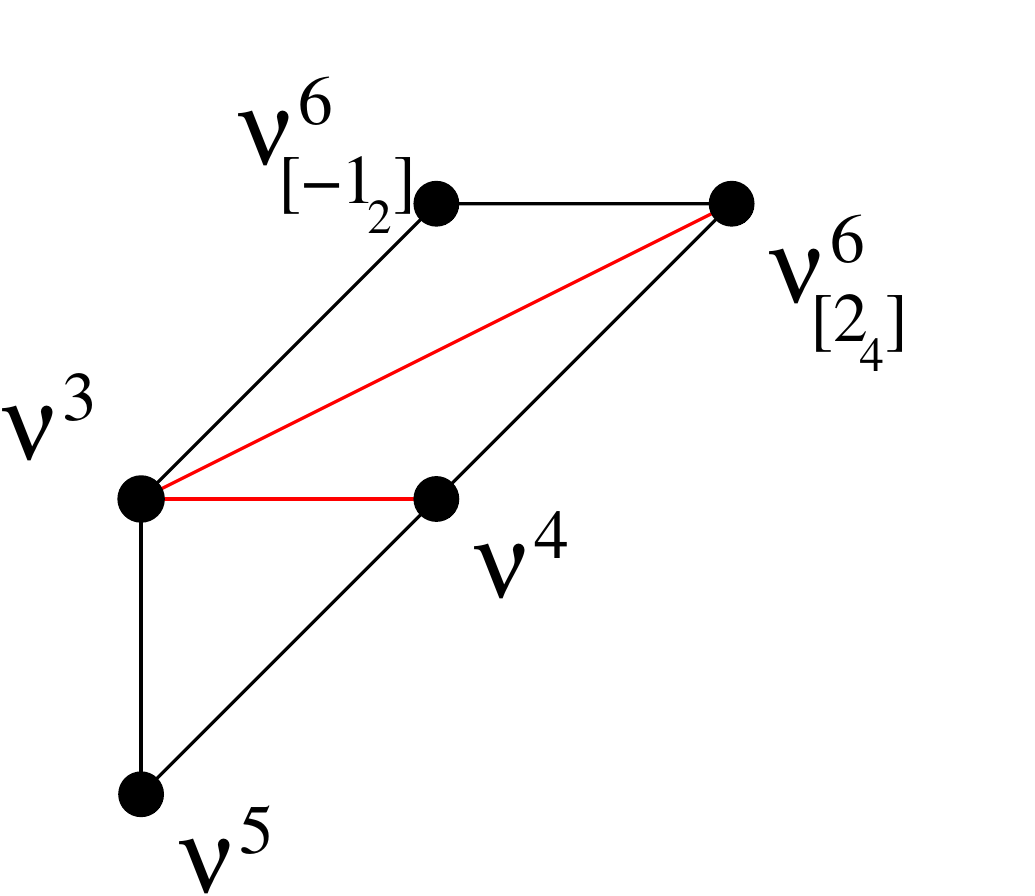}\hspace{1cm}
  \includegraphics[width=4cm]{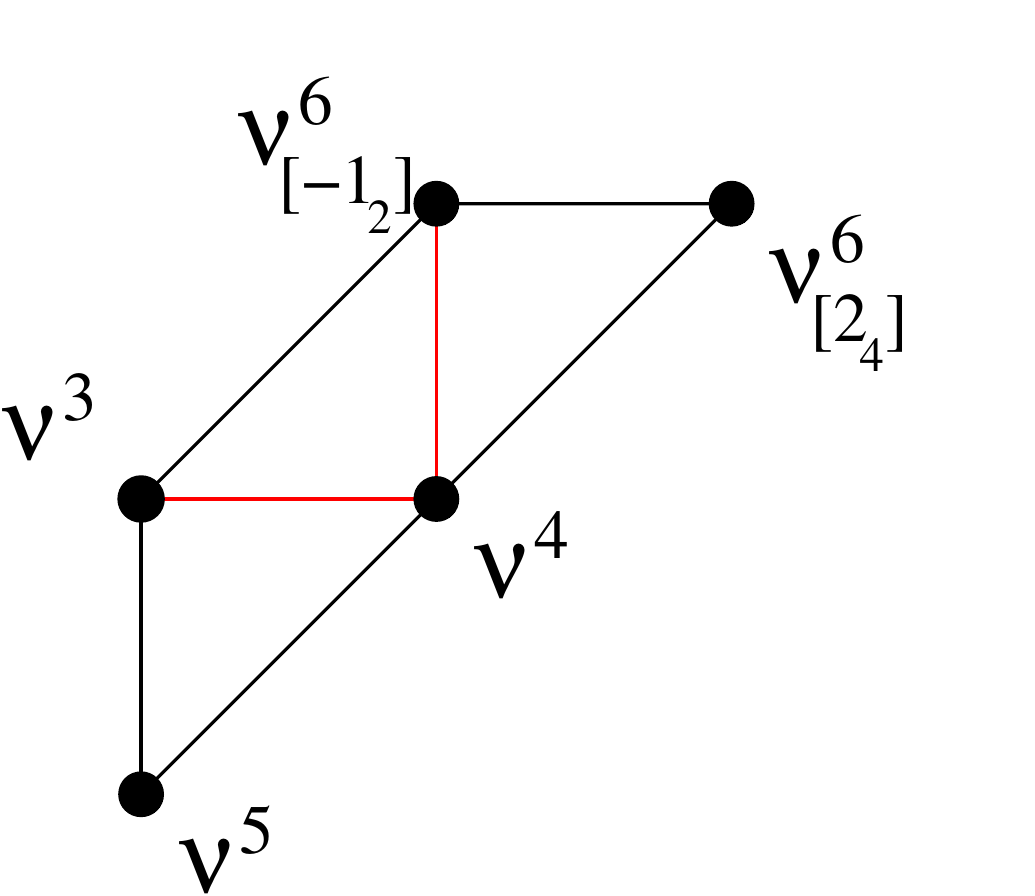}\hspace{1cm}
  \includegraphics[width=4cm]{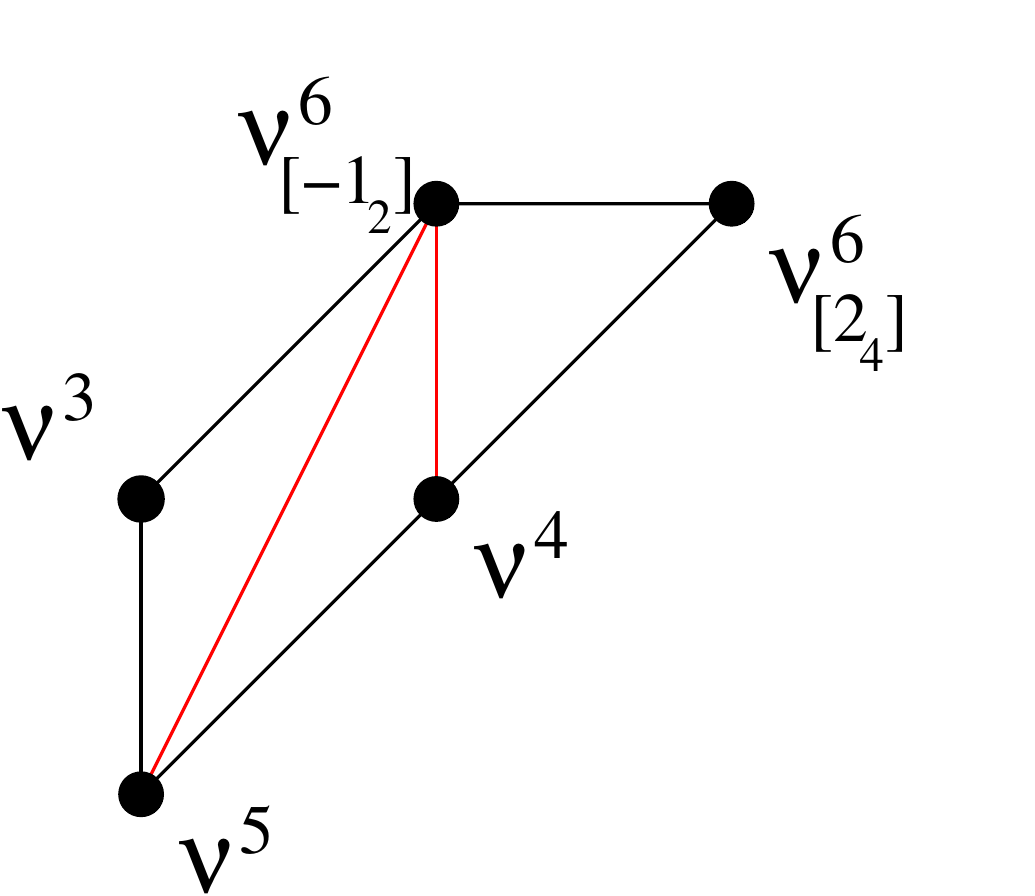}
 \end{center}
 \caption{Three different triangulations of a face giving rise to different phases of the Calabi-Yau threefold 
$M_{\vev{+2}}^{\{2_4,-1_2\}}$.  \label{fig:three_phases_11_13} }
\end{figure}

The polytope $\widetilde{\Delta}$ has a two-dimensional face which contains the lattice points $\nu^2$, $\nu^3$, 
$\nu^6_{2_4}$, $\nu^6_{-1_2}$ and $\nu^5$. The three different triangulations of this face 
(Figure \ref{fig:three_phases_11_13}) give rise to three different torically realized phases of $M_{\vev{+2}}^{\{2_4,-1_2\}}$. 
The geometries corresponding to the triangulation on the left and in the middle only differ in how the two components 
of the singular fibre are distributed among $\overline{D}_{6,2_4}$ and $\overline{D}_{6,-1_2}$. These two divisors are 
rational for any triangulation, and they obey
\begin{equation}
 \chi(\overline{D}_{2_4}) = 13 \hspace{1cm}  \chi(\overline{D}_{-1_2}) = 11
\end{equation}
for the triangulation shown on the left of Figure \ref{fig:three_phases_11_13} and
\begin{equation}
 \chi(\overline{D}_{2_4}) = 11 \hspace{1cm}  \chi(\overline{D}_{-1_2}) = 13
\end{equation}
for the triangulation shown in the middle of Figure \ref{fig:three_phases_11_13}. The two phases are connected by a flop which brings two $(-1)$ curves from one fibre component to the other. In each of the two cases, the divisor with $\chi = 11$ is a $dP_8$ realized as a hypersurface of degree $6$ in $\P^3_{3211}$, whereas the divisor with $\chi = 13$ is a blowup of $dP_8$ at two points. 

The phase corresponding to the triangulation shown on the right hand side of Figure \ref{fig:three_phases_11_13} does not respect the $K3$ fibration we intend to use for the duality between type IIA and heterotic string theory. Starting from the phase in the middle of Figure \ref{fig:three_phases_11_13}, we can reach the phase on the right hand side by passing through a wall of the K\"ahler cone. On the boundary of the K\"ahler cone in question, 
the curve $\overline{D}_{3}\cdot \overline{D}_{4}$ is collapsed, before another small resolution takes us to the phase corresponding to the triangulation shown on the right. This curve is projected surjectively on the base $\P_A^1$ of the K3 fibration, which means that we can get there only outside of the large base regime. 

The same fibre geometries, including the flops discussed above, are realized by threefolds connected along other edges of length $\sqrt{3}$ such as $M_{\vev{+2}}^{\{-1_i,-1_j\}}$ and $M_{\vev{+2}}^{\{1_i,1_j\}}$. The 3rd and 4th rows in Table \ref{tab:central-fibre} speak about that. However, the different phases cannot all be seen torically for all of these models. One sometimes has to go beyond toric hypersurfaces to realize the extended K\"ahler cone of the Calabi-Yau manifold $M$, as we have remarked already in footnote \ref{fn:extnd-kahler}. 

\begin{figure}[h!]
\begin{center}
 \includegraphics[width=5cm]{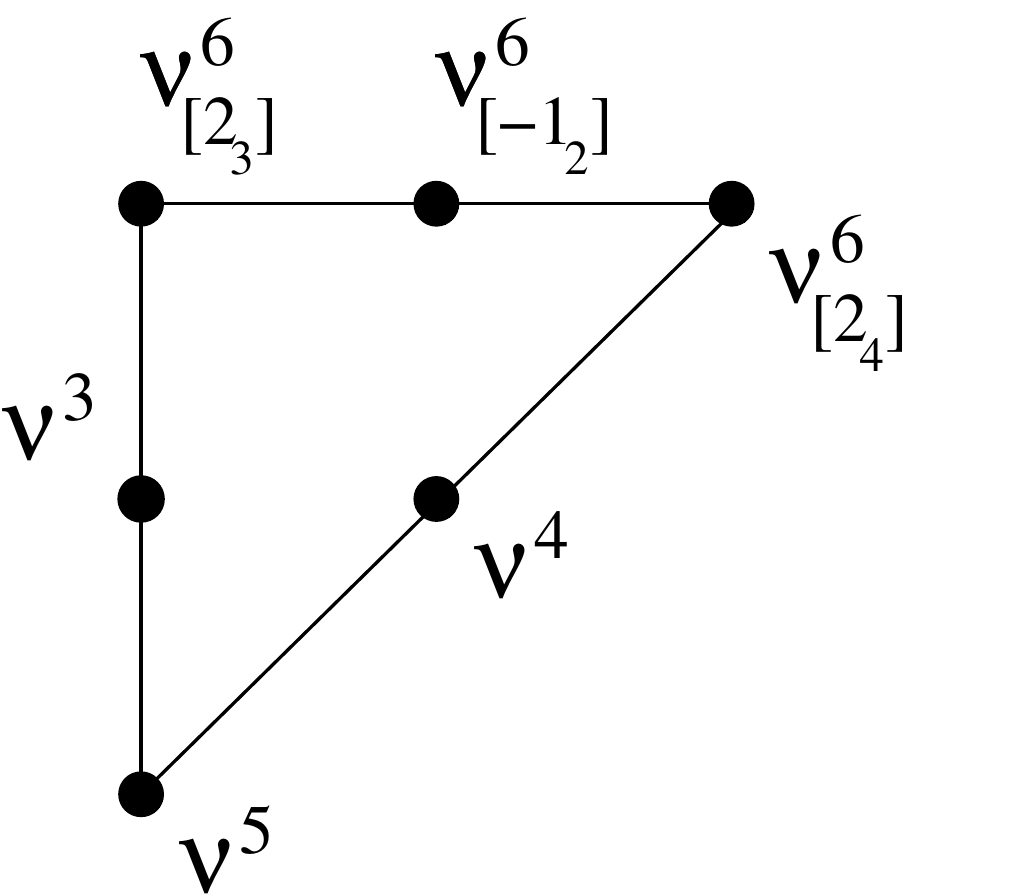}
\end{center}
\caption{The face of $M_{\vev{+2}}^{\{2_4,-1_2,2_3\}}$ giving rise to different phases. There are 4 different triangulations, three of which are compatible with the fibration structure. \label{fig:face_11_2_11}}
\end{figure}

In parallel to the models $M_{\vev{+2}}^{\{n,n-1,\cdots, m\} }$, we can construct a model with a longer chain of fibre components by including $\nu^6_{[2_3]}$ as a vertex along with $\nu^6_{[2_4]}$ and $\nu^6_{[-1_2]}$; the lattice point $\nu^6_{[-1_2]}$ ceases to be a vertex of $\widetilde{\Delta}$ then. This leads to a model with a fibre with three components, its Hodge numbers are 
\begin{equation}
h^{1,1}(M_{\vev{+2}}^{\{2_4,-1_2,2_3\}}) = 4 \, , \hspace{.5cm} h^{2,1}(M_{\vev{+2}}^{\{2_4,-1_2,2_3\}}) = 70 \, .
\end{equation}
In this case, there are three different triangulations respecting the K3 fibration. They originate from different triangulations of a two-dimensional phase containing the lattice points $\nu^2$, $\nu^3$, $\nu^5$ together with $\nu^6_{2_4}$, $\nu^6_{-1_2}$ and $\nu^6_{2_3}$. This face is shown in Figure \ref{fig:face_11_2_11}.
The Euler characteristics of the three fibre components for different triangulations are 
\begin{equation}
 \begin{array}{ccc}
  V_0 = \overline{D}_{6,2_4} & V_1 = \overline{D}_{6,-1_2} & V_2 = \overline{D}_{6,2_3} \\
  11  & 0 & 13 \\
  11 & 2 & 11 \\
  13 & 0 & 11
 \end{array}
.
\end{equation}
As before, the fibre components with Euler characteristic $11$ are $dP_8$ surfaces, and those with $13$ are $dP_8$ surfaces blown 
up in two points. The $\chi = 0$ component in the middle has a ruling over an elliptic curve $C = V_i \cap V_{i+1}$; 
it is in the form of $\P[{\cal O}_c \oplus {\cal L}]$ for a degree $(-1)$ line bundle $\mathcal{L}$ on $C$, because $(C)^2 = +1$ in 
$dP_8$. Starting from the first triangulation, a flop blows down two $(-1)$ curves in $V_2$, so both $V_0$ and $V_2$ turn 
into $dP_8$ in the second triangulation. At the same time, two points in $V_1$ along $V_1 \cap V_2$ are blown up, so 
the ``ruling'' $(\P^1$-fibration) in $V_1$ splits into $\P^1+\P^1$ over two points in $C$ and now $\chi(V_1)= 2$. The phase 
for the last triangulation is reached by a flop along the other $\P^1$ in the $\P^1+\P^1$ fibre (simultaneously at the 
two such fibres). Given the symmetry between the $[2_4]$ and $[2_3]$ vertices in Figure~\ref{fig:choicesofabc}, it is 
reasonable that these flops exist, so that there is no asymmetry between $V_0=\bar{D}_{6,2_4}$ and $V_2 = \bar{D}_{6,2_3}$.

There exists a fourth phase accessible via triangulation for which $\overline{D}_5 \cdot \overline{D}_{6,-1_2} \neq 0$. so that this phase cannot respect the K3 fibration. Again, this non-fibred phase can be reached outside of the large base regime. 

The general feature of all the examples with degenerate fibres discussed so far is that they correspond to Type II degenerations in the sense of Kulikov. 
The degenerate fibre is always in the form of $V_0 \cup V_1 \cup \cdots \cup V_\mu$. There are sometimes multiple geometric phases that are compatible with a K3-fibration, however, and details of the fibre geometry change from one phase to another. The monodromy of a generic fibre around the degeneration locus in the base $\P^1_A$, on the other hand, remains invariant under such birational transformations of the degenerate fibre. The Clemens--Schmid exact sequence extracts such an invariant part of information from the degenerate fibre (cf the appendix \ref{ssec:E7D10-CS}). Three out of four inequivalent Type II degenerations of degree-2 K3 surface in (\ref{eq:classification-deg2}), classified in terms of the lattice $(W_2 \cap \Lambda_T)/W_1$, have been realized in the compact models in this section. The one in section \ref{ssec:a17fibres} is for $(W_2\cap \Lambda_T)/W_1 = A_{17};\Z_3$, the one in sections \ref{sssec:corridor-E7D10}--\ref{sssec:kahler-E7D10} is for $(E_7 \oplus D_{10});\Z_2$ (see the appendix \ref{ssec:E7D10-CS} for derivation), while the $(E_8^{\oplus 2}\oplus A_1)$ case is realized in the examples in section \ref{sssec:root3}.
It is also possible to construct polytopes for which these different fibre types coexist within a single threefold. 

\subsection{Branches with Type III or Non-Kulikov Degenerations}
\label{ssec:corridor-III}

It is also known, in toric language, how to construct a compact K3-fibred Calabi--Yau threefold that develops 
a Type III degeneration \cite{short_tops}. The simplest example is to consider $M_{\vev{+2}}^{\{ 2_4, 1_4, -1_2 \}}$, where 
we collect three lattice points from Figure~\ref{fig:choicesofabc} to form a polytope $\widetilde{\Delta}$, a toric 
ambient space, and a Calabi--Yau hypersurface. A relevant toric graph is a two-dimensional face 
$\widetilde{\Theta}^{[2]}$ with $\nu^6_{2_4}$, $\nu^6_{1_4}$ and $\nu^6_{-1_2}$ as the vertices, but this comes 
with a multiplicity $\ell^*(\Theta^{[1]})+1=2$. The dual graph of this degenerate fibre is given by two copies 
of the triangle $\widetilde{\Theta}^{[2]}$ glued along the three edges, which topologically is a triangulation of a sphere $S^2$. 
Since this threefold should be regarded as a common subset of two different ``corridor'' branches of 
the complex structure moduli for the $[2_4]$--$[1_4]$ link and for the $[2_4]$--$[-1_2]$ link, we should expect this
degeneration to combine the Type II $(E_7\oplus D_{10});\Z_2$ and the Type II $(E_8^{\oplus 2} \oplus A_1)$ degenerations. 
In light of the stratification structure of the boundary components of the Baily--Borel compactification of 
lattice-polarized K3 surfaces (see the appendix \ref{sssec:BBcpt}), it is natural that a Type III degeneration 
develops in the common subset of the corridor branches.

It has been proved \cite{short_tops} that $M_{\Lambda_S}^{ \{ \cdots \}} \rightarrow \P^1_A$ has a Type III degeneration 
when the collection of lattice points $\{ \cdots \} \subset 2\widetilde{\Delta}_F \cap N_F$ forms a convex hull 
that is either 2-dimensional or 3-dimensional. The collection of vertices $\{ 2_4, 1_4, -1_2\}$ is a minimal collection 
to have a Type III degeneration. The other extreme is to have the collection $\{ \cdots \}$ all of 
$2\widetilde{\Delta}_F \cap N_F$. Despite this variety for construction of a threefold with a Type III degeneration 
(and the corresponding stratification of the moduli space) there is less richness in the classification of Type III 
degeneration of lattice-polarized K3 surface, primarily due to the indefinite signature of the lattice 
$(W_2 \cap \Lambda_T)/W_0$ (see the appendix \ref{sec:K3-degen-review}).

Degenerations of a lattice polarized K3 surface which correspond to Type I do not contribute 
to the story in this article although they are a very common phenomenon. To be more precise, \emph{all} 
the K3-fibred Calabi--Yau threefolds we have discussed in this article have many degenerations that 
are not semi-stable, which would become Type I degeneration after base change 
of order-2 (see the appendix \ref{sec:transcendental}). We call such degenerations ``would-be Type I'' in this article. 
In the threefolds $M_{\vev{+2}}^n$ discussed in section \ref{sssec:rho+1}, for example, there are $NL_{0,0}=300$ 
would-be Type I degenerations. These $NL_{0,0}=300$ degeneration points in the base $\P^1_A$ have also been known 
as the $NL_{2,1} = 300$ Noether--Lefschetz loci that contribute to the Gromov--Witten invariant $d_{\beta=2C_2}$ 
in (\ref{eq:prepotential-GW}) \cite{Klemm:2004km, MP, KMPS, HK}. There is nothing new in particular. 

In the context of string compactification over a compact Calabi--Yau threefold (that just happens to have 
a K3-fibration), we are not so happy to replace the threefold by its base change. Not all the degenerations in 
$\pi_M: M \rightarrow \P^1_A$ are semi-stable, or in Kulikov model, when we do not allow to ``replace'' them 
by their base changes. The would-be Type I degenerations above is the simplest example (cf. the 
appendix \ref{sec:transcendental}). Such degenerations still come with the notion of monodromy on the generic 
fibre $H^2(S_t;\Z)$. As one of the properties of Picard--Lefschetz monodromy of K3 fibration \cite{Landman}, 
the monodromy matrix $T$ is quasi-unipotent, in that there exists an integer $m$ so that 
\begin{align}
 \left(T^m - {\bf 1} \right)^3 = 0.
\end{align}
In the case of semi-stable degenerations, $m=1$ and the matrix $N$ defined by 
\begin{align}
  N := \ln \left[ T^m \right]
\end{align}
is a nilpotent matrix. We call these degenerations would-be Type I, Type II and Type III, when $N=0$, 
$N^2=0$ (but $N\neq 0$), and $N^3=0$ (but $N^2 \neq 0$), respectively, in this article. They may well be regarded 
as $1/m$-Type I (Type II, Type III, resp.) degenerations, similarly to fractional D-branes. We are also tempted 
to call them fractional Type I, Type II and Type III degenerations for this reason.

Such would-be Type II and would-be Type III degenerations will be constructed straightforwardly, given an observation 
in \cite{short_tops}. Recall that we started out in section \ref{ssec:fibr-option-general} by allowing to use a 
vertex of the form $\nu^6 = (\nu^6_F,+1)^T$ in (\ref{eq:fibre-toric}) to form a convex polytope 
$\widetilde{\Delta} \subset N \otimes \R$. It is the definition of a \emph{short} top in \cite{short_tops} to 
restrict the possibility of $\nu^6$ to this form (placed at height $+1$); this restriction guarantees that all the 
irreducible components in the degenerate fibre appear with multiplicity $+1$, which is one of the conditions of 
semi-stable degeneration \cite{short_tops}. Allowing to involve vertices that are placed at height $>1$, 
degenerations cease to be semi-stable as fibre components can now appear with multiplicity $>1$. In case the degenerate fibre in question has 
at least two components, they must be either would-be Type II or would-be Type III. Placing a vertex at height $>1$ this is guaranteed if there is at least a second lattice point `above $\tilde{\Delta}_F$' not contained in any face of dimension $<3$. Indeed this happens in all examples known to us, but we do not have a general proof securing this in general.

\subsection{Heterotic String Interpretation }
\label{ssec:Het-interpret}

We have seen many examples of $\Lambda_S$-polarized K3-fibred Calabi--Yau threefolds (that are non-singular) 
where the fibre K3 surface degenerates and forms multiple irreducible components over isolated points in the base 
$\P^1_A$. The adiabatic argument can be used to translate type IIA compactifications over such threefolds 
to heterotic string compactifications at points in $\P^1_A$ away from such degeneration points. 
In this section, we discuss the heterotic dual description of degenerations of K3 fibrations. 

There is an example of degenerations of lattice-polarized K3 surface whose heterotic dual is well-known. 
That is when we have an $E_8$-elliptic K3 surface in the fibre ($\Lambda_S=U$), and the fibre undergoes Type II degeneration 
with $(W_2 \cap \Lambda_T)/W_1 \cong E_8 \oplus E_8$. The local geometry of $M_{U}^{\{n_H,\cdots, -n_v\}}$ around a point 
of degeneration $t=0 \in \P^1_A$, discussed in page \pageref{pg:E8E8-ex}, corresponds to $(n_h+n_v)$ NS5-branes 
of heterotic string theory, wrapped on $T^2_{89}$, extending along $\R^{1,3}$ and localized at $t=0 \in \P^1_{\rm Het}$, 
the base of $T^2_{67}$-fibration of the K3$_{\rm Het}$ when we see it as an elliptic fibration. 

The most direct way to see this duality dictionary is in terms of monodromy around the degeneration point $t=0$. 
In type IIA language, period integrals of the generic fibre undergoes monodromy transformation 
$T = \exp[N]$, as a point in the base $t$ goes around $t=0$ by $t = t_* \times e^{2\pi i a}$, $a \in [0,1]$.
The nilpotent matrix $N$ is given by (\ref{eq:N0-def-II-pol}), with $\delta_1=\delta_2=1$ and $\mu = n_h+n_v$:
\begin{align}
  N = (n_h+n_v) \times  \left( \begin{array}{cc|cc} 
   & & & \\
   & & & \\
  \hline
   & -1 & & \\
 1 & & & 
 \end{array} \right).
\label{eq:N4muNS5}
\end{align}
The lower two components near the degeneration point are the period integrals over 2-cycles that are obtained 
by fibering 1-cycles of the elliptic fibre along the long cylinder axis in the base $\P^1$ of the elliptic K3$_A$.

The fibrewise heterotic--type IIA duality map simply replaces period integrals of a $\Lambda_S$-polarized K3 surface
in type IIA language by $\Lambda_T \otimes \C$-valued Narain moduli in heterotic string language. Being away 
from the degeneration point, the $\Lambda_T \otimes \C$-valued period integrals / Narain moduli are allowed 
to vary over the base $\P^1_A$ / $\P^1_{\rm Het}$. Now, the standard parametrization of Narain moduli in the 
case of $\Lambda_T = U^{\oplus 2} \oplus E_8^{\oplus 2}$ is  
\begin{align}
    \left( \begin{array}{c} 0 \\ \hline  -\tau \\ 1 \\ \hline  - \tilde{\rho} \\ - \tilde{\rho} \tau - (a)^2 \\ 
   \hline  a \end{array} \right), 
\label{eq:Narain-parametrize}
\end{align}
where the first row corresponds to the rank-2 $\Lambda_S = U$, the next four rows correspond to a basis 
$\{ \hat{e}^1, \hat{e}^2, \hat{e}^{'1}, \hat{e}^{'2} \}$ and the last row to $E_8 \oplus E_8 \subset \Lambda_T$.
$\tau$ and $\tilde{\rho}$ roughly correspond to the complex structure and complexified volume of the $T^2_{67}$ 
fibre of ${\rm K3}_{\rm Het}$, and $a$ the $E_8^{\oplus 2}$-valued Wilson lines along $T^2_{67}$. The monodromy matrix 
$N$ in (\ref{eq:N4muNS5}) is equivalent to shift $\tilde{\rho} \rightarrow \tilde{\rho} + (n_h+n_v)$ at the 
end of the monodromy.\footnote{$\tilde{\rho}$ behaves as $\tilde{\rho}(t) \simeq \left(\frac{1}{2\pi i}\right) \ln (t) + {\rm const}.$ near the degeneration point.} The degeneration of $E_8$-elliptic K3 surfaces is regarded in heterotic 
string theory as the presence of a magnetic source for the three-form field $dB$:
\begin{align}
 - \frac{1}{(2\pi)^2 \alpha'} \int_{S^1 \times T^2_{67}  } dB = \Delta {\rm Re} (\tilde{\rho}) = (n_h+n_v),
\end{align}
where $S^1$ is a circle around the $t=0$ point in $\P^1_{\rm Het}$ \cite{CHS-NS5}.


We have also seen other examples of Type II degenerations of lattice polarized K3 surfaces in this article.
All of the Calabi--Yau threefolds $M_{\vev{+2}}^{-\nu^1_F}$ in section \ref{ssec:a17fibres} and 
$M_{\vev{+2}}^{\{n,n-1,\cdots, m\}}$ and $M_{\vev{+2}}^{\{2_4,-1_2,2_3\}}$ in section \ref{ssec:corridor-II} have degree-2 
K3 surfaces in the fibre ($\Lambda_S = \vev{+2}$), but there are numerous examples of Type II degeneration of 
K3 surfaces with various choices of the polarizing lattice $\Lambda_S$ \cite{short_tops}. 
When the $\Lambda_S$-polarized K3 surface in the fibre exhibits Type II degeneration at a point $t=0 \in \P^1_A$, 
$\Lambda_T \otimes \C$-valued period integrals have monodromy around the degeneration point $t=0$.
Repeating the same argument as above, we find that such a degeneration in a threefold $M_{\Lambda_S}$ 
for type IIA compactification corresponds to a soliton in heterotic string theory localized at the $t=0$ point 
in $\P^1_{\rm Het}$. The monodromy matrix $T=\exp[N]$ now dictates how the heterotic string Narain moduli 
in the $T^4$-fibre over $\P^1_{\rm Het}$ are twisted. Let us parametrize the $\Lambda_T \otimes \C$ part of 
the Narain moduli as 
\begin{align}
  \left( \begin{array}{c} 
    -\tau / \delta_1 \\ 1 / \delta_2 \\ \hline 
    a \\ \hline
    - \tilde{\rho} \\ - \tilde{\rho}\tau - [(\tau, a)]
   \end{array} \right)
\end{align}
in the basis (\ref{eq:LambdaT-basis-general}). The contribution $[(\tau, a)]$---a quadratic in $\tau$ and $a$---needs to be 
determined by the unspecified part of the intersection form in (\ref{eq:int-form-II-pol-gen}),  but does not depend 
on $\tilde{\rho}$. The nilpotent matrix $N$ in (\ref{eq:N0-def-II-pol}) implies that only the $\tilde{\rho}$ parameter 
of the Narain moduli in the basis (\ref{eq:LambdaT-basis-general}) gets shifted by
\begin{align}
 \tilde{\rho} \longrightarrow \tilde{\rho} + \mu ,
\end{align}
and all other Narain moduli parameters remain intact around the soliton localized at $t=0 \in \P^1_{\rm Het}$.
Type II degenerations with different $\delta_1$, $\delta_2$ and $(W_2 \cap \Lambda_T)/W_1$ correspond to 
a shift (around a point in $\P^1_{\rm Het}$) of different Narain moduli. The shift depends on $\mu$ and the $\delta_i$ in the same way for each case. 
The value of $\mu$ in particular (the number of double curves in a Type II degeneration) is regarded as the number of coincident solitons of the same type. 

What is the Narain modulus $\tilde{\rho}$ that shifts in terms of the weakly coupled heterotic 
$E_8 \times E_8$ string theory for each one of those solitons? Let us take the $\Lambda_S = \vev{+2}$ case 
as an example, and provide an explicit answer to this question. 

\begin{figure}[tbp]
\begin{center}
\begin{tabular}{ccc}
\includegraphics[width=.26\linewidth]{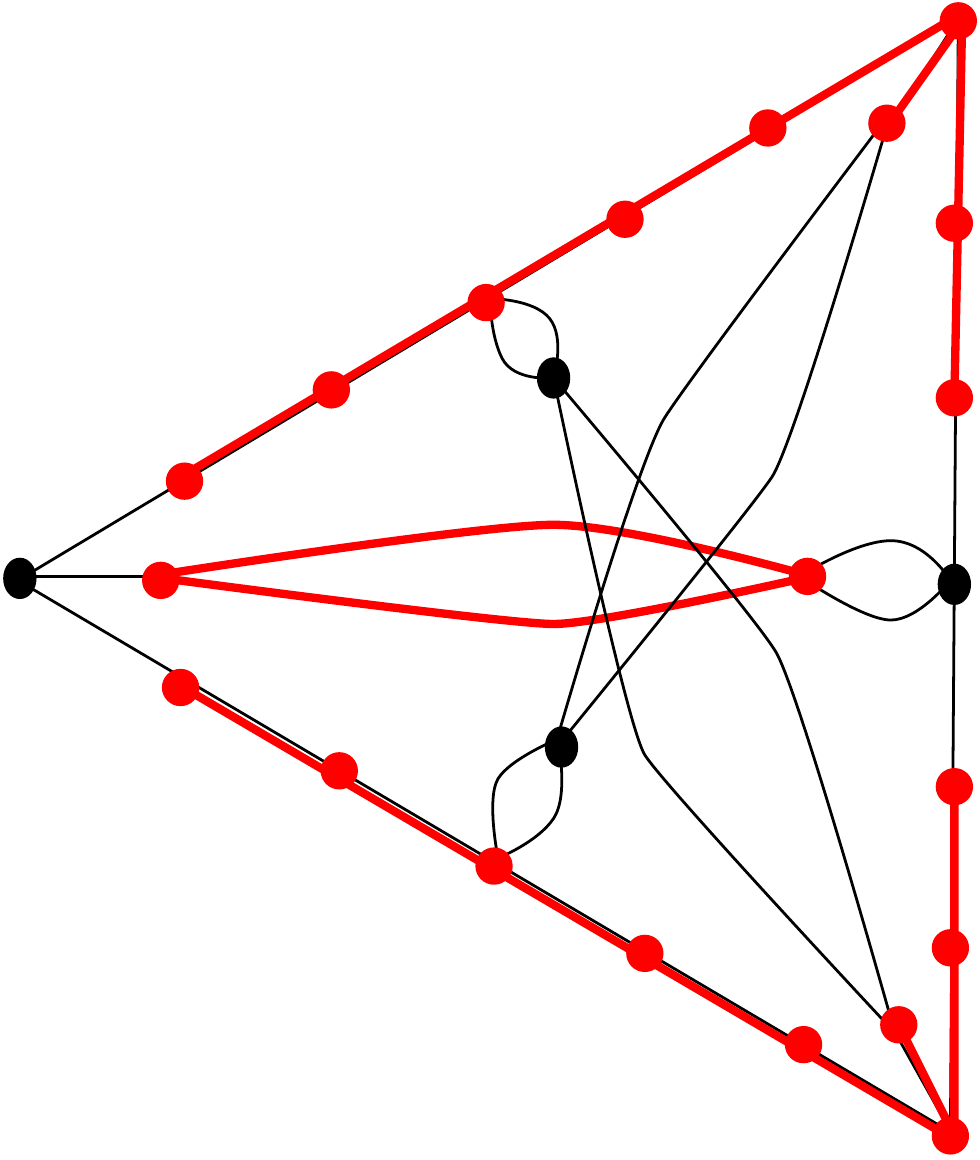} & $\qquad$ & 
\includegraphics[width=.26\linewidth]{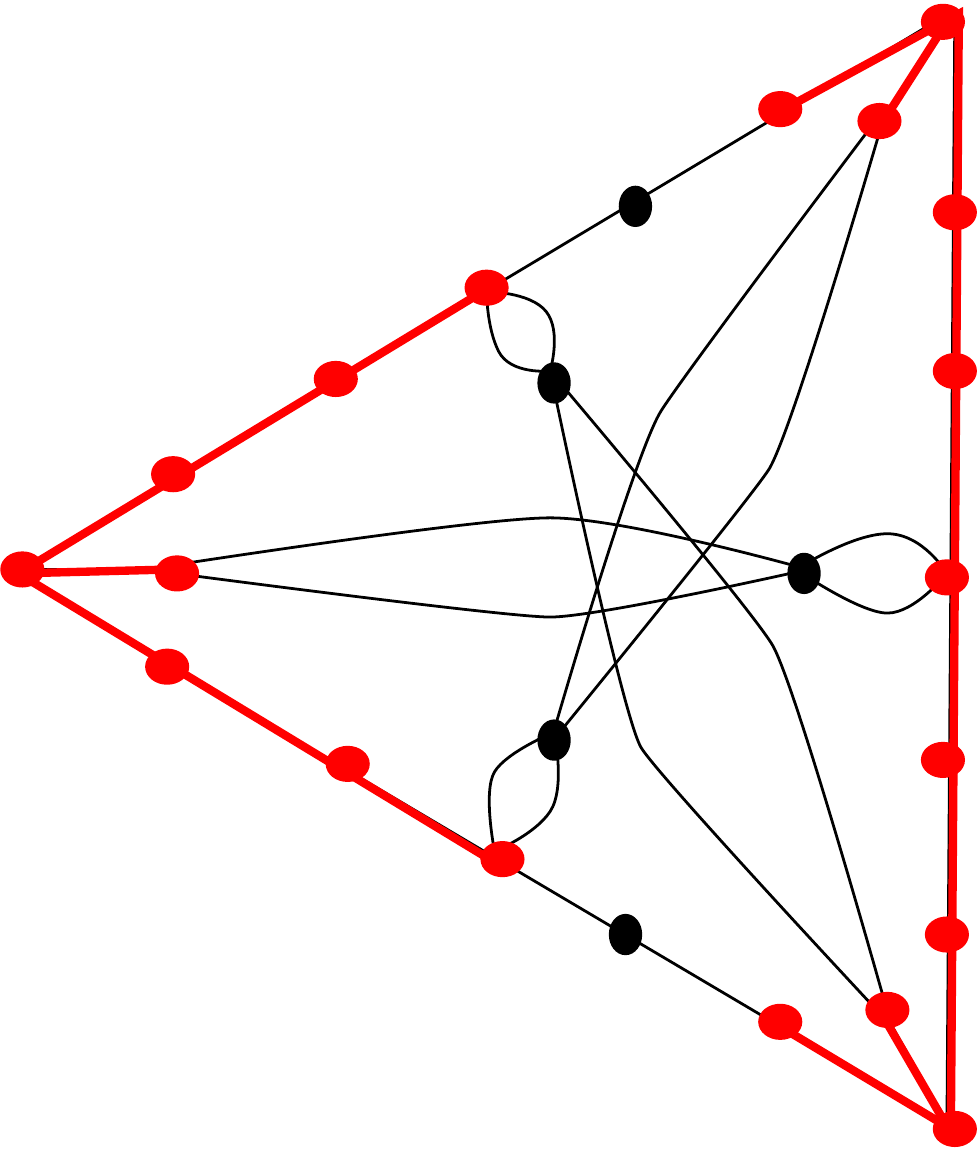} \\
 (a) & & (b)
\end{tabular}
\caption{The Coxeter Diagram of the signature $(1,18)$ lattice for degree-2 K3 surfaces 
(see text) \cite{Scattone}. This graph contains three nodes that are absent in Figure~\ref{fig:2DdualPolytope}~(b), 
while one point---the top vertex---of Figure~\ref{fig:2DdualPolytope}~(b) has been removed here. Links of mutual 
intersection number $+6$ in the Coxeter diagram of this lattice are omitted in this figure. Affine Dynkin diagrams 
of $A_1 \oplus E_8 \oplus E_8$ and $E_7 \oplus D_{10}$ are shown in red in (a) and (b), respectively, as a subgraph. 
These figures are taken from \cite{Scattone}, the graphs of embeddings of $A_{17}$ and $D_{16} \oplus A_1$ are also 
found in \cite{Scattone}.
\label{fig:Coxeter}}
\end{center}
\end{figure}

In the $\Lambda_S = \vev{+2}$ case, we can always take $\delta_1 = \delta_2 = 1$ (see \cite{Scattone}, or the 
appendix \ref{ssec:K3-degen-pol-review}), and the filtration structure 
$\{ 0 \} \subset W_1 \subset (W_2 \cap \Lambda_T) \subset \Lambda_T$ in (\ref{eq:filtr-TypeII}) can be transformed 
into a direct sum, 
\begin{align}
 \Lambda_T = U \oplus U \oplus [(W_2 \cap \Lambda_T)/W_1], 
 \label{eq:UU-W2/W1}
\end{align}
where 
\begin{align}
&
 W_1 = {\rm Span}_\Z \{ \hat{e}^{'1}, \hat{e}^{'2} \}, \quad W_1' = {\rm Span}_\Z \{ \hat{e}^1, \hat{e}^2 \}, \quad 
 W_1 \oplus W_1' = U \oplus U, \\
&
 ( \hat{e}^{'i}, \hat{e}^{'j} ) = (\hat{e}^i, \hat{e}^j) = 0, \qquad 
 (\hat{e}^{'i}, \hat{e}^j ) = \delta^{ij}.
\end{align}
At least as a question in mathematics, we can easily find how the structure (\ref{eq:UU-W2/W1}) fits into $\Lambda_T$
given by (\ref{eq:Tk3-rho=1}) in the case of $(W_2 \cap \Lambda_T)/W_1 = A_1 \oplus E_8 \oplus E_8$. One just needs 
to identify (\ref{eq:UU-W2/W1}) with (\ref{eq:Tk3-rho=1}). For three other choices of $(W_2 \cap \Lambda_T)/W_1$ 
in (\ref{eq:classification-deg2}), it is useful to note that all the four signature $(1,18)$ lattices 
$U \oplus [W_2 \cap \Lambda_T)/W_1]$ are mutually isometric, though the four $(0,17)$ lattices 
$[(W_2 \cap \Lambda_T)/W_1]$ are not. The Coxeter diagram of the common $(1,18)$ lattice can be used to describe how the 
three other $[(W_2 \cap \Lambda_T)/W_1]$ are embedded into this $(1,18)$ lattice, and hence into $\Lambda_T$ 
in (\ref{eq:Tk3-rho=1}). See \cite{Scattone} and references therein for more information. Figure~\ref{fig:Coxeter} 
contains all the information we need in this article. 

Now let us turn to the physics question. To get started, we fix the $S_3$ symmetry action. We have seen in 
section \ref{sssec:10-10-4-interpret} that the heterotic $ST$-model is dual to type IIA compactification 
on $M_{\vev{+2}}^{2_4}$, where the weak coupling $E_8 \times E_8$ in heterotic string theory corresponds to transcendental 
two cycles localized near $X_4 = X_2 = 0$ and $X_4=X_3=0$. In Figure~\ref{fig:Coxeter}~(a), we regard the upper right 
and lower right corners as those two locations in $\P^2$, and therefore, the upper half triangle and the lower half 
triangle can be regarded as the weak coupling $E_8 \times E_8$ in heterotic string. 

Now, at the transition from $M_{\vev{+2}}^{2_4}$ to $M_{\vev{+2}}^{\{2_4, 1_4\}}$, the $\tilde{E}_7$ singularity appears 
at $X_2 = X_3=0$, as we saw in (\ref{eq:E7-tilde}). This point corresponds to the left corner in 
Figure~\ref{fig:Coxeter}~(a). 
At the transition from $M_{\vev{+2}}^{\{2_4,1_4\}}$ to $M_{\vev{+2}}^{1_4}$, a curve of $A_1$ 
singularity forms along $X_4=0$, parametrized by $[X_2:X_3] \in \P^1$; this curve corresponds to the right edge 
in Figure~\ref{fig:Coxeter}~(a). Remembering that the $E_7$ algebra is contained in $V_1 = \bar{D}_{6,1_4}$ of 
$M_{\vev{+2}}^{\{2_4,1_4\}}$, which collapses at the transition to $M_{\vev{+2}}^{2_4}$, and that the $D_{10}$ algebra is in 
$V_0=\bar{D}_{6,2_4}$, which collapses at the transition to $M_{\vev{+2}}^{1_4}$, we conclude that the 
$E_7 \oplus D_{10}$ lattice in the Type II degeneration in $M_{\vev{+2}}^{\{2_4,1_4\}}$ is embedded into the 
$U \oplus E_8^{\oplus 2} \oplus A_1$ of the weak coupling heterotic string in a way that can be seen by superimposing
Figure~\ref{fig:Coxeter}~(a) and (b). In particular, the $E_7$ current algebra associated with the $E_7$-string 
at the $M_{\vev{+2}}^{2}$--$M_{\vev{+2}}^{\{2,1\}}$ transition (due to collapsed $dP_7$) is not from a subgroup of any one of 
the two weakly coupled $E_8$'s, but from somewhere in the middle of $E_8 \times E_8$ (cf \cite{Klemm:1996hh}).
The $(W_2 \cap \Lambda_T)/W_1 \cong E_7D_{10};\Z_2$ soliton squeezes instanton degrees of freedom in the $ST$-model 
from both of the two weakly coupled $E_8$'s, and the remaining $E_8$ factors do not have a free-choice in the 
instanton moduli anymore, as we saw in section \ref{sssec:10-10-4-interpret}. The soliton returns those 
degrees of freedom at the $M_{\vev{+2}}^{\{2_4,1_4\}}$--$M_{\vev{+2}}^{1_4}$ transition, but in a way that the free instanton 
interpretation is never restored in the $M_{\vev{+2}}^n$ branches with $n=1,0,-1$ (at least not in an obvious way for 
$n=-1$).

Similarly, in the $(W_2 \cap \Lambda_T)/W_1 = E_8^{\oplus 2}\oplus A_1$ soliton that appears in the heterotic duals of 
$M_{\vev{+2}}^{\{2_4,-1_2\}}$ or $M_{\vev{+2}}^{\{ 2_4, -1_2,2_3\}}$ type IIA compactifications, the two $E_8$ algebras in the 
degenerate fibre correspond to the lower left half triangle and upper right half triangle 
in Figure~\ref{fig:Coxeter}~(a). Imagine Figure~\ref{fig:Coxeter}~(a) rotated by $2\pi/3$ in a counter-clockwise direction.
This soliton sucks away instanton degrees of freedom from a skewed combination of the
two $E_8$'s (not diagonally as in the $(E_7\oplus D_{10});\Z_2$ soliton), and releases them somewhere else. 
Table \ref{tab:E8-instanton-count} summarizes the consequence of this chain of transitions. The $E_8$-string 
that emerges at the transition points is not associated simply with any one of the two weakly coupled $E_8$'s. 

By now, the question ``what is the $\tilde{\rho}$ modulus that shifts for these solitons'' is not more than a technical 
question that is not particularly illuminating. So, we are not presenting technical details here. Roughly speaking, 
the $\underline{U}$ factor of $\underline{U} \oplus (W_2 \cap \Lambda_T)/W_1 \cong U \oplus (E_8^{\oplus 2} \oplus A_1)$ 
picks up the nodes in the Coxeter diagram that have not been used for $(W_2 \cap \Lambda_T)/W_1$. 

A similar reasoning can be applied to Type III degenerations of K3 surfaces. When a type IIA compactification 
on a K3-fibred Calabi--Yau threefold $M$ has a Type III degenerate fibre at one point $t=0 \in \P^1_A$ in the base, 
the adiabatic argument (fibre-wise duality) can be applied to any points away from the degeneration point. The holomorphic 
dependence of the period integrals of the K3 fibre in type IIA over $\P^1_A$ is translated into the holomorphic dependence 
of the Narain moduli of the $T^4$ fibre in heterotic string theory over $\P^1_{\rm Het}$. Any monodromy action on the generic fibre K3 
surface $T: H^2(S_{t.A};\Z) \rightarrow H^2(S_{t.A};\Z)$ is directly translated into that on the Narain lattice. Parametrizing 
the Narain moduli / period integrals on the $\Lambda_T \subset {\rm II}_{4,20}$ part as 
\begin{align}
 \left( 1/\delta, \; X, \; 
    - \frac{1}{2} \left(\frac{a}{\delta^2} + 2 \frac{B \cdot X}{\delta} + X^T \cdot C \cdot X \right)
 \right)^T
\end{align}
in the basis (\ref{eq:LambdaT-basis-general-III}), we find that the monodromy due to 
$T=\exp[\mu N_0^{\rm III}(\delta, u,v,x)]$ in (\ref{eq:N0-def-III-pol}) amounts to 
\begin{align}
  X \rightarrow X + \mu v.
\end{align}
The heterotic dual of a Type III degeneration is to involve a soliton that is a magnetic source of the moduli 
field $X$ and $\mu$ is interpreted as the number of soliton of this type.

If we are to ignore the distinction between $\widetilde{\Lambda}_S$ and $\Lambda_T$ within ${\rm II}_{4,20}$ 
in choosing the parametrization of the Narain moduli, then we can always use the monodromy matrix 
$T = \exp[\mu N_0^{\rm III}(t_0)]$ in (\ref{eq:N0-def-III-nopol}) for (the heterotic dual of) a Type III degeneration.  
Using a parametrization $(1, \rho_1, \rho_2, - \rho_1\rho_2)^T$ for the Narain moduli in the basis adopted in 
(\ref{eq:N0-def-III-nopol}), the soliton in question is regarded as a magnetic source introducing a twist 
\begin{align}
 \rho_1 \rightarrow \rho_1 + \mu, \qquad \rho_2 \rightarrow \rho_2 + \mu t_0/2.
\end{align}

Degenerations of K3 fibration in type IIA compactification that are not in the Kulikov model are also regarded 
as solitons in heterotic string, and are magnetic sources of the Narain moduli fields precisely for the same 
reason as in the cases of Type II and Type III degenerations. A case-by-case study is necessary for the explicit 
form of the monodromy matrix in an integral basis for the would-be Type II and would-be Type III degenerations.
Fractional powers of $\exp[N_0^{\rm II}]$ and $\exp[ N_0^{\rm III}]$ need to be taken in an integral basis. 

\section{6D Perspectives}
\label{sec:6D}

Degenerations in the K3 fibre of type IIA compactifications and their heterotic duals are both regarded as solitons, localized in codimension two in the base $\P^1$. In this section, we attempt at recapturing those solitons in terms of 6D (1,1) supergravity by taking the decompactification limit of the base $\P^1$. This approach provides a more bottom-up (more general, less constructive) perspective, and makes it possible to extract the intrinsic nature of those solitons unaffected by anything associated with the compactness of the base $\P^1$. This section is therefore meant to provide a complementary perspective to the study in the previous section.

\subsection{6D (1,1) Supergravity and Half-BPS 3-branes}

Both $T^4$ compactification of the heterotic string and K3 compactification of the type IIA string leads to a 6D effective theory 
at low energy with $(1,1)$ supersymmetry. The massless field contents of the effective theory in supergravity consists of 
one supergravity multiplet and $n=20$ vector multiplets. 

The 32 bosonic degrees of freedom in the supergravity multiplet 
are represented by the 6D metric (9 DOF), $B_{\mu \nu}$ (3+3 DOF), 1 scalar\footnote{In 10D type IIA language, 
$e^{2\sigma} = e^{-2\phi_{10;A}} J^{\rm A; str}_{\rm K3}$.} $\sigma$ and four 6D vectors (16 DOF). The fermionic 
degrees of freedom consist of the gravitinos $\psi^{(+)i}_\mu$, $\psi^{(-)}_{\mu \; i'}$ as well as the 
dilatinos $\chi^{(+)}_{i'}$ and $\chi^{(-)i}$. All of the $\psi^{(+) i}_{\mu}$ ($i=1,2$, $\mu=0,\cdots,5$) and $\chi^{(+)}_{i'}$ ($i=1,2$) are in the spinor representation of $\SO(1,5)$ with the $\Gamma_7 = +1$ eigenvalue, while those with $^{(-)}$ are in the spinor representation with $\Gamma_7= -1$ eigenvalue. The spinors with $i=1,2$ (or with $i'=1,2$) combined form a symplectic--Majorana fermion in 6D. One vector multiplet consists of one 6D vector, four scalars and gauginos $\lambda^{(+)}_i$ and $\lambda^{(-)}_i$. The gauginos are subject to the symplectic Majorana condition.

The supersymmetry transformation parameters of 6D $(1,1)$ theories are $\epsilon^{(+)}_i$ with $i=1,2$ and 
$\epsilon^{(-)i'}$, $i'=1,2$, subject to the symplectic Majorana condition. In 10D type IIA language, 
two supersymmetry transformation parameters of 10D $(1,1)$ supergravity split under 
$\SO(1,5) \times \SU(2) \times \SU(2)$ as 
\begin{align}
 \Gamma_{11} \epsilon^{(+)} = \epsilon^{(+)} & \rightarrow  
   {\bf Spin}^{(+)} \otimes {\bf 2} \otimes {\bf 1} + 
   {\bf Spin}^{(-)} \otimes {\bf 1} \otimes \vev{\bf 2}, \\
 \Gamma_{11} \epsilon^{(-)} = - \epsilon^{(-)} & \rightarrow 
   {\bf Spin}^{(-)} \otimes {\bf 2} \otimes {\bf 1} + 
   {\bf Spin}^{(+)} \otimes {\bf 1} \otimes \vev{\bf 2}, 
\end{align}
where the last $\SU(2)$ factor corresponds to the holonomy group of a K3 surface. The SUSY transformation 
parameters $\epsilon^{(+)}_i$ are from ${\bf Spin}^{(+)} \otimes {\bf 2}$ in the first line, and 
$\epsilon^{(-)i'}$ from ${\bf Spin}^{(-)} \otimes {\bf 2}$ in the second line. Although it appears in 10D IIA sugra 
language that both $\epsilon^{(+)}_i$ and $\epsilon^{(-)i'}$ are doublets of a common $\SU(2)$ symmetry group, both 
are doublets of two separate $\SU(2)$ current algebras, one from the left movers and the other one from the right movers.

There are $4 \times n$ scalar fields, apart from $\sigma$, and they are known to parametrize a coset space 
\begin{align}
 M' :=  \SO(4,n)/\SO(4) \times \SO(n)
\end{align}
(modulo quotient). Let $\phi^x$ ($x=1,\cdots, 4n$) be a set of local coordinates in $M'$. 
\begin{align}
 f^{a \gamma} =  d\phi^x f_{x } ^{a \gamma}(\phi)
\end{align}
is a vielbein on $M'$, from which a metric $g_{xy} = f_{x}^{a \gamma} f_{y }^{a \gamma}$ is obtained. This metric is 
used in the 6D $(1,1)$ sugra action as the non-linear sigma model of $\phi^x$'s. 

Here is a little more about the geometry of the coset space $M'$. First, let $C^{IJ}$ be the intersection form of 
${\rm II}_{4,20}$. Each point in $M'$ corresponds to some choice of $\{ h_I^{\; \gamma} \}_{\gamma=1,2,3,4}$ satisfying 
\begin{align}
  h_I^\gamma C^{IJ} h_J^\delta = \delta^{\gamma \delta}, \qquad 
  h_I^\gamma C^{IJ} h_J^b = 0, \qquad 
  h_I^a C^{IJ} h_J^b = - \delta^{ab},
\end{align}
modulo $\SO(4)$ action $h_I^{\gamma} \rightarrow (h')_I^{\; \gamma} = h_I^{\; \delta} (H_{\SO 4})_\delta^{\; \gamma}$ for some 
$(H_{\SO 4})_\delta^{\; \gamma} \in \SO(4)$. For such a choice of $\{ h_I^{\; \gamma} \}_{\gamma=1,2,3,4}$, one can uniquely find 
one choice of $\{ h^a \in {\rm II}_{4,20} \otimes \R \; | \; a=1,\cdots, 20\}$ modulo the $\SO(20)$ 
action that satisfies the orthonormality conditions above. 

An $\SO(4) \times \SO(20)$ connection on $M'$ is defined by 
\begin{align}
 A^{\gamma \delta} = h^\gamma_I C^{IJ} (dh_J^\delta), \qquad 
 A^{ab} = - h^a_I C^{IJ} (dh_J^b).
\end{align}
For the vielbein on $M'$ introduced earlier, we used the following:
\begin{align}
  f^{a\gamma} = - \sqrt{2} h^a_I C^{IJ} (dh_J^\gamma) = \sqrt{2} (dh_I^a) C^{IJ} h_J^\gamma.
\end{align}
The dictionary between the $\SO(4)$ vector indices $\gamma, \delta$ and $\SU(2) \times \SU(2)'$ doublet 
indices $i, i'$ is given by  
\begin{align}
 ({\rm vect})_i^{ \; i'} := \frac{1}{\sqrt{2}} ({\rm vect})^\gamma (\sigma^\gamma)_i^{\; i'}, 
\qquad 
\sigma^\gamma = ({\bf 1}, i \vec{\tau}), \quad 
 \epsilon^{ij} \epsilon_{jk} = \delta^i_k, \quad 
 \epsilon^{12} = - \epsilon^{1'2'}.  
\end{align}
Therefore, the orthonormality condition becomes 
\begin{align}
 h_{I \; i}^{i'} C^{IJ} h_{J \; j}^{j'} = \epsilon_{ij} \epsilon^{i'j'}.
\end{align}
\begin{align}
  h_2^{2'} = (h_1^{1'})^{cc}, & \quad (h_1^{1'}, h_1^{1'}) = 0, \quad (h_{1}^{1'}, h_2^{2'}) = 1, \\
  h_2^{1'} = - (h_1^{2'})^{cc}, & \quad (h_1^{2'}, h_1^{2'}) = 0, \quad (h_1^{2'}, - h_2^{1'}) = 1.
  \label{eq:ortho-normal-d}
\end{align}

Using these geometric data, the supergravity transformation law is written down in eq.(3.2) of \cite{Townsend}.
The SUSY variation of fermionic fields is
\begin{align}
 \delta \psi^{(+)}_{\mu \; i} =& {\cal D}_\mu \epsilon^{(+)}_i + \cdots , \label{eq:SUSY-var-1} \\ 
 \delta \psi^{(-)i'}_{\mu} =& {\cal D}_\mu \epsilon^{(-)i'} + \cdots, \label{eq:SUSY-var-2} \\
 \delta \chi^{(-)}_i =& \frac{1}{2} \Gamma^\mu (\partial_\mu \sigma) \epsilon^{(+)}_i + \cdots , \\
 \delta \chi^{(+)i'} =& \frac{1}{2} \Gamma^\mu (\partial_\mu \sigma) \epsilon^{(-)i'} + \cdots,  \\
 \delta \lambda^{(+)a}_i=&  \frac{1}{2} \Gamma^\mu f^{ai'}_{x \; i} (\partial_\mu \phi^x) \epsilon^{(-)}_{i'} + \cdots , \\
 \delta \lambda^{(-)ai'} =& \frac{1}{2} \Gamma^\mu f^{ai'}_{x \; i} (\partial_\mu \phi^x) \epsilon^{(+)i} + \cdots , 
\end{align}
where ellipsis stands for terms that involve multiple fermions, or $H_{\mu\nu\rho}$ or $F_{\mu\nu}^I$.

We consider codimension$_\R$ = 2 defects of this 6D $(1,1)$ supergravity that preserves $\SO(1,3)$ Lorentz symmetry 
and half of the SUSY charges. In particular, we consider field configurations where the 6D metric and scalars have 
a non-trivial configuration in $(x,y) \in \R^2$. Under the unbroken $\SO(1,3)$ Lorentz symmetry, the supersymmetry 
transformation parameters $\epsilon^{(+)}_i$ and $\epsilon^{(-)i'}$ in 6D decompose as 
\begin{align}
 {\bf Spin}^{(+)} \otimes {\bf 2} \rightarrow & {\bf Spin}_L \otimes \uparrow \otimes
     \left[ {\bf 2} = \left\{ (\uparrow \uparrow)_{i=1}, (\downarrow \downarrow)_{i=2} \right\} \right] + \cdots, \\
 {\bf Spin}^{(-)} \otimes {\bf 2} \rightarrow & {\bf Spin}_L \otimes \downarrow \otimes 
     \left[ {\bf 2} = \left\{ (\uparrow \uparrow)^{i'=1}, (\downarrow \downarrow)^{i'=2} \right\} \right] + \cdots,
\end{align}
where ${\bf Spin}_L$ is a left-handed spinor of $\SO(1,3)$, and $+ \cdots$ is the other term involving 
a right-handed spinor ${\bf Spin}_R$ of $\SO(1,3)$. We are interested in defects where 
${\bf Spin}_L \otimes \uparrow \otimes (\uparrow \uparrow)_{i=1}$ in the first line and 
${\bf Spin}_L \otimes \downarrow \otimes (\downarrow \downarrow)^{i'=2}$ in the second line remain as 
transformation parameters of the unbroken supersymmetry. 

Let us first take 
\begin{align}
 ds^2 = (dx^2+dy^2) e^{2\varphi(x,y)} = e^{2\varphi} (dz \otimes d\bar{z} + d\bar{z} \otimes dz)/2
\end{align}
to be the metric configuration in the directions $\R^2$ transverse to the defect. 

The BPS conditions from dilatino variations are  
\begin{eqnarray}
 \delta \chi^{(-)}_{i=1} = e^{-\varphi} (\bar{\partial}_{\bar{z}} \sigma) {\bf Spin}_L \otimes \downarrow, & \qquad &
 \delta \chi^{(-)}_{i=2} = e^{-\varphi} (\partial_z \sigma) {\bf Spin}_R \otimes \uparrow, \\
 \delta \chi^{(+)i'=2} = e^{-\varphi} (\partial_z \sigma) {\bf Spin}_L \otimes \uparrow, & \qquad & 
 \delta \chi^{(+)i'=1} = e^{-\varphi} (\bar{\partial}_{\bar{z}} \sigma) {\bf Spin}_R \otimes \downarrow.
\end{eqnarray}
For all of these to vanish, we need $(\partial_z \sigma) = (\bar{\partial}_{\bar{z}} \sigma) = 0$.
That is, the value of $\sigma$ must remain constant.

The BPS conditions from gaugino variations are
\begin{eqnarray}
 \delta \lambda_i^{(+)a} & = & e^{-\varphi} f_{x \; i}^{a\; i'=1}(\partial_z \phi^x) {\bf Spin}_L \otimes \uparrow +  
 e^{-\varphi} f_{x \; i}^{a\; i'=2} (\bar{\partial}_{\bar{z}} \phi^x) {\bf Spin}_R \otimes \downarrow, \\
 \delta \lambda^{(-)ai'} & = & e^{-\varphi} f_{x \; i=1}^{a\; i'} (\bar{\partial}_{\bar{z}} \phi^x) {\bf Spin}_L \otimes \downarrow + 
 e^{-\varphi} f_{x \; i=2}^{a\; i'} (\partial_z \phi^x) {\bf Spin}_R \otimes \uparrow.
\end{eqnarray}
From the conditions in the first line, we find that 
\begin{align}
  (\partial_z \phi^x) f_{x \; i}^{a i'=1} = (\bar{\partial}_{\bar{z}} \phi^x) f_{x \; i}^{a \; i'=2} = 0,
  \qquad  i=1,2,
\end{align}
while the conditions on the second line yield 
\begin{align}
  (\bar{\partial}_{\bar{z}} \phi^x) f_{x \; i=1}^{a i'} = (\partial_z \phi^x) f_{x \; i=2}^{a \; i'} = 0, 
  \qquad i'=1,2.
\end{align}

Now, the 1-forms $f^{a \; i'=1}_{i=1}$ and $f^{a \; i'=2}_{i=2}$ on $M'$, pulled back by the scalar $\phi^x(z,\bar{z})$ 
field configuration of a half-BPS configuration, satisfy 
\begin{align}
 (d\phi^x) f_{x \; i=1}^{a \; i'=1} = (d\phi^x) f_{x \; i=2}^{a \; i'=2} = 0.
\label{eq:cond-h11-nearly-const}
\end{align}
These conditions are satisfied, when the values of $(h_I)^{i'=1}_{i=1} = [(h_I)^{i'=2}_{i=2}]^{cc}$ 
remain constant (and equal to their asymptotic values) in the transverse $(x,y) \in \R^2$ plane.
The authors are not sure if $h_1^{1'}=(h_2^{2'})^{cc}$ need to be constant for all possible half-BPS 
$\SO(1,3)$-preserving solitons, but we focus on solitons where $h_1^{1'}$ is constant here.

Let $\Lambda_T \subset {\rm II}_{4,20}$ be the lattice orthogonal to bot the constant values of 
$h^{i'=1}_{i=1}$ and $h^{i'=2}_{i=2}$, and $\widetilde{\Lambda_S} \subset {\rm II}_{4,20}$ be the orthogonal 
complement of $\Lambda_T$. Then $h^{i'=1}_{i=2} = -[h^{i'=2}_{i=1}]^{cc}$ takes its value in $\Lambda_T \otimes \C$. 
Moreover, the orthonormality condition (\ref{eq:ortho-normal-d}) implies that the space of $h_1^{2'}$'s is an 
$S^1$-fibration over the period domain of $\Lambda_T$, 
\begin{align}
D(\Lambda_T) := \P \left[ \left\{ \omega \in \Lambda_T \otimes \C \; | \; \omega^2 = 0, \;
    (\omega, \overline{\omega}) > 0 \right\}\right].
  \label{eq:period-domain}
\end{align}
The $S^1$ fibre corresponds to a complex phase multiplication for $h_1^{2'}$ ($SO(2)$ rotation on the 
2-plane), which does not change a point in $M'$. Thus, we can use a natural set of complex coordinates
of $D(\Lambda_T)$ for the subspace of $M'$ where the soliton has a non-trivial field configuration. 
The BPS condition  
\begin{align}
 (\bar{\partial}_{\bar{z}} \phi^x) f_{x \; i=1}^{a \; i'=2} = 0
\end{align}
means that the map $\phi: \R^2= \{z=(x+iy)\} = \C \rightarrow D(\Lambda_T)$ is holomorphic.

We are interested in the 6D $(1,1)$ supergravity where the target space $M'$ is replaced by 
${\rm Isom}({\rm II}_{4,20}) \backslash M'_{n=20}$, because ${\rm Isom}({\rm II}_{4,20})$ is 
the modular group of both heterotic/$T^4$ and type IIA/K3 compactifications \cite{Het-mod-grp, AM-94}.
Since the field configuration $\phi$ does not have to be well-defined at the centre of the soliton, 
the holomorphic map $\phi: [\C \backslash \{z=0\}] \rightarrow D(\Lambda_T)$ may have a branch cut 
emanating from the origin $\{z=0\}$, and the field configuration may be identified along the branch 
cut by some element $T \in {\rm Isom}^*(\Lambda_T)$.

\subsection{Recap and Speculations}

In the 6D $(1,1)$ supergravity with $M' = \SO(4,20)/\SO(4) \times \SO(20)$, strings are classified by 
their electric and magnetic charges under $B_{\mu \nu}$, while particles / 2-branes are classified by 
their electric / magnetic charges under the $4+20$ vector fields. Counting of BPS states has been carried out 
for those objects as a check of heterotic--type IIA duality. 3-branes (real-codimension-2 defects) 
are magnetic source of scalar fields. 

A half-BPS 3-brane comes with a choice of a pair of primitive sublattices $\widetilde{\Lambda}_S$ and 
$\Lambda_T$ of ${\rm II}_{4,20}$ that are mutually orthogonal in ${\rm II}_{4,20}$. $[(h)^{i'=1}_{i=1}]$ (modulo 
complex phase) takes its value in $D(\widetilde{\Lambda}_S)$ and remains constant over the real 
2-dimensional space transverse to the 3-brane.\footnote{We did not prove that this is necessary 
for (\ref{eq:cond-h11-nearly-const}), but certainly it is sufficient for (\ref{eq:cond-h11-nearly-const}).}
On the other hand, $[(h)^{i'=2}_{i=1}]$ (modulo complex phase) takes its value in\footnote{See Appendix \ref{ssec:K3-degen-pol-review} for more details.} 
${\rm Isom}^*(\Lambda_T) \backslash D(\Lambda_T)$, and is allowed to vary over the transverse space
$\C$ holomorphically.\footnote{In this class of solitons $\widetilde{\Lambda}_S$ remains constant and is not swapped with $\Lambda_T$ under monodromy.
This means that the monodromy twist under mirror symmetry is not included.} The 6D metric configuration in the real 2-dimensional transverse space $\R^2$ is assumed to be K\"{a}hler, and a holomorphic coordinate is introduced in $\R^2$ to turn it into $\C$.
The holomorphic configuration of $[h^{i'=2}_{i=1}]$ may be twisted around the defect (3-brane) by 
$T \in {\rm Isom}^*(\Lambda_T)$. Such 3-branes are therefore classified by a choice of lattices, 
${\rm Isom}({\rm II}_{4,20}) \backslash (\widetilde{\Lambda}_S, \Lambda_T)$, and conjugacy classes of 
${\rm Isom}^*(\Lambda_T)$ are to be used for the twist $T$ around the defect.
Note that we have not yet assumed that the transverse space is compact. 
 
The type IIA string compactified on an family of $\Lambda_S$-polarized K3 surfaces provides 
an example of such half-BPS 3-branes whenever the fibre K3 surface has a Type II, Type III, would-be Type II or 
would-be Type III degeneration. 
The classification of such degenerations, reviewed in the Appendix \ref{ssec:K3-degen-pol-review}, is regarded 
as a study of a subset of possible twists in ${\rm Inn}[{\rm Isom}(\Lambda_T)] \backslash {\rm Isom}^*(\Lambda_T)$. 
At least for the choices of $T$ that correspond to those degenerations we know that there is holomorphic solutions 
to $h^{i'=2}_{i=1}$.

It may turn out that the BPS 3-branes of the 6D $(1,1)$ theory from degenerations of lattice polarized K3 surface is only
a small subset of all possible BPS 3-branes characterized above. We leave it as an open problem for which 
$T \in {\rm Isom}^*(\Lambda_T)$ a holomorphic solution to $h^{i'=2}_{i=1}$ exists.\footnote{The half-BPS condition 
from (\ref{eq:SUSY-var-1}, \ref{eq:SUSY-var-2}) should also be implemented, although we did not study those conditions 
explicitly in this article.} It is not obvious to us purely from the perspective of solitons in 6D $(1,1)$ supergravity whether 
the monodromy matrix $T$ should have the quasi-unipotent property or not. 
We also note, from the 6D soliton perspective, that $\widetilde{\Lambda}_S = U \oplus \Lambda_S$ is not 
necessarily required. For more general choices of $\widetilde{\Lambda}_S \subset {\rm II}_{4,20}$, we cannot expect to 
obtain such a 3-brane in a family of K3 surface in the geometric phase. Whether such a 3-brane solution to supergravity 
has a UV completion is yet another open question.

\subsection{Analogy and Difference to 7-branes in F-theory}

The 3-branes in the 6D $(1,1)$ supergravity share many aspects with 7-branes in Type IIB string/F-theory.
Both are magnetic sources of scalar fields, a branch cut emanates from the centre of the soliton, 
and scalar fields are identified by an element of the modular group, ${\rm Isom}({\rm II}_{4,20})$ or 
${\rm SL}(2;\Z)$, along the branch cut.

As an isolated object, $(p,q)$ 7-branes are all alike, in that a $(p,q)$ 7-brane can be taken into 
a $(1,0)$ 7-brane by ${\rm SL}(2; \Z)$ transformation in Type IIB string theory (and if the asymptotic value 
of Type IIB axi-dilaton is not referred to). It is associated with a shift of a scalar field by an integer unit 
around the defect. 3-branes in 6D associated with a Type II degeneration are also all alike (if the asymptotic value 
of the scalar fields $\phi \in M'$ is not referred to), in that one and the same matrix (\ref{eq:N0-def-II-nopol}) 
is used in describing the monodromy. 

The crucial difference is that the 3-branes in 6D have a lot more variety. 3-branes associated with a Type III degeneration 
(monodromy $T = \exp[N_0^{\rm III}(t_0)]$) are labelled by an invariant $t_0$, and are not equivalent to the 3-branes 
associated with a Type II degeneration. Solitons associated with a would-be Type II or would-be Type III degeneration
also constitute a collection of solitons that are inequivalent from one another. 

Furthermore, when the information of the asymptotic value of the scalar fields $\phi \in M'$ is brought back into 
the discussion, there is a notion of $(\widetilde{\Lambda}_S, \Lambda_T)$ even for a 3-brane isolated in $\C = \R^2$. 
If a pair of 3-branes share the same $(\widetilde{\Lambda}_S, \Lambda_T)$, they can form a BPS configuration together. 
If they do not, they cannot be BPS together. Such a notion is absent in the case of $(p,q)$ 7-branes in F-theory. 

Ramond--Ramond 7-brane charge cancellation condition in a Type IIB orientifold is replaced by 
a condition in F-theory that 
\begin{align}
 \prod_i T_i = {\bf id}., 
 \label{eq:cond-multiplicative}
\end{align}
where $T_i$ is the ${\rm SL}(2;\Z)$-valued Picard--Lefschetz monodromy matrix associated with a 
discriminant point $z_i \in \P^1_F$ of the F-theory base. This condition is not additive anymore. 

Similarly, the heterotic string Bianchi identity for the $B$-field 
\begin{align}
 \delta_{NS5}^4 - \frac{1}{4} T_R^{-1} \tr {}_R \left[\left(\frac{F}{2\pi}\right)^2\right]
  = \frac{1}{4} \tr _{\SO{\rm vect.}}\left[ \left(\frac{R}{2\pi}\right)^2 \right]
\label{eq:Het-Bianchi}
\end{align}
is the condition to be imposed in the supergravity regime (the gauge field $F$ is Hermitian, here). This condition 
yields non-trivial constraints for any choice of compact four-cycle. Relevant to the present discussion is 
the four-cycle K3$_{\rm Het}$, which is a $T^2_{67}$-fibration over $\P^1_{\rm Het}$  (here, we use $\Lambda_S = U$, 
so that we are in the supergravity regime in heterotic language). This condition is replaced by (\ref{eq:cond-multiplicative}), but now with 
$T_i \in {\rm Isom}(\Lambda_T) \subset {\rm Isom}({\rm II}_{4,20})$.

The Type IIB additive condition is reproduced from (\ref{eq:cond-multiplicative}) in F-theory when we consider 
the orientifold limit. Two 7-branes come so close to one another, that we can collectively treat them as an O7-plane. 
The monodromy $\prod_i T_i$ around the two 7-branes (O7-plane as a whole) commutes with the monodromy matrix 
for a D7-brane. The multiplicative condition (\ref{eq:cond-multiplicative}) for 3-branes in 6D $(1,1)$ 
supergravity also becomes additive in the same way. To see this, note first that monodromy from NS5-branes 
is given by $T = \exp[N^{\rm II}_0]$ acting non-trivially on $(U \oplus U) \subset 
U^{\oplus 2} \oplus E_8^{\oplus 2}=\Lambda_T$.
Secondly, the instanton number is the same as the zero of $g_{\pm 1}$ in (\ref{eq:gm-E8ell}), which is a section of 
${\cal O}(\eta_\pm)$. The monodromy locus was worked out in \cite{Hayashi:2010zp}. Each one of the zeros 
of $g_{\pm 1}$ splits into multiple monodromy points, and the splitting remains small when we take $\epsilon_\eta$ and 
$\epsilon_K$ in section \ref{sssec:AetaBK} small, just like in the orientifold limit of F-theory. 
The monodromy $\prod_i T_i$ from a set of those monodromy points associated with a given zero of $g_{\pm 1}$ 
is block diagonal in $E_8 \oplus (U \oplus U) \oplus E_8  = \Lambda_T$; it is a Weyl reflection on one of 
the two $E_8$'s, while it is trivial on the other $E_8$, and it acts as $\exp[-N^{\rm II}_0]$ on $(U \oplus U)$,  
according to footnote 11 of \cite{Hayashi:2010zp}.\footnote{We adopt a heterotic--F dictionary 
$-\tilde{\rho} \leftrightarrow \pm \int_{C_\beta^2} \Omega_{\rm K3}$ and 
$(-\tilde{\rho}\tau-a^2) \leftrightarrow \pm \int_{C_\alpha^2} \Omega_{\rm K3}$, or 
$-\tilde{\rho} \leftrightarrow \pm \int_{C_\alpha^2} \Omega_{\rm K3}$ and 
$(-\tilde{\rho}\tau - a^2) \leftrightarrow \mp \int_{C_\beta^2} \Omega_{\rm K3}$. } 
The monodromy $\prod_i T_i$ from the NS5-branes and instantons
combined---the left-hand side of (\ref{eq:Het-Bianchi})---therefore splits into the two $E_8$'s and $(U \oplus U)$. 
The monodromy in the $E_8 \oplus E_8$ takes values in the Weyl group, and will probably cancel after all of the 
would-be Type I monodromies (cf the appendix \ref{sec:transcendental}) are taken into account. The monodromy on 
the $(U \oplus U)$ component has become additive, $\prod_i T_i = \exp[ (\#({\rm NS5}) - \#({\rm inst.})) N_0^{\rm II} ]$. 
We believe that there is a mistake somewhere and the correct result is $\prod_i T_i = \exp[ (\#({\rm NS5})+\#({\rm inst.})) N_0^{\rm II} ] = \exp[24 N_0^{\rm II}]$, since there can be trade-off between instantons and the NS5-branes \cite{GH}, but we have not managed to identify an error. 
This $ U\oplus U$ part of the monodromy will presumably be cancelled against contributions from other would-be Type I 
monodromy points that are attributed to the contribution on the right-hand side of (\ref{eq:Het-Bianchi}).

\section*{Acknowledgements}

We would like to thank P.~Berglund, T.~Eguchi, S.~Hellerman, D.~Morrison, Y.~Tachikawa and C.~Vafa for helpful discussions related to the present work.
TW thanks the Centre for Fundamental Laws of Nature at Harvard University for hospitality, as this work was done during his stay there.
This work is supported in part by STFC grant ST/L000474/1 and EPSCR grant EP/J010790/1 (APB) and by WPI Initiative and a Grant-in-Aid for Scientific Research on Innovative Areas 2303, MEXT, Japan, and JSPS Brain Circulation program (TW).

\appendix

\section{\texorpdfstring{$K3$}{Lg}-fibred Calabi-Yau Threefolds as Toric Hypersurfaces}
\label{sec:K3-fib-tor-review}

Famously, Calabi-Yau threefolds can be constructed as hypersurfaces in toric varieties by starting from a pair of reflexive polytopes
$\Delta,\widetilde{\Delta}$ \cite{1993alg.geom.10003B} (see e.g. \cite{Kreuzer:2006ax} for a quick review). Such Calabi-Yau threefolds 
may admit a fibration by K3 surfaces which can be spotted already at the level of the 
polytopes \cite{KLM,Candelas:1996su,Avram:1996pj,Kreuzer:1997zg,Kreuzer:2000qv}. As most of this material is in principle well-known, 
we restrict ourselves to highlight those facts which are relevant to our discussion. 

Assume we are given a four-dimensional reflexive polytope $\widetilde{\Delta} \subset (N \otimes \R)$ such that, for a three-dimensional hyperplane $N_F \otimes \R$ passing 
through the origin, $\widetilde{\Delta}_F = \widetilde{\Delta} \cap (N_F \otimes \R)$ is again a reflexive lattice polytope. Note that this means in particular 
that the vertices of $\widetilde{\Delta} \cap (N_F \otimes \R)$ must be lattice points. A Calabi-Yau hypersurface $M$ constructed from $\Delta,\widetilde{\Delta}$ 
then admits a fibration by K3 surfaces $S$ over a base $\P^1$. Let $f$ be a unit vector in $N^\vee$ (the dual lattice of $N$) orthogonal to $N_F \subset N$. More precisely, we have to use a triangulation (fan $\Sigma$) of $\widetilde{\Delta}$ such that the fibration morphism is realized as a toric morphism of the ambient space, i.e. there is a projecting to the fan of $\P^1$ such that every cone in $\Sigma$ is mapped to a unique cone of the fan of $\P^1$. For the examples discussed in this work, this is easy to verify, Section \ref{ssec:corridor-II} contains several interesting examples.

With a triangulation admitting a fibration morphism, we may then describe the coordinates of the base $\P^1$ by
\begin{equation}\label{eq:coords_base_general}
[z_{0}:z_{1}] =  \left[\prod_{\nu^i|\langle f, \nu^i\rangle > 0} X_i^{\langle f, \nu^i\rangle} : \prod_{\nu^i|\langle f, \nu^i\rangle < 0} X_i^{-\langle f, \nu^i\rangle} \right]
\end{equation}
One can think of all but one of the coordinates $\nu^i$ for which $\langle f, \nu^i\rangle > 0$ (and similarly for $<0$) as corresponding to the exceptional divisors of blow ups of singular fibres. 

\subsection{Geometry of Generic Fibres}

Fixing the coordinates at a generic point of the base $\P^1$, we find a generic fibre $S_{t}$ described as an algebraic hypersurface. 
The defining polynomial is found from the defining polynomial of $M$ upon fixing all $X_i$ for which $\langle \nu^i, f \rangle \neq 0$ and 
this hypersurface is embedded in an ambient toric variety with rays $\nu^i_F$ on $\widetilde{\Delta}_F$. Equivalent to $\widetilde{\Delta}_F = \widetilde{\Delta} \cap (N_F \otimes \R)$ being a lattice polytope is the existence of a projection $P:\Delta \rightarrow \Delta_F$ induced by translations along $f$ such that $\Delta_F$ is the polar dual to $\widetilde{\Delta}_F$ \cite{Avram:1996pj}. This means that a generic fibre $S_{t}$ is described in the usual way by a pair of reflexive polytopes $\Delta_F, \widetilde{\Delta}_F$. In particular, the Picard lattice of a generic fibre is the same as the Picard lattice of a generic toric K3 hypersurface.

\subsection{Singular Fibres}

Over specific points in the base $\P^1_A$, the K3 fibre may become reducible. Individual components of such reducible fibres contribute to $h^{1,1}$, which can also be computed combinatorially. We hence expect to be able to describe in terms of combinatorial data when reducible fibres occur. From the point of view of $\widetilde{\Delta}$, these come in two types, which we discuss now.  
The first type can already be seen from \eqref{eq:coords_base_general}: whenever $\widetilde{\Delta}$ contains\footnote{As usual, points interior to facets do not count as they do not give rise to divisors on a Calabi-Yau hypersurface.} more than one lattice point with $\langle f, \nu \rangle > 0$ (more than one lattice point with $<0$), there is a reducible fibre over $z_{0}=0$ ($z_{1}=0$).
In particular, we may write the total fibre class 
\begin{equation}
 [S_{t}]  = \sum_{\nu^i|\langle f, \nu^i\rangle > 0} \langle f, \nu^i\rangle [D_i] =
   - \sum_{\nu^i|\langle f, \nu^i\rangle < 0} \langle f, \nu^i\rangle [D_i]
\end{equation}
Note that some of these fibre components will contribute with a multiplicity greater than one. This means in particular that fibres for which this happens are not reduced. A closely related discussion is given in \cite{short_tops}. 

The second type of singular fibres stems from interior points of facets of $\widetilde{\Delta}_F$ which do not lie in facets of $\widetilde{\Delta}$. Denoting such a facet of $\widetilde{\Delta}_F$ by $\widetilde{\Theta}^{[2]}_F$, this means that $\widetilde{\Theta}^{[2]}_F$ is also a face of $\widetilde{\Delta}$. As $\widetilde{\Delta}$ is reflexive, there is hence a dual one-dimensional face $\Theta^{[1]}$ on $\Delta$ and each interior point of $\widetilde{\Theta}^{[2]}$ gives rise to a divisor that has $(\ell^*(\Theta^{[1]})+1)$ irreducible components. Those $(\ell^*+1)$ irreducible pieces, however, do not form a single reducible fibre of the K3 fibration but are distributed among several reducible fibres separated in the base $\P^1$, as we now explain. First of all, calling the dual vertex (under polar duality of $\Delta_F,\widetilde{\Delta}_F$) of $\widetilde{\Theta}^{[2]}_F$ by $m_F$, the dual one-dimensional face $\Theta^{[1]}$ of $\widetilde{\Theta}^{[2]}_F$ is contained in the line $ m_F + l \cdot f$ ($l \in \R$).   
For any point $\nu$ interior to a two-dimensional face $\widetilde{\Theta}^{[2]}_F$, the defining equation of the associated divisor 
is hence of the form 
\begin{equation}
\left( \prod_{\nu^i \in \widetilde{\Delta}_F} (X_{F;i})^{\langle m_F,\nu^i \rangle+1} \right) P(z_{0},z_{1}) = 0
\end{equation}
It follows from the theory of \cite{DK} that $P(z_{0},z_{1})$ has $\ell^*(\Theta^{[1]})+1$ roots $p_i$, so that $D_\nu$ has $\ell^*(\Theta^{[1]})+1$ components. As is apparent from the above equation, these components are sitting over $\ell^*(\Theta^{[1]})+1$ different locations in the base $\P^1$.
Note that the same monomials (the ones related to $\Theta^{[1]}$) will appear in the defining equation of each of the divisors corresponding to interior points of the face $\widetilde{\Theta}^{[2]}_F$, so that the same $P(z_{0},z_{1})$ will appear for each of them.

We can turn the argument around and investigate the geometry of the K3 fibre over the points $p_i$ in the base. As we have learned above, the defining equation of $M$ must have the form 
\begin{equation}
R(X_i) \prod_{\nu \supset \widetilde{\Theta}^{[2]}_F} X_\nu + Q(X_i) P(z_{0},z_{1}) = 0
\end{equation}
where only $\nu$ interior to $\widetilde{\Theta}^{[2]}_F$ are considered. Hence the singular fibre over any of the $p_i$ has the components  
\begin{equation}
[S_{t}] = \sum_{\nu \supset \widetilde{\Theta}^{[2]}_F} D_\nu^{p_i} + [R]^{p_i}
\end{equation}
We cannot exclude that $R$ is reducible and gives an non-trivial multiplicity to some of the $D_\nu$.
Examples of this second type of singular fibre are found in \cite{Candelas:1996su}.

\subsection{Hodge Numbers of Divisors of a Calabi--Yau Threefold}
\label{ssect:hodge-num-div}

Let $M$ be a Calabi--Yau $n$-fold obtained as a hypersurface of a toric $(n+1)$-dimensional ambient space, 
$\nu$ a lattice point $\widetilde{\Delta} \cap N$, and $\overline{D}_\nu$ the corresponding divisor of $M$.
The Hodge numbers of the $(n-1)$-fold $\overline{D}_\nu$ can be derived using the methods of \cite{DK}. 
Formulas for $h^{1,1}(\overline{D}_\nu)$ are found in \cite{BW-H22}, but they are applicable only to cases 
with $n\geq 4$. While the same reasoning can be applied to the $n=3$ case, the formula looks different. 
This appendix provides a summary of the result for $n=3$. 

Let $\widetilde{\Theta}^{[k]}, k = 0,1,2$ be the face containing $\nu$ in its relative interior (a vertex 
corresponds to a zero-dimensional face and our convention is to consider it as its own relative interior) 
and let us denote the dual face of $\widetilde{\Theta}^{[k]}$ by $\Theta^{[3-k]}$. We can then summarize the 
Hodge numbers $h^{0,i}(\overline{D}_\nu)$ by \cite{DK}
\begin{equation}
 \begin{array}{c|ccc}
k & h^{0,0} & h^{1,0} & h^{2,0} \\
\hline
0 & 1 & 0 & \ell^*(\Theta^{[3]}) \\
1 & 1 & \ell^*(\Theta^{[2]}) & 0 \\
2 & \ell^*(\Theta^{[1]}) & 0 & 0 
 \end{array}\, ,
\end{equation}
where $\ell^*(\Theta^{[l]})$ counts points in the relative interior of the face $\Theta^{[l]}$. Note that these are already determined without having to specify the details of the fan of the ambient space (triangulation). 

Since divisors of a threefold are surfaces, we only need to determine $h^{1,1}(\overline{D}_\nu)$ now. It 
depends on the triangulation data of $\widetilde{\Delta}$ and can be described as
\begin{align}
 h^{1,1}(\overline{D}_\nu) & = \sum_{\widetilde{\Theta}^{[1]}} \ell^1_{\nu} (\widetilde{\Theta}^{[1]}) +  \sum_{(\widetilde{\Theta}^{[2]},\Theta^{[1]})} \ell^1_{\nu} (\widetilde{\Theta}^{[2]})\ell^*(\Theta^{[1]}) - 3  \nonumber \\ 
 & + \delta_{k,0}\left( \ell^*(2 \Theta^{[3-k]}) - 4\ell^*(\Theta^{[3-k]}) - \sum_{\Theta^{[2-k]}\supset \Theta^{[3-k]}} \ell^*(\Theta^{[2-k]})\right).
\label{eq:h11-formula}
\end{align}
Here, $\ell^1_\nu(\widetilde{\Theta}^{[l]})$ counts the number of $\nu$-containing one-simplices in the relative 
interior of a face $\widetilde{\Theta}^{[l]}$. The last line only contributes if $\nu$ is a vertex.

\section{Mathematics of Degenerations of K3 Surfaces}
\label{sec:K3-degen-review}

A lot is known about the degeneration of K3 surface in the mathematics literature. Here is a quick summary of what we use 
in this article, for convenience of readers. The largest fraction of material in this appendix \ref{sec:K3-degen-review}
originates from \cite{FM, Mor-CS, Scattone}. Whenever we do not refer to a reference for a non-trivial statement, 
at least some clue is provided in one of these papers.

\subsection{Degenerations of K3 Surfaces}

\subsubsection{Kulikov models and the geometry of the central fibre}

[{\bf Def.}] A one parameter {\it family of K3 surface} consists of $({\cal X}, \pi, {\rm Disc})$, where ${\cal X}$ is a complex
threefold, ${\rm Disc} := \{ t \in \C \; | \; |t|<1 \}$, and $\pi: {\cal X} \rightarrow {\rm Disc}$ is a morphism 
such that $S_t := \pi^{-1}(t)$ for ${}^\forall t \neq 0$ is a non-singular K3 surface. A {\it central fibre} of a 
degeneration is $\pi^{-1}(t=0)$ which is often denoted by $S_0$. 

One can think of a degeneration over a multi-dimensional parameter space, where ${\rm Disc} \subset \C$ is replaced 
by ${\rm Disc} \subset \C^n$. We do not deal with multi-parameter degenerations, and we will drop ``one parameter'', 
though it is always assumed implicitly in this article. 
  
\noindent  
[{\bf Def.}] A degeneration of a K3 surface $({\cal X},\pi, {\rm Disc})$ is {\it semi-stable}, if the following conditions 
i)--iii) are satisfied. 
\begin{itemize}
 \item[i)] ${\cal X}$ is non-singular
 \item[ii)]  the central fibre $S_0$ consists of irreducible 
components $S_0 = V_0 \cup V_1 \cup \cdots \cup V_\mu$, and all the singularity in the variety $S_0$ corresponds 
to normal crossing loci of the divisors $V_i$'s in ${\cal X}$
\item[iii)]each one of $V_i$'s appear in $S_0$ with 
multiplicity 1 ($S_0$ is reduced)
\end{itemize}

\noindent  
[{\bf Def.}] A semi-stable degeneration of a K3 surface $({\cal X}, \pi, {\rm Disc})$ is a {\it Kulikov model}, if 
$K_{\cal X}=0$.
  
\noindent    
[{\bf Theorem} (Kulikov \cite{Kulikov}, Persson--Pinkham \cite{Persson_Pinkham})] For a semi-stable degeneration of a K3 surface 
$({\cal X}, \pi, {\rm Disc})$, one can always find a chain of birational transformations and base changes
of the degeneration so that the resulting degeneration $({\cal X}', \pi', {\rm Disc})$ is a Kulikov model..

\noindent  
[{\bf Def.}] A degeneration of $({\cal X}',\pi', {\rm Disc})$ is a birational transformation of another degeneration 
of a K3 surface $({\cal X}, \pi, {\rm Disc})$ and vice versa, if there is a birational morphism between ${\cal X}'$ and 
${\cal X}$ that commutes with the projections $\pi$ and $\pi'$, and the birational morphism induces isomorphism between 
${\cal X}' \backslash (\pi')^{-1}(0)$ and ${\cal X} \backslash \pi^{-1}(0)$.

\noindent  
[{\bf Def.}] When $({\cal X}, \pi, {\rm Disc})$ is a degeneration of a K3 surface, one can construct another degeneration of a
K3 surface, $({\cal X}', \pi', {\rm Disc})$, by using a base change of order $n$. 
Let $f: {\rm Disc} \ni t' \mapsto (t')^n = t \in {\rm Disc}$; then ${\cal X}'$ is the fibre product 
${\cal X} \times_{\rm Disc} {\rm Disc}$ of $\pi:{\cal X} \rightarrow {\rm Disc}$, $f$ is the base change morphism, and $\pi'$ 
is the projection to the second factor. A degeneration $({\cal X}', \pi', {\rm Disc})$ constructed in this way 
is a {\it base change} of a degeneration of a K3 surface.

The theorem above makes Kulikov models into a well-motivated class of degenerations to study. Kulikov model degenerations 
of K3 surface are {\bf classified into three types}. Let $({\cal X}, \pi, {\rm Disc})$ be a Kulikov model degeneration 
of a K3 surface, and $S_0$ the central fibre. The classification is stated in terms of the geometry of the central fibre 
and also in terms of the monodromy of a generic fibre. Let $T$ be the monodromy matrix acting on $H^2(S_t; \Z)$ of a generic 
non-singular K3 fibre at $t \neq 0$ around the point of degeneration $t=0$, from which a matrix $N$ is defined 
as its log, $T = \exp[N]$. Now, 
\begin{itemize}
\item Type I: the central fibre $S_0$ consists of a single irreducible component. $N=0$. 
\item Type II: the central fibre $S_0 = V_0 \cup V_1 \cup \cdots \cup V_\mu$ consists of $\mu + 1 > 1$ irreducible 
components; $V_i \cdot V_j$ is non-empty if and only if $|j-i|=1$; the graph of intersection (the dual graph) has 
$\mu$ edges connecting $\mu+1$ nodes, and forms an interval as a whole (e.g. Figure~\ref{fig:NS5-Ftheory}~(b)). 
$N \neq 0$, but $N^2=0$.
\item Type III: the central fibre $S_0$ consists of multiple irreducible components, and their dual graph is 
a triangulation of $S^2$ (two-dimensional sphere). $N^2 \neq 0$, and $N^3=0$.
\end{itemize}
Even in Type III degeneration, it is known \cite{FS} that the matrix $N$ is integer valued, when represented 
in the integral basis of $H^2(S_t; \Z)$. 

A choice of Kulikov model is not necessarily unique, in that there may be two Kulikov model degenerations of a K3 surface 
$({\cal X}, \pi, {\rm Disc})$ and $({\cal X}', \pi', {\rm Disc})$ that are birational transforms of one another. 
Those Kulikov models are always classified into the same type, because birational morphism between ${\cal X}$ and 
${\cal X}'$ do not modify the properties of the monodromy matrix of a generic fibre. 

In this article, examples of Type II degeneration are discussed in sections \ref{ssec:a17fibres}, \ref{ssec:corridor-II}, 
those of Type III degeneration in section \ref{ssec:corridor-III}, while a brief discussion is given on a ``cousin'' of 
Type I degeneration in the appendix \ref{sec:transcendental}.

The geometry of the central fibre of a Type II degeneration has the following properties.
\begin{itemize}
\item [$\{a\}$] $V_0$ and $V_\mu$ at the ends of the dual graph are both rational surfaces, while 
the surfaces in the middle, $V_1, \cdots, V_{\mu-1}$, have a minimal model that is ruled over an elliptic curve. 
\item [$\{b\}$] The curve $V_i \cap V_{i+1} =: C_{i,i+1}$ of a pair of adjacent irreducible pieces $V_i$ and $V_{i+1}$ 
is often referred to as the {\it double curve}. Within ${\cal X}$, there is a normal crossing singularity at each $C_{i,i+1}$. 
The double curve $C_{i,i+1}$ is always an elliptic curve. 
\item [$\{c\}$] On a surface $V_i$, $(-K_{V_i}) = C_{i,i+1}+C_{i-1,i}$; if $i=0$ or $i=\mu$, just keep one of them.
For any double curve, there is a relation $(C_{i,i+1})^2|_{V_i} + (C_{i,i+1})^2|_{V_{i+1}}=0$.
\item [$\{d\}$]  All the double curves $C_{i,i+1}$ with $i=0,\cdots, \mu-1$ in a Type II degeneration share the same 
complex structure, which is ensured by the ruling of the surfaces $V_{i+1}$. 
\label{itm:c-fib-TypeII}
\end{itemize}
When there is a pair of Type II degenerations of K3 surfaces that are birational transformations of one another, 
the number of irreducible components $\mu+1$ is common to both. The rational surfaces $V_0$ and $V_\mu$ of 
one degeneration may not be isomorphic to those of the other degeneration. The value of $(C_{i,i+1})^2|_{V_i}$ is 
not necessarily preserved in the birational transform either. 

The geometry of the central fibre of a Type III degeneration has the following properties. 
\begin{itemize}
\item [$\{a\}$] Each one of the irreducible components, $V_i$, is a rational surface.
\item [$\{b\}$] $V_i \cap V_j =: C_{i,j}$, if not empty, is a rational curve.
\item [$\{c\}$] On a surface $V_i$, $(-K_{V_i})=\sum_j C_{i,j}$. For any curve $C_{i,j}$, there is a relation 
$(C_{i,j})^2|_{V_i} + (C_{i,j})^2|_{V_j}=-2$, where ``2'' is the number of triple points ($V_i \cdot V_j \cdot V_k$ 
for some $V_k$) on the curve $C_{i,j}$. 
\label{itm:c-fib-TypeIII}
\end{itemize}
The number of triple points $(V_i \cdot V_j \cdot V_k)$---the number of triangles in the dual graph---remains 
invariant under flops. This invariant is denoted by $t$.  

\subsubsection{Monodromy action}

The monodromy group action $T=\exp[N]: H^2(S_t;\Z) \rightarrow H^2(S_t; \Z)$ and the geometry of the central 
fibre $S_0$ are related by the Clemens--Schmid exact sequence:
\begin{equation}
 \vcenter{\xymatrix{
 & 0        \ar[r]^{N} & H^0(S_t) \ar[r]^{\beta} & H_4(S_0) \ar[r]^{\alpha} & H^2(S_0) \ar[r]^-{i^*}
   &  \ar`r[d] `[l] `[llllld] `[lllld] [lllld] \\
 & H^2(S_t) \ar[r]^{N} & H^2(S_t) \ar[r]^{\beta} &  H_2(S_0) \ar[r]^{\alpha} & H^4(S_0) \ar[r]^-{i^*} 
   &  \ar`r[d] `[l] `[llllld] `[lllld] [lllld] \\
 &  H^4(S_t) \ar[r]^{N} & 0.                   &                        &
   & 
}}
 \label{eq:CS-sequence}
\end{equation}
Reference \cite{Mor-CS} provides background material for the Clemens--Schmid exact sequence 
including the definition of other homomorphisms (such as $\alpha$, $\beta$ and $i^*$). 
The monodromy matrix $N$ introduces a filtration (called the monodromy weight filtration) into the 
cohomology groups of a generic fibre. A filtration is also introduced into the cohomology and homology 
groups of the central fibre by using the Mayer--Vietoris spectral sequence computation. 
In the Clemens--Schmid exact sequence for the degeneration of a K3 surface, the morphisms $\alpha$, 
$i^*$, $N$ and $\beta$ respect the filtration structure while shifting the weight by $+6$, $+0$, $-2$ and $-4$, 
respectively.

The monodromy weight filtration on the middle dimensional cohomology $H^2(S_t;\Z)$ is given by 
\begin{align}
  \{ 0 \} \subset W_0 \subset W_1 \subset W_2 \subset W_3 \subset W_4 = H^2(S_t; \Z),
\end{align}
where 
\begin{align}
  W_0 = {\rm Im}(N^2), \quad W_1 = {\rm Ker}(N) \cap {\rm Im}(N), \quad 
  W_2 = {\rm Ker}(N) + {\rm Im}(N), \quad W_3 = {\rm Ker}(N^2).
\end{align}
The following properties are useful: $N(W_j) \subset W_{j-2}$, and $W_{3-j} \subset [W_j^\perp \subset H^2(S_t;\Z)]$.
The monodromy is trivial on $H^4(S_t; \Z)$ and $H^0(S_t; \Z)$ and the filtration is formally defined 
here by $H^4(S_t) = W_4 \supset W_3 = \{0\}$ and $H^0(S_t) = W_0 \supset W_{-1} = \{0\}$.

In a Type II degeneration of a K3 surface, where $N^2=0$, $W_0=\{0\}$, $W_3 = W_4=H^2(S_t;\Z)$, 
\begin{align}
 \{ 0\} \subset [W_1 ={\rm Im}(N)] \subset [W_2 = {\rm Ker}(N)] = W_1^\perp \subset H^2(S_t;\Z),
\end{align}
The $W_1$ subspace is always of rank-2, and $W_2$ always of rank-20, within the rank-22 space $H^2(S_t; \Z)$. 
Restriction of the intersection form of $H^2(S_t;\Z)$ to its primitive sublattice $W_1$ is trivial, because 
an element of $W_1 \subset W_2$ is orthogonal to any element in $W_1$. That is, $W_1$ is a rank-2 isotropic 
primitive sublattice of $H^2(S_t; \Z)\cong {\rm II}_{3,19}$. 

Because of the self-dual nature of ${\rm II}_{3,19}$, one can always find a sublattice 
\begin{align}
W_1 \subset (U \oplus U) \subset (U \oplus U) \oplus U \oplus E_8^{\oplus 2} = {\rm II}_{3,19}
\end{align}
so that 
\begin{align}
  W_1 \oplus W'_1 \cong U \oplus U, \qquad 
  W_1 = {\rm Span}_\Z \left\{ \hat{e}^{'1}, \hat{e}^{'2} \right\}, \qquad 
  W'_1 = {\rm Span}_\Z \left\{ \hat{e}^1, \hat{e}^2 \right\}, 
\label{eq:N-basis-H2-nopol}
\end{align}
and 
\begin{align}
   (\hat{e}^{'i}, \hat{e}^j ) = \delta^{ij}, \qquad (\hat{e}^{'i}, \hat{e}^{'j}) = 0, \qquad 
(\hat{e}^i, \hat{e}^j) = 0. 
\end{align}
It follows that $W_2/W_1 \cong (U \oplus E_8^{\oplus 2})$, and $(W_3/W_2) \cong W_1'$. 
Because the matrix $T = \exp[N] = {\bf 1} + N$ needs to be an isometry of the lattice ${\rm II}_{3,19}$, 
the nilpotent matrix $N$ for a Type II degeneration of K3 surface is always in the form of 
\begin{align}
 N = \mu N_0^{\rm II} = \mu \times 
\left(\begin{array}{cc|cc}
  & & & \\
  & & & \\
\hline 
  &-1& & \\
 1& & & \end{array} \right)
\label{eq:N0-def-II-nopol}
\end{align}
for some integer $\mu$, in the basis of $\{ \hat{e}^1, \hat{e}^2, \hat{e}^{'1}, \hat{e}^{'2} \}$; 
$N$ acts trivially on $W_2$, because $N(W_2)\subset W_0 = \{0\}$.
The integer $\mu$---taken always positive---is called ${\rm index}$ of a Type II degeneration of K3 surface. 

It is known that the integer $\mu$ is the same as the number of double curves $C_{i,i+1}$ ($i=0,\cdots,\mu-1$) in the 
geometry of the central fibre \cite{friedman:new_proof}. It is reasonable that a birational invariant of the 
geometry of the central fibre is also captured in the language of monodromy acting on a generic fibre. 
 
The central fibre is regarded as a limit of complex structure of the fibre K3 surface in such a way that 
the period integral is dominated by $W_1 \otimes \C \subset {\rm II}_{3,19} \otimes \C$. The limiting value 
$\Omega \in W_1 \otimes \C$ of the period integrals obviously satisfies $\Omega^2 = 0$, because the intersection 
form on $W_1$ is trivial.

In a Type III degeneration of K3 surface, where $N^3=0$, 
\begin{align}
   \{ 0 \} \subset \left( W_0 = W_1 \right) \subset \left( W_2 = W_3 = W_0^\perp \right) \subset H^2(S_t; \Z).
\end{align}
The $W_0$ subspace is always of rank-1 and $W_2$ always of rank-21 within the rank-22 space $H^2(S_t;\Z)$.
The restriction of the intersection form of $H^2(S_t;\Z)$ to its primitive sublattice $W_0$ is trivial, because 
an element of $W_0 \subset W_3$ is orthogonal to any element in $W_0$. 

The self-dual nature of the lattice ${\rm II}_{3,19}$ can be exploited to find a sublattice 
\begin{align}
 W_0 \subset U \subset U \oplus (U^{\oplus 2} \oplus E_8^{\oplus 2}) = {\rm II}_{3,19}
\end{align}
so that 
\begin{align}
 W_0 = {\rm Span}_\Z \{ \hat{e}' \}, \quad W'_0 = {\rm Span}_\Z \{ \hat{e} \}, \quad 
 W_0 \oplus W'_0 \cong U, \qquad (\hat{e}',\hat{e}')=(\hat{e},\hat{e})=0, \quad (\hat{e}',\hat{e})=1.
\end{align}
It follows that $W_2/W_0 \cong (U^{\oplus 2} \oplus E_8^{\oplus 2})$ and $(W_4/W_2) \cong W_0'$.

In the case of a Type III degeneration, both $N: W_2/W_0 \rightarrow W_0$ and $N:W_4/W_2 \rightarrow W_2$ are 
non-trivial. Because of the self-dual nature of $W_2/W_0 \cong U^{\oplus 2} \oplus E_8^{\oplus 2}$, one can always find 
a sublattice $U' = {\rm Span}_\Z \{ \hat{f}, \hat{f}' \}$ isometric to $U$ such that $N(W_4) \subset U' \subset W_2/W_0$. In the basis 
of $\{ \hat{e}, \hat{f}, \hat{f}', \hat{e}' \}$, the monodromy matrix $N$ is always in the form of 
\begin{align}
 N = \mu N_0^{\rm III}(t_0) = \mu \times 
\left(\begin{array}{c|cc|c}
    & 0 & 0 & \\
 \hline
  1 & & & 0 \\
  t_0/2 & & & 0 \\
 \hline
    & -t_0/2 & -1 & \end{array} \right), 
\label{eq:N0-def-III-nopol}
\end{align}
for some integer $\mu$. Here, $\mu^2 t_0 = t = (N(\hat{e}), N(\hat{e}))$. 
It is known that this $t$ is the same as the birational invariant $t$ of the central fibre geometry 
explained earlier. The period integral in this degeneration limit is dominated by the components in $W_0 \otimes \C$.

\subsection{Degeneration of Lattice-Polarized K3 Surfaces}
\label{ssec:K3-degen-pol-review}

[{\bf Def.}] A degeneration of a K3 surface $({\cal X}, \pi, {\rm Disc})$ is called {\it lattice-polarized}, 
if the restriction of divisors $D_{i=1,\cdots, \rho}$ of ${\cal X}$ to a generic fibre 
$S_t$ generates a subset of ${\rm Pic}(S_t)$ that is isometric to a lattice $\Lambda_S$. Such a degeneration 
is also called a {\it degeneration of $\Lambda_S$-polarized K3 surface}.

The isotropic sublattice $W_1$ in a Type II degeneration and $W_0$ in a Type III degeneration is a primitive 
sublattice of $\Lambda_T := \left[ \Lambda_S^\perp \subset {\rm II}_{3,19}\right]$, because this is where the 
limiting values of period integrals reside. The lattice $\Lambda_S$ sits within the $W_2$ component for 
a Type II (resp. Type III) degeneration, because the algebraic component in $\Lambda_S$ must be orthogonal 
to $W_1$ (resp. $W_0$). The monodromy matrix $T$ (and hence $N$) acts non-trivially on the lattice $\Lambda_T$, 
and trivially on $\Lambda_S$. 

Scattone \cite{Scattone} formulated a classification problem of Type II and Type III degenerations of 
$\Lambda_S$-polarized K3 surface as follows:
\begin{itemize}
\item Type II: Classify rank-2 primitive isotropic sublattice $W_1$ in $\Lambda_T$, modulo $\Gamma$,
\item Type III: Classify rank-1 primitive isotropic sublattice $W_0$ in $\Lambda_T$, modulo $\Gamma$.
\end{itemize}
Two well-motivated choice of the quotient group $\Gamma$ are ${\rm Isom}(\Lambda_T)$ and its normal subgroup
\begin{align}
  {\rm Isom}^*(\Lambda_T) := {\rm Ker} \left[ {\rm Isom}(\Lambda_T) \longrightarrow 
   {\rm Isom}(G_{\Lambda_T}, q_{\Lambda_T}) \right];
\end{align}
Classification under $\Gamma = {\rm Isom}^*(\Lambda_T)$ achieves a finer classification than 
that under the choice $\Gamma = {\rm Isom}(\Lambda_T)$. It is often easier to think of classification by 
$\Gamma = {\rm Isom}(\Lambda_T)$ first, and then to refine the classification later. 
In the rest of this appendix, we only refer to classification under $\Gamma={\rm Isom}(\Lambda_T)$.

The ${\rm Isom}(\Lambda_T)$ classification for Type II degenerations can be worked out in this way \cite{Scattone}.
Note first, that one can always find a basis 
\begin{align}
 \left\{ \hat{e}^{'1}, \hat{e}^{'2}, \hat{f}^1, \cdots, \hat{f}^{18-\rho}, \hat{e}^1, \hat{e}^2 \right\}
  \label{eq:LambdaT-basis-general}
\end{align}
of $\Lambda_T$ in such a way that 
\begin{align}
   W_1 =& {\rm Span}_\Z \{ \hat{e}^{'1}, \hat{e}^{'2} \}, \nonumber \\
   (W_2 \cap \Lambda_T)/W_1 =& {\rm Span}_\Z \{ [\hat{f}^1], \cdots, [\hat{f}^{18-\rho}] \}, \nonumber \\ 
   \Lambda_T/(W_2 \cap \Lambda_T) =& {\rm Span}_\Z \{ [\hat{e}^1], [\hat{e}^2] \},
  \label{eq:filtr-TypeII}
\end{align}
and the intersection form of $\Lambda_T$ is given by\footnote{The presentation in \cite{Scattone} corresponds 
to $\delta_1 = 1$ and $\delta_2 = e$. Since concrete examples of Type II degeneration treated in \cite{Scattone} were 
all for $\rho=1$ lattice polarization, the discriminant group $G_{\Lambda_S}$ is always a cyclic group. It was thus safe 
to set $\delta_1 = 1$ for that reason. The presentation here is a straightforward generalized of that. }   
\begin{align}
   \left( \begin{array}{cc|cc|cc}
      * & * & * & * & \delta_1 & \\
      * & * & * & * & & \delta_2 \\
     \hline 
      * & * & * & * & & \\
      * & * & * & * & & \\
     \hline
      \delta_1 & & & & & \\
      & \delta_2 & & & & 
   \end{array} \right) 
 \label{eq:int-form-II-pol-gen}
\end{align}
for some positive integers $\delta_1$ and $\delta_2$ satisfying $\delta_1 | \delta_2$.
The two integers $\delta_1$ and $\delta_2$ (with a constraint $\delta_1 | \delta_2$) and a lattice 
$(W_2 \cap \Lambda_T)/W_1$ (modulo isometry) are uniquely determined for a given ${\rm Isom}(\Lambda_T)$-equivalence 
class. 

Once a pair $(\Lambda_S, \Lambda_T)$ is given, possible choices of $\delta_1$, $\delta_2$ and an isometry 
class of $(W_2\cap \Lambda_T)/W_1$ can be worked out systematically as follows. 
The discriminant group $G_{\Lambda_T}$ is supposed to allow this substructure, first of all:
\begin{align}
 G_{\Lambda_T} = (\Z_{\delta_1} \times \Z_{\delta_2}) . G_{(W_2 \cap \Lambda_T)/W_1} . (\Z_{\delta_1} \times \Z_{\delta_2}).
\end{align}
The $(\Z_{\delta_1} \times \Z_{\delta_2})$ subgroup is an isotropic subgroup of $(G_{\Lambda_T}, q_{\Lambda_T})$, and 
furthermore the $(\Z_{\delta_1} \times \Z_{\delta_2}) . G_{(W_2 \cap \Lambda_T)/W_1}$ subgroup is orthogonal to 
the $(\Z_{\delta_1} \times \Z_{\delta_2})$ subgroup under the discriminant bilinear form $b(\bullet, \bullet)$.
Since $G_{\Lambda_T} \cong G_{\Lambda_S}$ is a finite group, there are only finitely many options for such a substructure 
in $(G_{\Lambda_T}, q_{\Lambda_T})$. In particular, there are only finitely many choices of $\delta_1$, $\delta_2$ and isometry 
classes of the signature $(0,18-\rho)$ lattice $(W_2 \cap \Lambda_T)/W_1$. There can be multiple isometry classes 
for a given discriminant form $(G_{(W_2\cap \Lambda_T)/W_1}, q)$ because of negative definite signature.

The nilpotent matrix $N$ for the monodromy matrix $T = \exp[N]$ is determined uniquely. When it is presented 
in the basis (\ref{eq:LambdaT-basis-general}),  
\begin{align}
  N = \mu N^{\rm II}_0(\delta_1, \delta_2) = \mu \times 
\left( \begin{array}{cc|cc|cc}
      & & & & & \\
      & & & & & \\
   \hline
      & & & & & \\
      & & & & & \\
   \hline
      & -\delta_2 & 0 & & & \\
     \delta_1 & & & 0 & & 
      \end{array} \right). 
\label{eq:N0-def-II-pol}
\end{align}
Here, $\mu \in \Z$ is the {\it index}. This $\mu$ is the same as that in (\ref{eq:N0-def-II-nopol}), when we allow 
to choose a basis without respecting the distinction between $\Lambda_S$ and $\Lambda_T$. 

Here are some examples. The first one is for $\Lambda_S=U$, the $E_8$-elliptic K3 surface. In this case, 
there are only two Type II degenerations of $\Lambda_S=U$-polarized K3 surface in the ${\rm Isom}(\Lambda_T)$
classification. $\delta_1 = \delta_2=1$ (obviously because $\Lambda_T$ is self-dual), and 
\begin{align}
   (W_2 \cap \Lambda_T)/W_1 \cong E_8 \oplus E_8 \quad {\rm or} \quad D_{16};\Z_2.
\end{align}

For $\Lambda_S= \vev{+2}$ and $\vev{+4}$ (i.e., degree-2 and quartic K3 surface), there is no choice but 
$\delta_1=\delta_2=1$, and there are four choices 
\begin{align}
  (W_2 \cap \Lambda_T)/W_1 \cong E_8^{\oplus 2}\oplus A_1, \quad (D_{16};\Z_2)\oplus A_1, \quad 
  (E_7\oplus D_{10});\Z_2, \quad A_{17};\Z_3
\label{eq:classification-deg2}
\end{align}
for $\Lambda_S = \vev{+2}$, whereas there are nine choices
\begin{align}
 (W_2 \cap \Lambda_T)/W_1 \cong & E_8^{\oplus 2} \oplus \vev{-4}, \quad D_{16};\Z_2 \oplus \vev{-4}, \quad 
  E_8\oplus D_9, \quad (E_7^{\oplus 2}\oplus A_3);\Z_2,  \nonumber \\ 
 & D_{17}, \quad (D_{12}\oplus D_5);\Z_2, \quad (\vev{-4} \oplus D_8^{\oplus 2});(\Z_2 \times \Z_2), \nonumber \\ 
 & (A_1^{\oplus 2}\oplus A_{15});(\Z_4 \times \Z_2), \quad (E_6 \oplus A_{11});\Z_3.
\label{eq:classification-deg4}
\end{align}
for $\Lambda_S= \vev{+4}$. For the $\rho=1$ cases $\Lambda_S = \vev{+2k}$, $\delta_1 = \delta_2=1$ are the only possibility, if $k$ is not 
divisible by a square of an integer. See \cite{Scattone} for more information.

Similarly, the $\Gamma = {\rm Isom}(\Lambda_T)$-classification of Type III degenerations can be worked out 
as follows. Note first that one can choose a basis 
\begin{align}
  \left\{ \hat{e}', \hat{f}^1,\cdots, \hat{f}^{20-\rho}, \hat{e} \right\}
  \label{eq:LambdaT-basis-general-III}
\end{align}
of $\Lambda_T$ so that 
\begin{align}
  W_0 =& {\rm Span}_\Z \left\{ \hat{e}' \right\}, \nonumber \\
 (W_2 \cap \Lambda_T)/W_0 =& {\rm Span}_\Z \left\{ [\hat{f}^1], \cdots, [\hat{f}^{20-\rho}] \right\}, \nonumber \\
 \Lambda_T/(W_2 \cap \Lambda_T) =& {\rm Span}_\Z \left\{ [\hat{e}] \right\},
\end{align}
and the intersection form of $\Lambda_T$ is given in this basis as 
\begin{align}
 \left(  \begin{array}{c|c|c}
    a & B & \delta \\
 \hline 
    B^T & C & \\
 \hline 
    \delta & & 
  \end{array} \right)
 \label{eq:int-form-III-pol-gen}
\end{align}
for some positive integer $\delta$. Other parts of the intersection form, $a$, $B$, $B^T$ and $C$ are also 
integer valued. 

Once a pair $(\Lambda_S, \Lambda_T)$ is given, one can systematically work out possible values of $\delta$ and isometry classes 
of the lattice $(W_2 \cap \Lambda_T)/W_0$ as follows. First, the discriminant group needs to allow the substructure
\begin{align}
 G_{\Lambda_T} = \Z_\delta . G_{(W_2 \cap \Lambda_T)/W_0} . \Z_\delta .
\end{align}
The $\Z_\delta$ subgroup is isotropic under the discriminant form, and the $(\Z_\delta. G_{(W_2 \cap \Lambda_T)/W_0})$
subgroup is orthogonal to the $\Z_\delta$ subgroup under the discriminant bilinear form $b$.
There are only a finite number of such options for a given $(G_{\Lambda_T}, q_{\Lambda_T}) = (G_{\Lambda_S}, -q_{\Lambda_S})$. 
In the classification of Type III degenerations, the lattice $(W_2 \cap \Lambda_T)/W_0$ has signature 
$(1,19-\rho)$, which is not negative definite. Due to a theorem of Nikulin \cite{Nikulin} (Thm 1.14.2), 
any two even lattices $(W_2 \cap \Lambda_T)/W_0$ that reproduce the same discriminant form $(G_{(W_2 \cap \Lambda_T)/W_0}, q)$ 
are mutually isometric provided that $\rho \leq 11$ \cite{Scattone}.
The fact that the monodromy matrix $T = \exp[N]$ is an isometry translates to the skew-symmetry condition on 
$N$ with respect to the intersection form above. Therefore, 
\begin{align}
 N = \mu N_0^{\rm III}(\delta,u,v,x) = \mu \times \left( \begin{array}{c|c|c}
    & & \\
 \hline 
  \delta v & & \\
 \hline
  \delta x & u^T &
  \end{array}\right), \qquad u = - C \cdot v, \quad 2 \delta \cdot x = - B \cdot v,
\label{eq:N0-def-III-pol}
\end{align}
where $u, v, x$ are assumed to be integral. 
Allowing to choose a basis that does not respect the distinction between 
$\Lambda_S$ and $\Lambda_T$, this $\mu$ here becomes the index $\mu$ in (\ref{eq:N0-def-III-nopol}), and 
$t_0 = v^T \cdot C \cdot v = (v,v)|_{(W_2 \cap \Lambda_T)/W_0}$.

\subsubsection{Baily--Borel compactification}
\label{sssec:BBcpt}

The period domain of a $\Lambda_S$-polarized K3 surface is given by (\ref{eq:period-domain}), 
and the moduli space of $\Lambda_S$-polarized K3 surface is the quotient of this space by 
$\Gamma = {\rm Isom}^*(\Lambda_T)$. This group mods out unphysical marking without touching the 
lattice polarization divisors in $\Lambda_S$.

This moduli space $D(\Lambda_T)/\Gamma$ is not compact. There are multiple different ways to make it compact 
by adding boundary components. The Baily--Borel compactification $\overline{D(\Lambda_T)/\Gamma}$ is a minimal one. 
The boundary components $\overline{D(\Lambda_T)/\Gamma} \; \backslash \; D(\Lambda_T)/\Gamma$ form different strata, each 
of which corresponds to one of the Type II or Type III degenerations of a $\Lambda_S$-polarized K3 surface. 
A stratum corresponding to a Type II degeneration comes with a variety of one complex dimension, while one 
corresponding to a Type III degeneration is a point. This is because $\P[W_1 \otimes \C]$ for Type II is 
of one dimension, while $\P[W_0 \otimes \C]$ for Type III is of zero dimension. 

Multiple strata for Type II degenerations labelled by various choices of $\delta_1$, $\delta_2$ and 
$(W_2 \cap \Lambda_T)/W_1$ can meet at a point (stratum) for a Type III degeneration. 
The structure of such a stratification of the boundary components is studied for $\rho=1$ polarized K3 surfaces 
in \cite{Scattone}. In the case of $\Lambda_S=\vev{+2}$, for example, there is just one Type III stratum and 
four Type II strata (appearing in (\ref{eq:classification-deg2})), and all the four Type II curve strata meet 
at the Type III stratum point.

Other compactifications of the moduli space make it possible to retain more information of a K3 surface at a
degeneration limit \cite{K3-moduli-bdry,Kondo_typeII}. Possibly interesting in the context of heterotic--type IIA duality is the one 
discussed in \cite{Kondo_typeII}, which retains information in the $[(W_2 \cap \Lambda_T)/W_1]\otimes \C$ component of 
the complex structure in the degeneration limit. The heterotic string ``instanton'' moduli can be 
translated into these moduli at the degeneration limit. A version for $[(W_2 \cap \Lambda_T)/W_1] = E_8^{\oplus 2}$
in $\Lambda_S=U$ is well-known in string theory community through \cite{FMW}, but this story may be generalized 
for other $\Lambda_S$ and $(W_2 \cap \Lambda_T)/W_1$-Type II degenerations.

\section{Picard--Lefschetz Monodromy and Collapsing \texorpdfstring{$dP_7$}{Lg}}
\label{sec:transcendental}

One of the simplest forms of degenerations of a K3 surface is for an $A_1$ singularity to be formed. In a local geometry, 
${\cal X} \rightarrow {\rm Disc}$ may be given by 
\begin{equation}
  {\cal X} = \left\{ x^2 + y^2 + z^2 + t = 0 \right\} \rightarrow {\rm Disc} \subset \left\{ t \in \C \right\}. 
\end{equation}
This degeneration at $t = 0$ is not semi-stable, since the fibre at $t=0$ has an $A_1$ singularity, which is not a 
normal crossing singularity. We can turn this into a Kulikov model by a base change followed by a resolution. In the present case, this means 
to replace the coordinate of ${\rm Disc}$ from $t$ to $t=s^2$, and further replace ${\cal X}$ by a small resolution of the conifold singularity 
at $(x,y,z,s)=(0,0,0,0)$. We then arrive at a Kulikov model of Type I. 

In the context of string compactifications, however, we are interested in a \emph{compact} threefold $M$ fibred over 
a compact space $\P^1_A$, instead of ${\cal X} \rightarrow {\rm Disc}$. We are usually not happy to replace 
$M \rightarrow \P^1_A$ by its base change, either. We would rather think of the degeneration above as a ``would-be'' Type I. 

$A_1$-singularities in the fibre, i.e. would-be Type I degenerations, are quite a common phenomenon. In fact, for any 
Calabi--Yau threefold $M$ with a $\Lambda_S= \vev{+4}$-polarized K3-fibration (quartic K3 in the fibre), 
there are 216 such would-be Type I fibres; the topological Euler characteristic of $M$ is understood in a simple way then:
\begin{align}
 \chi(M_{\vev{+4}}) =& 2(h^{1,1}-h^{2,1}) = 2(2-86)=-168,   \nonumber \\
  =&  (\chi(\P^1_A) - 216) \chi(K3) + 216 \times 23,
\end{align}
where the singular fibre of each one of those would-be Type I degenerations has $\chi=23$. The number 
of would-be Type I fibres---216---remains the same for any one of $\nu^6_F$ chosen from 
$2\widetilde{\Delta}_F \cap N_F$. Similarly, for any Calabi--Yau threefold $M$ with $\Lambda_S= \vev{+2}$-polarized 
K3-fibration (degree-2 K3 is in the fibre) discussed in section \ref{sssec:rho+1}, there are 300 
would-be Type I singular fibres. Here is how the counting goes, then:
\begin{align}
 \chi(M_{\vev{+2}}) =& 2(h^{1,1}-h^{2,1}) = 2(2-128)=-252,  \nonumber \\
  =&  (\chi(\P^1_A) - 300) \chi(K3) + 300 \times 23,
\end{align}
which holds for all the choices of $\nu^6_F$ in Figure~\ref{fig:choicesofabc}. The number of would-be Type I 
singular fibres can be determined by using the discriminant of the K3 fibre (similarly to elliptic fibration). 
See \cite{GKZ} for how to compute the discriminant, from which we can derive such values as 216 and 300 above.  

Let $C_p$ be the two-cycle in the K3-fibre that shrinks at a discriminant point $z_p \in \P^1_A$ of a would-be Type I 
singular fibre. The Picard--Lefschetz monodromy $T_p$ on $H_2(S_{t.A};\Z)$ around $z= z_p$ is given by 
\begin{align}
 T_p: H_2(S_{t.A};\Z) \ni x \longmapsto x + (C_p,x) C_p \in H_2(S_{t.A};\Z). 
\end{align}
This is a reflection, 
\begin{align}
(T_p)^2 = {\bf Id}\, ,
\end{align}
and the monodromy becomes trivial after base change.

Consider tuning the complex structure moduli of $M_{\vev{+2}}^{2}$ so we approach the transition point to the 
branch of $M_{\vev{+2}}^{\{2,1\}}$. The hypersurface equation (\ref{eq:E7-tilde}) for the local geometry of $M_{\vev{+2}}^2$ 
at the transition can be deformed to
\begin{align}
 X_1^2 + X_2^4 + X_3^4 + \prod_{i=1}^4(X_6-z_i) = 0
\end{align}
by introducing four parameters $z_i$. When all of the $z_i$'s are set to zero, this hypersurface equation approaches 
(\ref{eq:E7-tilde}) at the transition point. 
At each one of $z_i$'s, this equation is in the form of a deformation of a parabolic singularity $X_9$ \cite{Arnold}.
In the local geometry of the fibre K3 captured in this equation (deformed $X_9$), nine compact two-cycles and 
two non-compact two-cycles are identified \cite{gabrielov}. Seven of them---$\alpha_{1,2,\cdots,7}$---form $E_7$, 
the two remaining compact two-cycles are denoted by $e_{1,2}$, and the two non-compact ones by $e'_{1,2}$.
The intersection form is $(e_i, e'_j)= \delta_{ij}$, $(e_i, e_j)=0$. The singular fibre at a given $z_i$ is regarded 
as nine would-be Type I fibres coming on top of another, and the Picard--Lefschetz monodromy 
$T_i := \prod_{p=1}^9 T_p$ can be computed by using the information in \cite{gabrielov}. The monodromy from all of
the four $z_i$'s combined, $T = \prod_{i=1}^4 T_i$ acts trivially on the $E_7$ part of the two-cycles in the fibre, and on the remaining cycles as 
\begin{align}
 T: \left( \begin{array}{c} e_1 \\ e_2 \\ e'_1 \\ e'_2 \end{array} \right) \longmapsto 
    \left( \begin{array}{cc|cc}
      1&   & & \\
       &1  & & \\
    \hline 
       &-1 &1& \\
      1&   & &1
    \end{array} \right)
    \left( \begin{array}{c} e_1 \\ e_2 \\ e'_1 \\ e'_2 \end{array} \right).
\end{align}
This computation---purely transcendental---reproduces the monodromy matrix for a Type II degeneration 
$T=\exp[N^{\rm II}_0(1,1)]$. The $E_7$ sublattice of $dP_7$ in the central fibre of the degeneration 
in $M_{\vev{+2}}^{\{2,1\}}$ has also been captured. 

The Type II degeneration of a degree-2 K3 surface, with $(W_2 \cap \Lambda_T)/W_1 \cong (E_7 \oplus D_{10});\Z_2$, 
can be regarded as a certain limit of the complex structure where $4 \times 9 = 36$ would-be Type I degenerations 
come on top of each other.  With the remaining $(300-36=264)$ would-be Type I degenerations, the topological 
Euler characteristic of $M_{\vev{+2}}^{\{2,1\}}$ can be understood as 
\begin{align}
 \chi(M_{\vev{+2}}^{\{2,1\}}) =& 2(h^{1,1}-h^{2,1}) = 2(3-111)=-216,  \nonumber \\
  =&  (\chi(\P^1_A) - 264 -1) \chi(K3) + 264 \times 23 + 1 \times \chi(V_0 \cup V_1).
\end{align}
Here, the contribution of the degenerate fibre is $\chi(V_0 \cup V_1)=24$, the same as a generic K3 fibre.

\section{Technical Details of the Degeneration of the Degree-2 K3 Surface in \texorpdfstring{$M_{\vev{+2}}^{\{n,n-1\}}$}{Lg}}
\label{ssec:E7D10-CS}

\subsection{Geometry of Degenerate Central Fibre in \texorpdfstring{$M_{\vev{+2}}^{\{n,n-1\}}$}{Lg}}

Let $S_0$ be the central fibre in a Type II semi-stable degeneration of a degree-2 K3 surface found 
in $M_{\vev{+2}}^{n,n-1}$. The central fibre $S_0$ consists of two irreducible components. Let us 
use the notation $V_0 = \bar{D}_{6,n}$ and $V_1 = \bar{D}_{6,n-1}$.

The surface $V_1$ has the property 
\begin{align}
 h^{i,0}(V_1) = 0 \quad (i=1,2), \qquad h^{1,1}(V_1) = 8,
 \label{eq:V1-property}
\end{align}
which we found by using the techniques of \cite{DK} (and an additional formula (\ref{eq:h11-formula})) available 
for divisors of a toric hypersurface. The restriction of divisors of $M_{\vev{+2}}^{\{n,n-1\}}$ to $\bar{D}_{6,n-1}$, 
the first line of (\ref{eq:h11-formula}), accounts for only 1 generator within 
$H^{1,1}(V_1)$ and we can take $\bar{D}_{6,n}|_{V_1}$ as this generator.\footnote{For $M_{\vev{+2}}^{\{n,n-1\}}$, rational 
equivalence relations are $\bar{D}_1|_{V_1} \sim 2 \bar{D}_2|_{V_1}$ and 
$\bar{D}_2|_{V_1} \sim \bar{D}_3|_{V_1} \sim \bar{D}_{6,n}|_{V_1}$.}
The double curve $C = \bar{D}_{6,n}|_{V_1}$ has $(\bar{D}_{6,n}|_{V_1})^2 = +2$ in this surface. 
This curve class is also the restriction of the polarization divisor $\bar{D}_2$ of the generic K3 surface 
(see footnote).
 
The surface $V_1 = \bar{D}_{6,n-1}$ is a $dP_7$. To see this, note first that it is a hypersurface of 
$W\P^3_{[2:1:1:1]}$, because of the relation 
\begin{align}
 2\nu^1 + \nu^2+\nu^3 + \nu^{6,n}=\nu^{6,n-1}
\end{align}
among the toric vectors. Secondly, $V_1$ belongs to the class 
\begin{align}
 \left(\sum_i D_i\right)|_{D_{6,n-1}} \sim (D_1+D_2 + D_3)|_{D_{6,n-1}} \sim \left[(4) \subset W\P^3_{[2:1:1:1]}\right].
\end{align}

The cohomology group $H^2(V_1; \Z)$ is a unimodular lattice. It is an index-2 overlattice of 
\begin{align}
 \vev{+2} \oplus E_7,
\end{align}
where the rank-1 lattice $\vev{+2}$ is generated by $C = \bar{D}_{6,n}|_{V_1}$. There are elements of 
$H^2(dP_7; \Z)$ that correspond to a $\Z_2$ subgroup of the discriminant group $\Z_2 \times \Z_2$ of 
the lattice above ($H^2(dP_7;\Z)$ is not an even lattice, however). 

Let us now turn our attention to $V_0$. Blowing down this irreducible component $V_0$ in $M_{\vev{+2}}^{\{n,n-1\}}$, 
we obtain a threefold $M_{\vev{+2}}^{n-1}$ with a singularity which may be deformed so that we find a smooth 
Calabi-Yau threefold $M_{\vev{+2}}^{n-1}$ branch. The singularity in $M_{\vev{+2}}^{n-1}$
right after this transition is given by 
\begin{equation}\label{eq:D6,3}
 X_1^2 + X_4^2 F^{(4)} G^{(2)} + X_4 X_{6} F^{(5)} G^{(4-n)} + X_{6}^2 F^{(6)} G^{(6-2n)} \simeq 0 \, ;
\end{equation}
$F^{(d)}$'s and $G^{(d)}$'s are homogeneous functions of $[X_2:X_3]$ and $[X_5:X_6]$, respectively, with 
the degree specified in the superscript. The singular locus is along the curve $X_{6} = X_1 = X_4 = 0$, 
which is a $\P^1_{[X_2:X_3]}$. The $A_1$ singularity in the directions transverse to this curve gets worse 
at 10 points in this $\P^1$; that is where the Hessian of the quadratic form in $(X_4, X_{6})$ degenerates. 
The divisor $V_0 = \bar{D}_{6,n}$ in $M_{\vev{+2}}^{\{n,n-1\}}$ before the blow-down is obtained as the exceptional 
locus of this singularity. 

A computation using toric techniques (as in $V_1$) indicates that 
\begin{align}
 h^{i,0}(V_0) = 0 \quad (i=1,2), \qquad h^{1,1}(V_0) = 12.
\end{align}
Two generators of $H^{1,1}(V_0)$ are realized as restriction of toric divisors in $M_{\vev{+2}}^{\{n,n-1\}}$ 
(the first line of (\ref{eq:h11-formula}) and we can use $\bar{D}_{6,n-1}|_{V_0}$ and $\bar{D}_3|_{V_0}$ for now.\footnote{
There are three rational equivalence relations among restriction of the five toric divisors in 
$M_{\vev{+2}}^{\{n,n-1\}}$: 
$\bar{D}_3|_{V_0} \sim \bar{D}_2|_{V_0}$, 
$\bar{D}_4|_{V_0} \sim \bar{D}_{6,n-1}|_{V_0}+D_3|_{V_0}$ and 
$\bar{D}_1|_{V_0} + 2 \bar{D}_{6,n-1}|_{V_0} \sim 3 \bar{D}_4|_{V_0}$. }

The complete linear system of the divisor $\bar{D}_3|_{V_0}$ can be used to construct a projection 
$\Phi_{|\bar{D}_3|_{V_0}}: V_0 \longrightarrow \P^1$. This $\P^1$ can be identified with the curve of 
$A_1$ singularities in $M_{\vev{+2}}^{n-1}$; $[X_2:X_3]$ is the homogeneous coordinate of this $\P^1$ and
$\bar{D}_3|_{V_0} \sim \bar{D}_2|_{V_0}$ is the fibre class in this projection. This fibre 
is generically a conic in $\P^2_{[X_1:X_4:X_{6,n-1}]}$. Both of the divisor classes $\bar{D}_4|_{V_0}$ and 
$\bar{D}_{6,n-1}|_{V_0}$ are 2-sections in the fibration corresponding to the projection, they differ only by the fibre class. The fibre 
conic degenerates into $\P^1+\P^1$ whenever the Hessian degenerates. The 2-section $\bar{D}_{6,n-1}|_{V_0}$ 
intersects once with one $\P^1$ and also once with the other $\P^1$ in such singular fibres. Let one of 
those two $\P^1$'s be $E_i^\pm$ ($i=1,\cdots, 10$ labels singular conic fibres and $\pm$ distinguishes the 
two components). We have that $\bar{D}_3|_{V_0} \sim E_i^+ + E_i^-$ for any $i$. From $(\bar{D}_3|_{V_0})^2 = 0$ 
and $E_i^+ \cdot E_i^- = \delta^{ij}$ it follows that $(E_i^\pm)^2 = -1$. Then the intersection form of $V_0$ 
in the basis  $(\bar{D}_{6,n-1}|_{V_0}$, $\bar{D}_3|_{V_0}$, $E_1^+, \cdots, E_{10}^+)$ is in the form 
\begin{align}
 \left( \begin{array}{cc|ccc}
    -2 & 2 & 1 & \cdots & 1 \\
    2 &    &   &        &   \\
    \hline
    1 &    & -1 &       &   \\
    \vdots & &  & \cdots &   \\
    1 &    &    &       & -1 
   \end{array} \right). 
\end{align}
The discriminant of this intersection form is $- 4$. Hence the unimodular lattice $H^2(V_0; \Z)$ must be 
an index-2 overlattice of the lattice generated by the basis above. The intersection form 
above indicates that there must be an element that is topologically regarded as $\frac{1}{2} \bar{D}_3|_{V_0}$ 
The basis above, with $\bar{D}_3|_{V_0}$ replaced by $\frac{1}{2}\bar{D}_3|_{V_0}$ can be regarded 
as a generator set of $H^2(V_0; \Z)$.

This surface $V_0$ is rational, because $V_0$ ends up with a Hirzebruch surface $F_2$ after blowing down 
all the $E_i^+$'s. The double curve $C = \bar{D}_{6,n-1}|_{V_0}$ has self-intersection $(-2)$ in $V_0$.
The polarization divisor $\bar{D}_2$ of a generic fibre K3 surface is restricted on this surface to be 
$\bar{D}_2|_{V_0} \sim \bar{D}_3|_{V_0}$.

The intersection form can be presented in any choice of basis one likes; we do so as preparation for 
study in Clemens--Schmid exact sequence later. 
When we choose $\bar{D}_{6,n-1}|_{V_0}$, $(\bar{D}_{6,n-1} + \bar{D}_3)|_{V_0}$, $\bar{D}_3|_{V_0} - E_1^+-E_2^+$ 
and $(E_i^+-E_{i+1}^+)$ $(i=1,\cdots, 9)$ as a set of generators, it becomes
\begin{align}
 \left( \begin{array}{cc|cccccc}
   -2 & & & & & & & \\
      & 2 & & & & & & \\
\hline 
      & & -2 & & 1 & & & \\
      & & & -2 & 1 & & & \\
      & & 1 & 1 & -2 & 1 & & \\
      & & & & 1 & \cdot & \cdot & \\
      & & & & & \cdot & -2 & 1 \\
      & & & & & & 1 & -2
     \end{array} \right).
\end{align}
This intersection form is that of the lattice $\vev{-2}\oplus \vev{+2} \oplus D_{10}$. 
The unimodular lattice $H^2(V_0; \Z)$ should be an index-4 overlattice of this one.

\subsection{Cohomology and Homology Groups of the Central Fibre}

Homology and cohomology groups of the central fibre, $S_0 = V_0+V_1$, can be determined by 
using the Mayer--Vietoris spectral sequence. Let us begin with the homology groups. 

First of all, $H_4(S_0;\Z) \cong H_4(V_0;\Z) \oplus H_4(V_1;\Z) \cong \Z \oplus \Z$. The filtration 
$H_4(S_0;\Z) = W_{-4} \supset W_{-5} = \{0\}$ corresponds to the convention in \cite{Mor-CS}.

On $H_2(S_0; \Z)$, the spectral sequence introduces a filtration 
\begin{align}
 \{0 \} = W_{-3} \subset W_{-2} \subset W_{-1} = H_2(S_0; \Z); 
\end{align}
where $W_{-2}$ is obtained by identifying the double curve $E_{2,1}$ in $H_2(V_0;\Z)$ and $H_2(V_1;\Z)$;
\begin{align}
  W_{-2}(H_2(S_0;\Z)) \cong \left[E_7 \oplus (\vev{+2} \oplus D_{10}); \Z_2 \oplus \Z \vev{[C]}\right]
     ;\Z_2 \times \Z_2.
\end{align}
The $W_{-1}/W_{-2}$ part, which is isomorphic to $H_1(C; \Z) \cong \Z^{\oplus 2}$, are the two-cycles 
that are obtained by gluing discs in $V_0$ and $V_1$ along $\alpha$ or $\beta$ cycle in the double 
curve $C$.

The cohomology groups of the central fibre are also worked out similarly. 
We have $H^4(S_0;\Z) \cong H^4(V_0;\Z) \oplus H^4(S_0;\Z) \cong \Z \oplus \Z$ and the
convention on the filtration in \cite{Mor-CS} is to take $\{0 \} = W_3 \subset W_4 = H^4(S_0;\Z)$.

The filtration in $H^2(S_0;\Z)$ is 
\begin{align}
 \{ 0 \} \subset W_1 \subset W_2 = H^2(S_0;\Z).
\end{align}
The $W_1 \cong H^1(C;\Z)\cong \Z^{\oplus 2}$ part vanishes on $W_{-1} \subset H_2(S_0;\Z)$. 
$W_2/W_1$ is identified with the subspace of $H^2(V_0;\Z) \oplus H^2(V_1;\Z)$ that evaluates the double 
curve class $[C]$ in $H_2(V_0; \Z)$ and $H_2(V_1;\Z)$. That is, 
\begin{align}
 W_2/W_1 \cong  \left( E_7 \oplus \left[ \vev{+2} \oplus D_{10} \right];\Z_2 \oplus
     \Z\vev{\bar{D}_{6,n}|_{V_1} - \bar{D}_{6,n-1}|_{V_0}} \right) ; \Z_2 .
\label{eq:G2-H2-central-fib}
\end{align}
%

\subsection{Clemens--Schmid Exact Sequence}

Let us use the technical results obtained earlier in this appendix and see how the
material fits into the Clemens--Schmid exact sequence (\ref{eq:CS-sequence}) to confirm that the 
degeneration in $M_{\vev{+2}}^{\{n,n-1\}}$ corresponds to the $(W_2 \cap \Lambda_T)/W_1 
= [E_7\oplus D_{10}];\Z_2$ case in the classification of (\ref{eq:classification-deg2}).

Following the definition of the morphisms $\beta$ and $\alpha$ explained in \cite{Mor-CS}, one 
finds that 
\begin{align}
\beta: H^0(S_t;\Z) \ni 1 \longmapsto [V_0]+[V_1] \in H_4(S_0;\Z)  
\end{align}
and 
\begin{align}
  \alpha([V_0]) = \alpha(- [V_1]) = (\bar{D}_{6,n}|_{V_1}, \; -\bar{D}_{6,n-1}|_{V_0}) \in H^2(S_0; \Z)
\end{align}
in the first line of (\ref{eq:CS-sequence}). Thus, the exact sequence (\ref{eq:CS-sequence}) in the 
first line, 
\begin{equation}
 \vcenter{\xymatrix{
    0 \ar[r] & W_0(H^0(S_t))^{[1]} \ar[r]^{\beta} & W_{-4}(H_4(S_0))^{[2]} \ar[r]^{\alpha}
 & W_2/W_1(H^2(S_0))^{[19]}  \\
             & & 0 \ar[r]^\alpha & W_1(H^2(S_0))^{[2]}     
 }}
\end{equation}
continues to the second line, with the $W_1$ part of $H^2(S_0)$ and the $\left[E_7 \oplus 
\left[ \vev{+2} \oplus D_{10} \right];\Z_2\right];\Z_2$ factor in (\ref{eq:G2-H2-central-fib}). 
Here, the superscript $^{[n]}$ is the rank (dimension) of a given space. 

The morphism $\alpha: H_2(S_0;\Z) \rightarrow H^4(S_0;\Z)$ in the second line of (\ref{eq:CS-sequence}) 
vanishes on $E_7 \oplus (\vev{+2} \oplus D_{10});\Z_2$ that are orthogonal to the double curve $C$.
\begin{align}
 \alpha:  W_{-2}(H_2(S_0;\Z)) \ni [C] \mapsto (-2 [1_{V_1}], \; +2 [1_{V_0}]) \in H^4(S_0;\Z).  
\end{align}
The cokernel of this map generated by $(1_{V_1},0) \sim (0,1_{V_0})$ is isomorphic to $H^4(S_t;\Z)$ under $i^*$. 

The heart of the Clemens--Schmid exact sequence is this.
\begin{align}
 \vcenter{\xymatrix{
  H^2(S_0;\Z) \ar[r]^{i^*} & H^2(S_t;\Z) \ar[r]^{N} & H^2(S_t;\Z) \ar[r]^\beta & H_2(S_0;\Z) \\
   & (W_3/W_2)^{[2]} = {\rm PD}(E) \ar[rdd] & (W_3/W_2)^{[2]} = {\rm PD}(E) \ar[rd] & \\
 (W_2/W_1)^{[19]} \ar[r] & (W_2/W_1)^{[18]} & (W_2/W_1)^{[18]} \ar[rd] & (W_{-1}/W_{-2})^{[2]} \\
 W_1^{[2]} \ar[r] & W_1^{[2]} = {\rm PD}(E') & W_1^{[2]} = {\rm PD}(E') & W_{-2}^{[19]}
}}
\end{align}
Here, $E' := {\rm Span}_\Z\{e'_1, e'_2\}$ is a rank-2 space of those two-cycles of $S_{t.A}$ where the period integral 
of $S_{t.A}$ near the degeneration limit dominates. The limit of the two-cycles $e'_{1,2}$ in $S_0$ are in the 
form of a pair of discs in $V_0$ and $V_1$ glued along a one-cycle in the double curve $C$. 
$E:={\rm Span}_\Z\{ e_1, e_2\}$, on the other hand, is a rank-2 space of two-cycles of $S_{t.A}$, where 
the two-cycles $e_{1,2}$ become topologically trivial in $V_0$ and in $V_1$ in the degeneration limit.
${\rm PD}$ stands for Poincar\'e duality. 
Both ${\rm Coker}\left( \alpha: W_{-4}(H_4(S_0)) \rightarrow (W_2/W_1)(H^2(S_0)) \right)$ and 
${\rm Ker}\left( \alpha: W_{-2}(H_2(S_0)) \rightarrow W_4(H^4(S_0)) \right)$ are the same and give
\begin{align}
  (W_2/W_1)(H^2(S_t;\Z)) \cong \left( E_7 \oplus (\vev{+2} \oplus D_{10});\Z_2 \right) ; \Z_2.
\end{align}
The matrix $N$ is essentially in ${\rm Hom}(W'_1, W_1)$, where $W'_1\cong {\rm PD}(E)$.

The $\vev{+2}$ part of this $W_2/W_1$ should be regarded as the Neron--Severi lattice of the 
degree-2 K3 surface. So, 
\begin{align}
 (W_2 \cap \Lambda_T)/W_1 \cong \left( E_7 \oplus D_{10} \right) ; \Z_2 .
\end{align}

\end{document}